\documentclass[12pt,preprint]{aastex} 
\usepackage{epsfig}

\usepackage{amsmath}
\usepackage{url}
\usepackage{txfonts}
\usepackage{natbib}
\usepackage{graphicx}
\usepackage{longtable}
\usepackage{fltpage}

\newcommand{\obsWithXMM}{178}
\newcommand{\obsWithCXO}{213}

\newcommand{\MsecObsTime}{8.6 Msec}
\newcommand{\psrTempUL}{263}
\newcommand{\psrLxUL}{262}

\newcommand{\identified}{20}

\shorttitle{{A Search for X-ray Counterparts of Pulsars}}
\shortauthors{Tobias Prinz, Werner Becker}

\begin{document}

\title{A Search for X-ray Counterparts of Radio Pulsars}
\author{T. Prinz\altaffilmark{1} and W. Becker\altaffilmark{1,2}}

\altaffiltext{1}{Max-Planck Institut f\"ur extraterrestrische Physik, Giessenbachstrasse 1, 85741 Garching, Germany}
\altaffiltext{2}{Max-Planck Institut f\"ur Radioastronomie, Auf dem H\"ugel 69, 53121 Bonn, Germany}

\date{Submitted for publication to ApJ on 2015 November 14. / revised and updated 1. Juli 2016}

\begin{abstract}
 We describe a systematic search for X-ray counterparts of radio pulsars. The
search was accomplished by cross-correlating the radio timing positions of all
radio pulsars from the ATNF pulsar database (version 1.54) with archival
XMM-Newton and Chandra observations publicly released by July 1$^{st}$~2016. In
total, 178 of the archival XMM-Newton observations and 213 of the archival
Chandra datasets where found to have a radio pulsar serendipitously in the field
of view. From the 288 radio pulsars covered by these datasets we identified
\identified~previously undetected X-ray counterparts. For 6 of them the
statistics was sufficient to model the energy spectrum with one- or
two-component models. For the remaining new detections and for those pulsars for
which we determined an upper limit to their counting rate we computed the energy
flux by assuming a Crab-like spectrum.  Additionally, we derived upper limits on
the neutron stars' surface temperature and on the non-thermal X-ray efficiency
for those pulsars for which the spin-down energy was known. The temperature 
upper limits where compared with predictions from various neutron star cooling 
models and where found to be in agreement with the minimal cooling paradigm. 
\end{abstract}


\section{Introduction}

 Neutron stars (NSs) are the most dense objects accessible in the universe by direct 
 observations. They represent unique astrophysical laboratories which allow us to
 explore the properties of matter under the most extreme conditions observable
 in nature. Studying neutron stars is therefore an interdisciplinary field, where 
 astronomers and astrophysicists work together with a broad community of physicists. 
 Particle, nuclear and solid-state physicists are strongly interested in the internal
 structure of neutron stars which is largely uncertain because of the unknown nuclear 
 interaction potential at matter densities above the nuclear density 
 $\rho_{\rm nuc} = 2.8\times 10^{14} \mbox{g cm}^{-3}$.  The structure of the star is 
 usually described by the equation of state (EOS) which is a set of non-linear 
 coupled differential equations respecting general relativity and linking physical 
 parameters like star composition, pressure and density as well as the 
 energy balance, radiative energy transport, opacity and luminosity to each other.
 
 Neutron stars are formed at very high temperatures, $\sim$$10^{11}$\,K, in the
imploding core of supernova explosions. Much of the initial thermal energy is
radiated away from the interior of the star by various processes of neutrino
emission (mainly, Urca processes and neutrino bremsstrahlung), leaving a
one-day-old neutron star with an internal temperature of about
$10^9$--10$^{10}$\,K. After $\sim 100$\,yr (typical time of thermal relaxation),
the star's interior is supposed to become nearly isothermal. The details of
neutron star cooling, however, depend very strongly on the chemical composition
and the physical behavior of the super-dense neutron star matter, described by
its EOS.  A very promising approach to constrain that EOS is therefore to look
at the thermal evolution of neutron stars by comparing theoretical predictions
from various cooling models (i.e.~various EOSs) with measured surface
temperatures \citep[e.g.][and references therein]{2009ASSL..357...91B}. In this 
scenario, \citet{2004ApJS..155..623P} introduced
the minimal cooling paradigm, which extends and replaces the so-called standard
cooling scenario \citep{1998PhR...292....1T} to include neutrino emission from
the Cooper pair breaking and formation process for temperatures close to the
critical temperature of superfluid pairing.

 As far as non-thermal emission processes in neutron stars are concerned, several
attempts have been made in recent years to explore the fraction of spin-down
energy which is emitted into the X-ray band. With every new X-ray observatory
and improvement in sensitivity this ratio was reinvested. Often, different authors
used different approaches and different energy bands which makes it difficult to 
compare the various results available in the literature. Different point-spread 
functions and energy responses used in the various X-ray telescopes sample more 
or less of an eventually existing diffuse compact pulsar wind nebula. This adds 
to the complexity of this problem too.

\citet{1988ApJ...332..199S} used the Einstein observatory to explore the X-ray 
effeciency of rotation-powered pulsars. Based of only a handful of mostly young 
and powerful pulsars they found $L_\text{X}(0.2-4\,\text{keV}) \approx 10^{1.39} 
\dot E$. With ROSAT data and on a much larger dataset, which for the first time 
included a variety of rotation-powered pulsars at various spin-down ages and energies,
\citet{1997A&A...326..682B} fitted $L_\text{X} (0.1-2.4\,\text{keV}) \approx
10^{1.03} \dot E$. Adopting the soft energy spectra measured by ROSAT and
assuming that those spectra continue unchanged up to 10 keV,
\citet{2002A&A...387..993P} derived $L_\text{X} (2-10\,\text{keV}) \approx
10^{1.34} \dot E$. Results from XMM-Newton and Chandra soon showed that the
emission in the soft and hard X-ray band is often based on different emission
processes and hence have different efficiencies, invalidating the X-ray
efficiencies derived by \citet{2002A&A...387..993P}. The most recent results
which are based on XMM-Newton and Chandra data yield
$L_\text{X}(2-10\,\text{keV}) \approx 10^{0.92} \dot E$
\citep{2008ApJ...682.1166L} and $L_\text{X}(0.1-2\,\text{keV}) \approx
10^{0.997} \dot E$ \citep{2009ASSL..357...91B} for the hard and soft X-ray band,
respectively.

 Thanks to the collecting power and sensitivity of todays X-ray satellites the
sample of rotation-powered pulsars detected in X-rays is steadily increasing. At
the end of the ROSAT mission 33 rotation-powered pulsars were detected
\citep[e.g.][]{1997A&A...326..682B, 1999A&A...341..803B}. By July 1$^{st}$~2016 
and about 16 years after the launch of Chandra and XMM-Newton we count 188 X-ray 
detected neutron stars which are reported in the literature. 123 pulsars are 
detected in the $\gamma$-regime of which 73 are seen in both wavelength bands. Their 
emission properties are seen to scale with their spin-down age $\tau=P/(2\dot P)$. 
Among the X-ray detected pulsars 23 turn out to be Crab-like (age $\tau \leq 10^4~\text{yrs}$), 
41 Vela-like ($10^4~\text{yrs} < \tau \leq 10^5~\text{yrs}$), 17 are classified as
cooling neutron stars ($10^5~\text{yrs} \leq \tau \leq 10^6~\text{yrs}$), 13 are so 
called old field pulsars ($10^6~\text{yrs} \leq \tau \leq 10^8~\text{yrs}$) and 62 are
millisecond pulsars (MSPs, $\tau \geq 10^8~\text{yrs}$ and $P \leq 20$ ms). Of the X-ray 
detected pulsars 17 are in the category of Anomalous X-ray Pulsars (AXPs) or Soft 
Gamma-ray Repeaters (SGRs) and 7 are classified as Central Compact Objects (CCOs).  
A high-energy overview of the soft $\gamma$-ray pulsar population was given
recently by \citet{2015MNRAS.449.3827K}.

 Whereas with Einstein, ROSAT, ASCA  and BeppoSax mostly the brighter pulsars
and those which are seen under a favorable beaming geometry got detected, we 
now have also apparently less efficient X-ray emitting pulsars in the sample.  
E.g.~PSR J2022+3842 has an X-ray efficiency $\mu_\text{X}=L_\text{X}/\dot E=6 \times
10^{-5}$ in the 2--10 keV range \citep{2011ApJ...739...39A} and
\citet{2012ApJS..201...37K} found that $\mu_\text{X}$ of PSR J0940--5428 and PSR
J1913+1011 is below $4.0\times 10^{-6}$ and $3.3\times 10^{-6}$ in the 0.5--8
keV energy range, respectively. Whether these pulsars are intrinsically less
efficient X-ray emitters because of differences in the emission process or
because of an unfavorable beaming geometry is still unclear. Further
indications of apparently underluminos X-ray pulsars is therefore of high
importance.

 In this paper we report on X-ray counterpart searches for radio pulsars. The
search was accomplished by correlating the ATNF Pulsar Catalogue 
\citep[version 1.54,][]{2005AJ....129.1993M}\footnote{\url{
http://www.atnf.csiro.au/research/pulsar/psrcat/}} with the XMM-Newton and
Chandra data archives to identify those observations which had a radio pulsar
serendipitously in the field of view. The structure of the paper is as follows:
in section \ref{sec:obs} we describe the XMM-Newton and Chandra observations
used in our correlation and provide the details of the data processing, source
detection as well as counting rate upper-limit determination for undetected
pulsars. The spatial and  spectral analysis of the detected pulsars as well as
flux estimates for those pulsars for which we just could determine a count rate
upper limit are described in section \ref{sec:results_psrSearch}. A discussion of
the results and a concluding summary is given in section
\ref{sec:discussion_psrSearch} and \ref{sec:summary_psrSearch}, respectively.

\section{Observations and data reduction}\label{sec:obs}

 In the past 16 years, Chandra\footnote{\url{http://cda.harvard.edu/chaser/}}
and XMM-Newton\footnote{\url{http://xmm.esac.esa.int/xsa/index.shtml}} have
proven to be extremely reliable and successful observatories
(e.g.~\cite{2014RPPh...77f6902T}, \cite{2012MmSAI..83...97S}). Both missions
release their data to a public archive after a one year proprietary time. End of
2013 the XMM-Newton observatory covered already about 877 deg$^2$ of the sky
\citep{2013yCat.9044....0X}. For Chandra the sky coverage is predicted to be
about 550 deg$^2$ by the end of 2015 (cf.~The Chandra Source 
Catalog\footnote{\url{http://cxc.harvard.edu/csc2/preliminary/}} Rev.~2, 
planned to be released in 2016, B.~Wilkes, priv.~com.)
 Many of these observations where targeted at sources located in the galactic
plane, where the majority of radio pulsars is located. The chance probability
that a pulsar has been covered serendipitously in one of these observations is
therefore quite high. We made use of all pointed observations publicly
available in these archives by July 1$^{st}$~2016 and searched for datasets which
had one of the $\sim$2500 known radio pulsars listed in the ATNF Pulsar Catalogue
\citep[version 1.54,][]{2005AJ....129.1993M} in the field of view. Pulsars which
were previously already detected in X-rays where neglected in our search. In
total we found \obsWithXMM~XMM-Newton datasets and \obsWithCXO~Chandra datasets
which had the position of a pulsar previously undetected in X-rays covered. All
these datasets sum up to an on-axis observing time of \MsecObsTime. A summary of
these data, listing instruments, observing modes and filters, off-axis angels
of pulsar locations as well as on- and off-axis exposure times of the relevant
observations is given in Table \ref{tab1:used_observations}. 

 For the reduction of XMM-Newton data we used the XMM-Newton Science Analysis
 Software (SAS), version 15.0.0. In case of the Chandra data we used standard
 processed level-2 data and the Chandra Interactive Analysis of Observations
 (CIAO) software, version 4.8.2. In both cases we had to remove flares from the
 observation. As for XMM-Newton these flares are mostly coming from soft protons
 hitting the detector. To identify and remove these time intervals of high particle
 background we created light curves from MOS1/2 and PN event files at energies 
 above 10 keV and of bin size 100s. We detected flares in almost all XMM-Newton 
 data sets and rejected time intervals in which the count rate per bin was 
 $3\sigma$ higher than the mean count rate without flares. Regarding Chandra 
 observations we made use of the CIAO \sffamily deflare\normalfont~script, 
 which applies the criterion mentioned above by default.

 The effective exposure time $t(\theta)$ ($\theta$ labels the off-axis angle a
pulsar was observed under) was obtained by generating an exposure map with the
SAS-tool \sffamily eexpmap\normalfont~or the CIAO-tool \sffamily
fluximage\normalfont~after correcting for times with high background and dead
time correction. In both cases the task takes into account the telescope
vignetting at the respective off-axis angle. The net exposure on the source
$t(\theta)$ and on the background region $t_\text{BG}(\theta)$ was then
calculated by averaging the exposure map over the source and background region,
respectively. This is required to include the effect of the CCD gaps on the
exposure time. The on-axis exposure $t_\text{on-axis}$ and $t(\theta)$  are both
listed in Table \ref{tab1:used_observations}. 

\subsection{Spatial and image analysis}

 In order to search for X-ray sources in the selected datasets we applied a sliding box source 
 detection algorithm for XMM-Newton data (SAS-tool \sffamily edetect\_chain\normalfont) using 
 the five standard bands $(0.2-0.5$ keV, $0.5-2$ keV, $2-4.5$ keV, $4.5-7.5$ keV, and
 $7.5-12$ keV). For Chandra data we applied a wavelet source detection algorithm (CIAO-tool 
 \sffamily wavdetect\normalfont) in the energy range 0.1--10 keV. Both procedures turn out
 to be standard for the source detection in XMM-Newton and Chandra data. We accepted a source 
 to be detected if its significance was $\ge5\sigma$ above the background.
 
 The angular separation in arcseconds between the radio timing position of a pulsar and the
 centroid of the nearest X-ray source were compared. An X-ray source was considered to be a 
 potential counterpart of a radio pulsar if the angular distance $\Delta$ between them was
 
 \begin{equation}\label{eq:delta}
 \Delta = \sqrt{ (\Delta RA\cdot\cos(DEC_\text{X}))^2 + \Delta DEC^2} \leq 3\sigma.
 \end{equation}

 Here $\Delta RA=RA_\text{R}-RA_\text{X}$ and $\Delta DEC=DEC_\text{R}-DEC_\text{X}$. 
 ($RA_\text{X}$, $DEC_\text{X}$) and ($RA_\text{R}$, $DEC_\text{R}$) denote the positions
 of the X-ray source and the radio pulsar's timing position, respectively.

  The uncertainty $\sigma$ is obtained by deducing the error in $\Delta$ applying error propagation:

 \begin{equation}\label{eq:delta_error}
 \sigma = \frac{\sqrt{ (\Delta RA \cos(DEC_\text{X}))^2(\delta RA_\text{R}^2 + \delta RA_\text{X}^2) + 
 \Delta DEC^2 \delta DEC_\text{R}^2+
 (\Delta DEC - \Delta RA \sin(DEC_\text{X}))^2 \delta DEC_\text{X}^2 
 }}{\Delta}
 \end{equation}

 The uncertainties of the X-ray source positions in right ascension and
declination  $\delta RA_X, \delta DEC_X$ were determined by combining the
statistical position errors and the absolute astrometric accuracy of the
corresponding X-ray observatory squared. For Chandra ACIS-S, ACIS-I and HRC-S
data the 68\% confidence level is $\approx 0.21''$,  $\approx 0.4''$ and
$\approx 0.36''$, respectively\footnote{See
\url{http://cxc.harvard.edu/cal/ASPECT/celmon/} for more details about Chandra
absolute astrometric accuracy.}. The absolute astrometric accuracy for
XMM-Newton data is $\approx 2''$ (r.m.s.)\footnote{More information can be found
at \url{xmm.vilspa.esa.es/docs/documents/CAL-TN-0018.pdf}.}. We note that the
uncertainties in the pulsars' radio timing position $\delta RA_R, \delta DEC_R$
are generally at the level of milli-arcseconds. Thus, the dominant source of
uncertainty in the X-ray counterpart correlation is the error in the X-ray source
position. 

 To check whether an X-ray source has an optical counterpart we correlated all X-ray sources 
 with the USNO-B catalog. It covers the whole sky and is thought to be complete down to a 
 visible magnitude of $V=21$ \citep{2003AJ....125..984M}.  The correlation was done using the 
 online tool VizieR\footnote{\url{http://vizier.u-strasbg.fr/viz-bin/VizieR}} which also allowed 
 us to constraint the X-ray-to-visual flux ratio \citep{1988Maccacaro}  
 \begin{equation} \label{eq:logfx}
 \log(f_\text{X}/f_\text{V})=\log(f_\text{X}) + V/2.5 + 5.37.                            
 \end{equation}
 for each of the detected sources. Here $f_\text{X}$ is the X-ray flux in the
0.3--3.5 keV band in erg cm$^{-2}$ s$^{-1}$. If no optical source could be
detected within an error circle of radius $3\sigma$ we took the limiting
magnitude of the USNO-B catalog as an estimate for $V$. Within the $3\sigma$ 
error box around a potential pulsar X-ray counterpart no optical source
was found. 

 In order to compute the probability $P_\text{coin}$ of a chance identification with a background
 source
 \begin{equation} \label{eq:pcoin}
 P_\text{coin}=\frac{N_\text{X}}{l_\text{RA} l_\text{Decl.}}\times \pi\,\,\, \delta RA_R \,\,\, \delta DEC_R
 \end{equation}
 we computed the source density in the field of view and multiplied it with the area of the pulsar's 
 error ellipse, where $N_\text{X}$ is the number of sources detected within the field of view of 
 area $l_\text{RA} \times l_\text{Decl.}$ To exclude a by-chance identification of an X-ray source 
 with the X-ray counterpart of a radio pulsar we set the criteria $P_\text{coin} \le 3 \cdot 10^{-3}$.  

\subsection{Spectral analysis}\label{sec:spectral_analysis}

 Up to the High-Resolution Channel Plate (HRC) detector all focal plan instruments 
 aboard Chandra and XMM-Newton provide spectral information. However, the quality of 
 spectral fits and the constraints on the spectral model parameters are strongly dependent
 on the photon statistics and thus vary among the detected potential pulsar counterparts.

 For sources detected by XMM-Newton the corresponding response and effective area files 
 for the source and background emission were extracted using the SAS-script 
 \sffamily especget\normalfont. For the PN camera we used only those events which 
 have a detection pattern between 1 and 4 (single and double events). For the MOS 
 camera we included triple events as well, having a pattern between 5 and 12. The 
 pattern value indicates which detector pixels surrounding a CCD pixel hit by a 
 X-ray photon are taken into account for the photon energy integration.
 In case of Chandra observations we extracted the source and background spectra 
 of a potential counterpart and created the corresponding response and effective 
 area files with the CIAO-script \sffamily specextract\normalfont. 

For potential pulsar counterparts (cf.~Table \ref{tab2:detected_psrs_results})
which were detected with more than 70 source counts the statistics just turned
out to allow a very brief spectral analysis with mostly one-component power 
law or blackbody model spectra. The small photon statistics, though, strongly 
limited the fitting results also in those cases. Often we had to fix the hydrogen 
column absorption to the one deduced from the radio pulsar's dispersion measure
$DM$ which is listed in the ATNF Pulsar Catalogue \citep{2005AJ....129.1993M}. 
This value was estimated by assuming a 10\% ionization degree of the interstellar 
medium (ISM) along the line of sight towards the pulsar. The latter corresponds 
to a ratio of one free electron per ten neutral hydrogen atoms in the ISM, as 
commonly assumed in the literature.

 For a consistency check we deduced $N_\text{H}^\text{LAB}$ from the LAB Survey of 
 Galactic HI \citep{2005A&A...440..775K}. $N_\text{H}^\text{LAB}$ is based  on HI emission
 line measurements at a radio frequency of 21 cm and refers to the entire hydrogen column
 density along the line of sight. For all pulsars for which it turned out that $N_\text{H}$ 
 towards the source is higher than $N_\text{H}^\text{LAB}$ we used the latter.

 The used $DM$ and $N_\text{H}$ for the potential X-ray counterparts are listed in 
 Table \ref{tab4:detected_psrs_list}. Additionally, $N_\text{H}$ and $N_\text{H}^\text{LAB}$ 
 are given in Table \ref{tab5:upper_limits_results} for all analyzed potential counterparts.

As mentioned before, for sources observed with the HRC cameras no spectral
information can be obtained from the pulsar. The only informations we have are
the position and the counting rate in the energy range of the camera where it is
sensitive to incoming photons. In this case and in the case of observations with
less than 70 source counts we obtained constraints on the X-ray flux and
luminosity by converting the pulsar counting rate into an equivalent flux.
First, we obtained the net counting rate for these pulsar counterparts by
dividing the source counts derived with the source detection software by the
effective exposure time $t(\theta)$. The X-ray flux was then deduced for a
typical pulsar spectrum using the method described in Section
\ref{sec:upper_limits}. As a typical pulsar spectrum we assumed a power law
model with photon index 1.7 \citep[][]{2009ASSL..357...91B} and a hydrogen
column density based on the radio pulsar's dispersion measure as mentioned 
previously. 
As the assumption of a spectral shape introduces an extra uncertainty when
converting the photon flux to an energy flux we doubled the statistical error 
in the flux numbers off all potential counterparts which we detected with 
less than 70 source counts.

\subsection{Timing analysis}

The timing resolution of the ACIS and the EPIC-MOS cameras ($\sim 1$ s) were
insufficient to study the pulsed emission of the pulsars. Additionally, the time
resolution of the EPIC-PN CCDs depends on the used observation mode. For all
observations that had a high enough time resolution to study the temporal
properties of a rotating neutron star we converted the event time tags to the
Barycentric Dynamical Time using the SAS-tool \sffamily barycen\normalfont. For
this we used the radio timing position. After folding the events with the
pulsar's rotation frequency given by the ATNF Pulsar Catalogue
\citep{2005AJ....129.1993M} we searched for a periodic signal by applying the
H-test \citep{1989A&A...221..180D}. However, for none of the detected
counterparts pulsed emission could be found with a significance higher than a
confidence level of 95\%. 

\subsection{Upper limits}\label{sec:upper_limits}

For all observations in which no source was detected sufficiently close to the
radio position we derived an upper limit on the count rate as follows: First, we
extracted the counts $N_{i}$ including the background contribution for every
single data set from XMM-Newton or Chandra within a circle of radius
$r=30\arcsec$ or $r=2\arcsec$ around the radio position in the energy band
$E_\text{min}$ to $E_\text{max}=4.5$ keV. $E_\text{min}$ is 0.2 keV for
XMM-Newton and 0.1 keV for Chandra events. For Chandra data sets were the pulsar
was observed far off-axis the radius had to be adjusted, because above $\approx
4\arcmin$ the point-spread-function broadens significantly. Hence, for an
off-axis angle $4\arcmin < \theta \le 8 \arcmin$ the applied radius $r=5\arcsec$
and for $\theta > 8 \arcmin$ the radius $r=10\arcsec$. The background counts
$B_{i}$ were extracted from an annulus of inner radius $r_\text{inner}=1.5 r$
and outer radius $r_\text{outer}=2.5 r$. To obtain the background counts in the
source region, we had to normalize these counts to the source region by the
fraction $c_1$ between source region area and annulus area and by the fraction
$c_2$ between $t(\theta)$ and $t_\text{BG}(\theta)$ . Because we used a finite
region, we had to correct for the missing flux in the unselected wings of the
point-spread function. This encircled energy fraction ($EEF_{i}$) was calculated
with the help of the \sffamily eregionanalyse\normalfont- (XMM-Newton) and
\sffamily psf\normalfont-tool~(Chandra).

If $\sum_i^n N_i> 20$ counts, where $n$ is the number of X-ray observations of a
single pulsar, we derived the $x\cdot\sigma$ upper limit on the count rate with
the following equation:
\begin{equation}
 U\!L(x\cdot\sigma)=\frac{\max\left(\sum_{i}^{n} \frac{N_{i}- c_1c_2 \cdot B_{i}}{EEF_{i}}, 0 \right) + x \cdot \sum_{i}^{n} \frac{\sqrt{ N_{i}+c_1^2c_2^2B_{i}}}{EEF_{i}} }{\sum_{i}^{n} t_i(\theta_i)}.
\end{equation}
Otherwise, the upper limit was derived using the Bayesian approach
introduced by \citet{1991ApJ...374..344K} which has its advantage
for observations with very small numbers of counts in the presence 
of background when compared with the classical method.

Assuming a pure blackbody (BB) spectrum we deduced the $3\sigma$ upper
limit on the effective temperature $T_\text{e}=T_\infty (1+z)$ \footnote{$T_\infty$ denotes
the temperature measured by a distant observer.} by minimizing the difference
$U\!L(3\sigma)-U\!L_\text{BB}(T_\text{e})$
\citep{1993ispu.conf..104B,1995PhDTBecker}, where
\begin{equation}
 U\!L_\text{BB}(T_\text{e})={const.}\cdot \int^{E_\text{max}}_{E_\text{min}} \frac{E^2 A_\text{eff}(E)\cdot \exp{(-\sigma(E)\cdot N_\text{H}^\text{ul})}}{\exp{(E(1+z)/k_\text{B}T_\text{e})}} dE.
\end{equation}
Here ${const.}=\frac{2\pi(1+z)^2R_0^2}{h^3c^2d_\text{ul}^2}$ with $z$ the
gravitational red shift for a neutron star with radius $R_0=11.43$ km and mass
$M=1.4M_\odot$, $h$ the Planck constant, $c$ the speed of light and
$d_\text{ul}$ the upper limit of the distance to the pulsar. Furthermore,
$\sigma(E)$ is the interstellar photoelectric absorption cross section
\citep{1983ApJ...270..119M}, $N_\text{H}^\text{ul}$ the upper limit of the
hydrogen column density, $k_\text{B}$ is the Boltzmann constant and
$A_\text{eff}$ is the weighted effective area of all used instruments with which
a certain source was observed: 
\begin{equation}
 A_\text{eff}(E)=\frac{\sum_i^n A_i(E) t_i(\theta_i)}{\sum_i^n t_i(\theta_i)},
\end{equation}
where $A_i$ is the on-axis effective area of the instrument used in the $i$\,th
observation of the pulsar. Because the count rate is already vignetting
corrected we used the on-axis effective area obtained with the SAS-task
\sffamily calview\normalfont~for the XMM-Newton EPIC cameras with different
filters. The Chandra ACIS and HRC effective area were taken from the Portable,
Interactive Multi-Mission Simulator (WebPIMMS), version 4.6.\footnote{See 
\url{http://heasarc.nasa.gov/docs/software/tools/pimms.html}}
$N_\text{H}^\text{ul}$ is calculated as described in Section
\ref{sec:spectral_analysis}, except that we add the error of $DM$ to obtain the
upper limit of $N_\text{H}$.

The distance estimate is based on the work by \citet{2012ApJ...755...39V}, where
they present a detailed analysis of all measured distances to pulsars. For all
other sources we used the NE2001 model of \citet{2002astro.ph..7156C} to
transform the measured $DM$ to a distance. We apply a distance error of 20~\% to
120~\% depending on their Galactic coordinates. According to this work the 
large difference in the used uncertainties is because the error in $d$ is larger
for a pulsar in a region of small electron density than in a region with higher
density \citep{2002astro.ph..7156C}.

To estimate the $2\sigma$ upper limit of the non-thermal X-ray efficiency we
derived an expression for $U\!L_\text{model}$ in a similar way as for the
blackbody spectrum, except that we inserted a power law model (PL) with photon
index $\Gamma=1.7$ \citep[][]{2009ASSL..357...91B} and minimizing
$U\!L(2\sigma)- U\!L_\text{PL}(K)$ with 
\begin{equation}\label{eq:fluxFromRate}
 U\!L_\text{PL}(K)=K \int^{E_\text{max}}_{E_\text{min}} E^{-\Gamma} \exp{(-\sigma(E) \cdot N_\text{H}^\text{ul})} \cdot A_\text{eff}(E)dE.
\end{equation}

Using the resulting value for the normalization constant $K$ and the assumed
photon index we can derive the flux upper limit in a certain energy range by
integrating the power law model with the derived normalization over this energy
band. 

In the following, all given uncertainties represent the 1$\sigma$ confidence
range for one parameter of interest, unless stated otherwise.

\section{Potential X-ray counterparts of radio pulsars}\label{sec:results_psrSearch}

\begin{figure*}[!tbp]
\centering
  \resizebox{0.215\hsize}{!}{\includegraphics[clip]{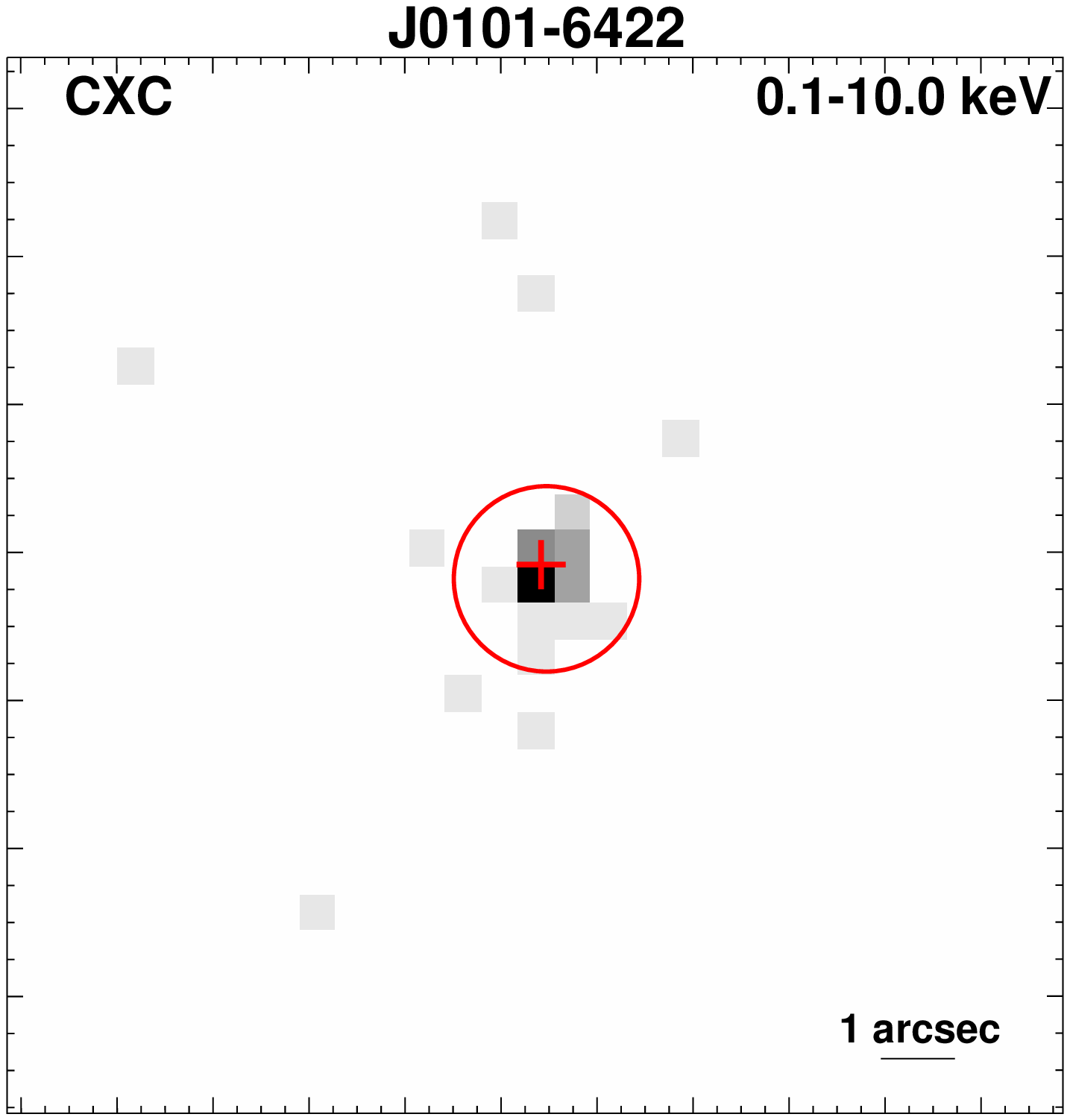}}
  \resizebox{0.215\hsize}{!}{\includegraphics[clip]{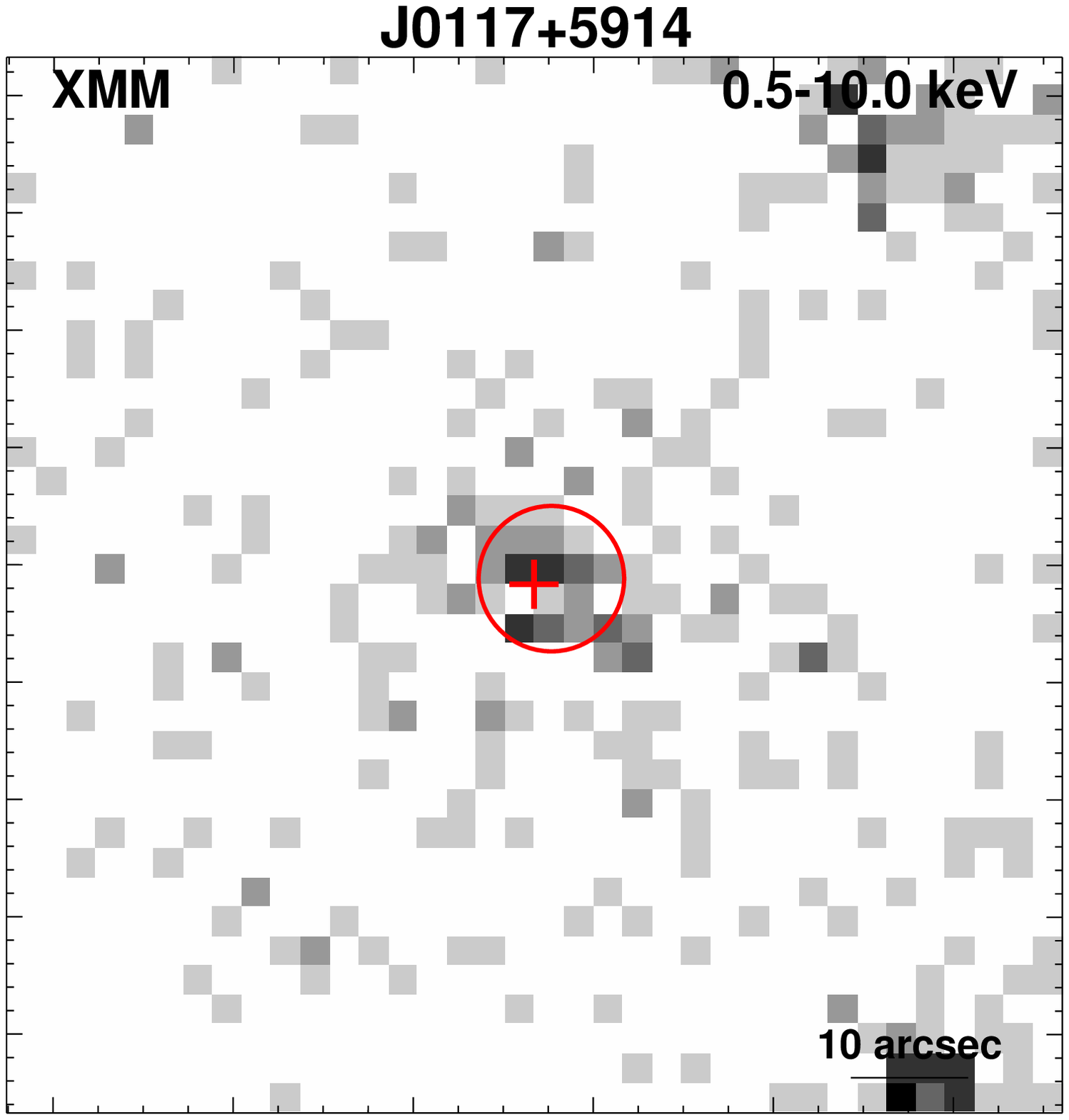}}
  \resizebox{0.215\hsize}{!}{\includegraphics[clip]{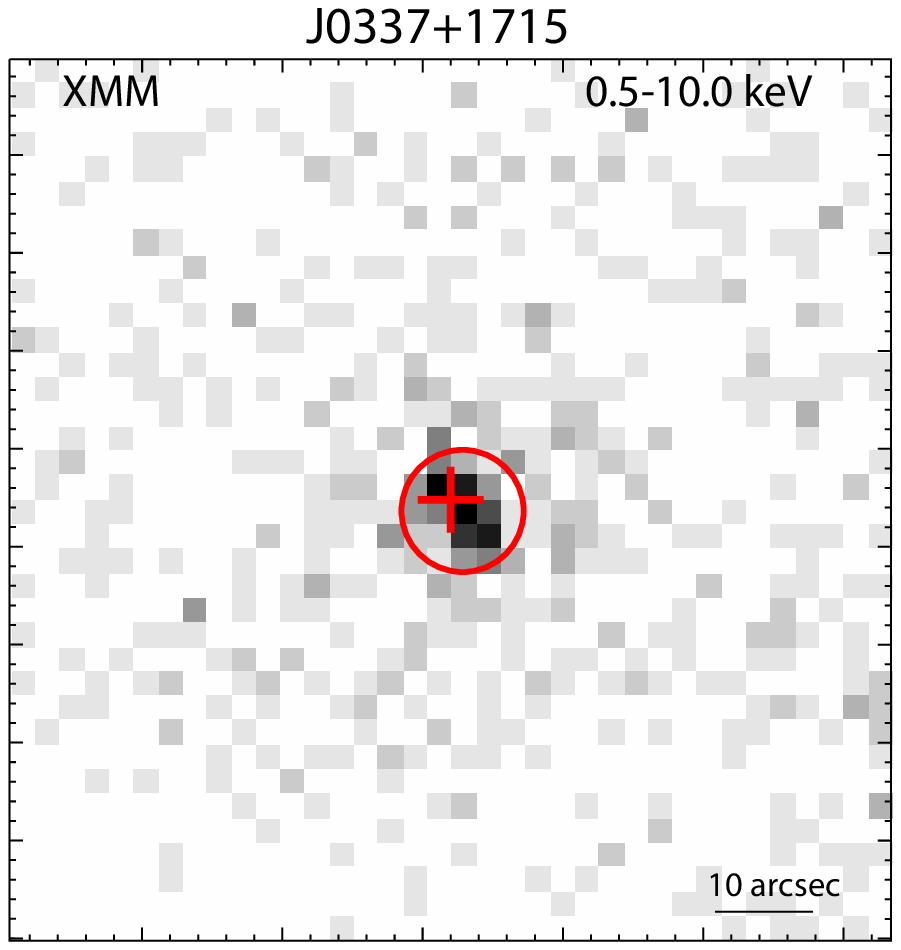}}
  \resizebox{0.215\hsize}{!}{\includegraphics[clip]{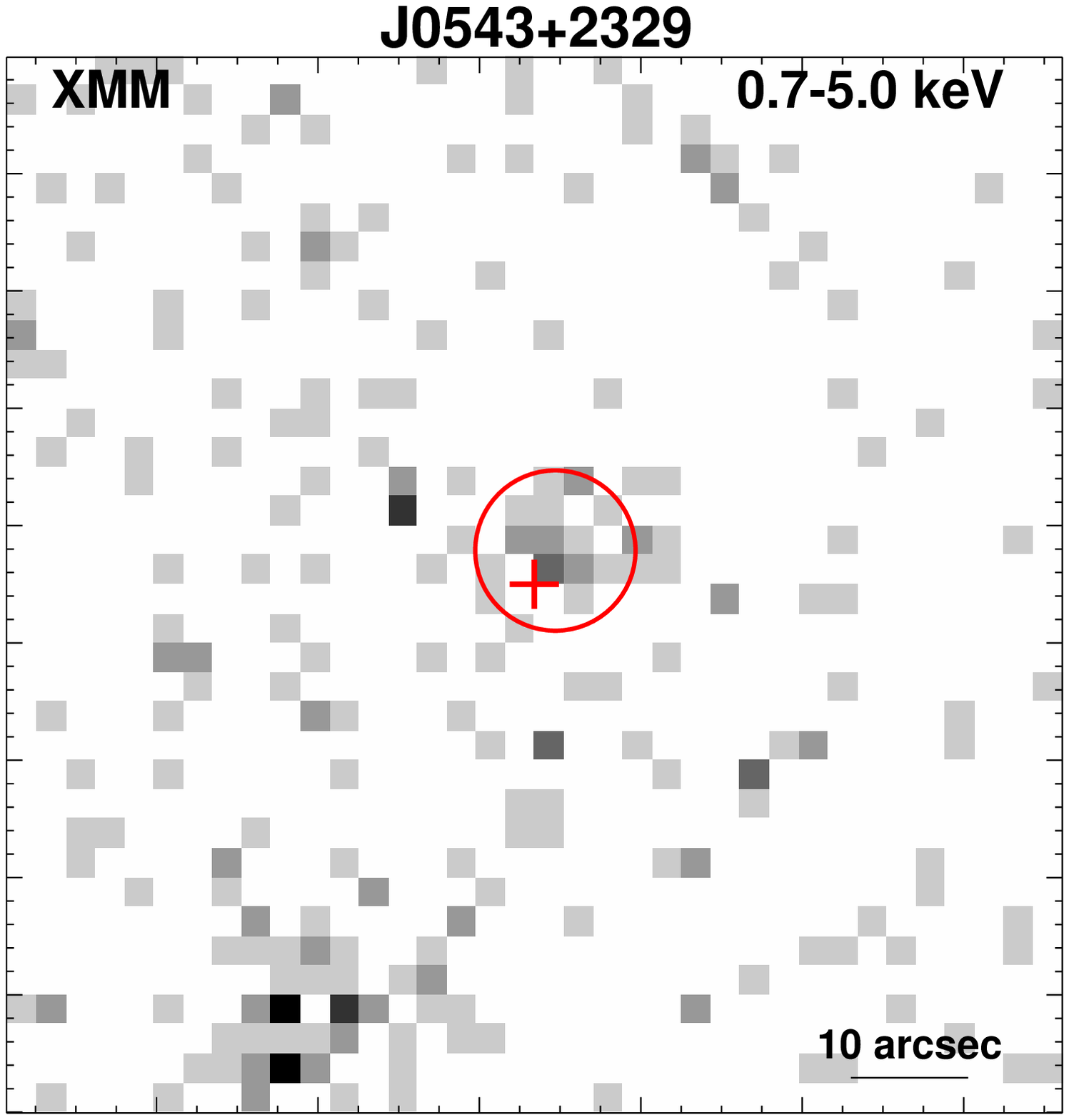}}
\\[0.5ex]
  \resizebox{0.215\hsize}{!}{\includegraphics[clip]{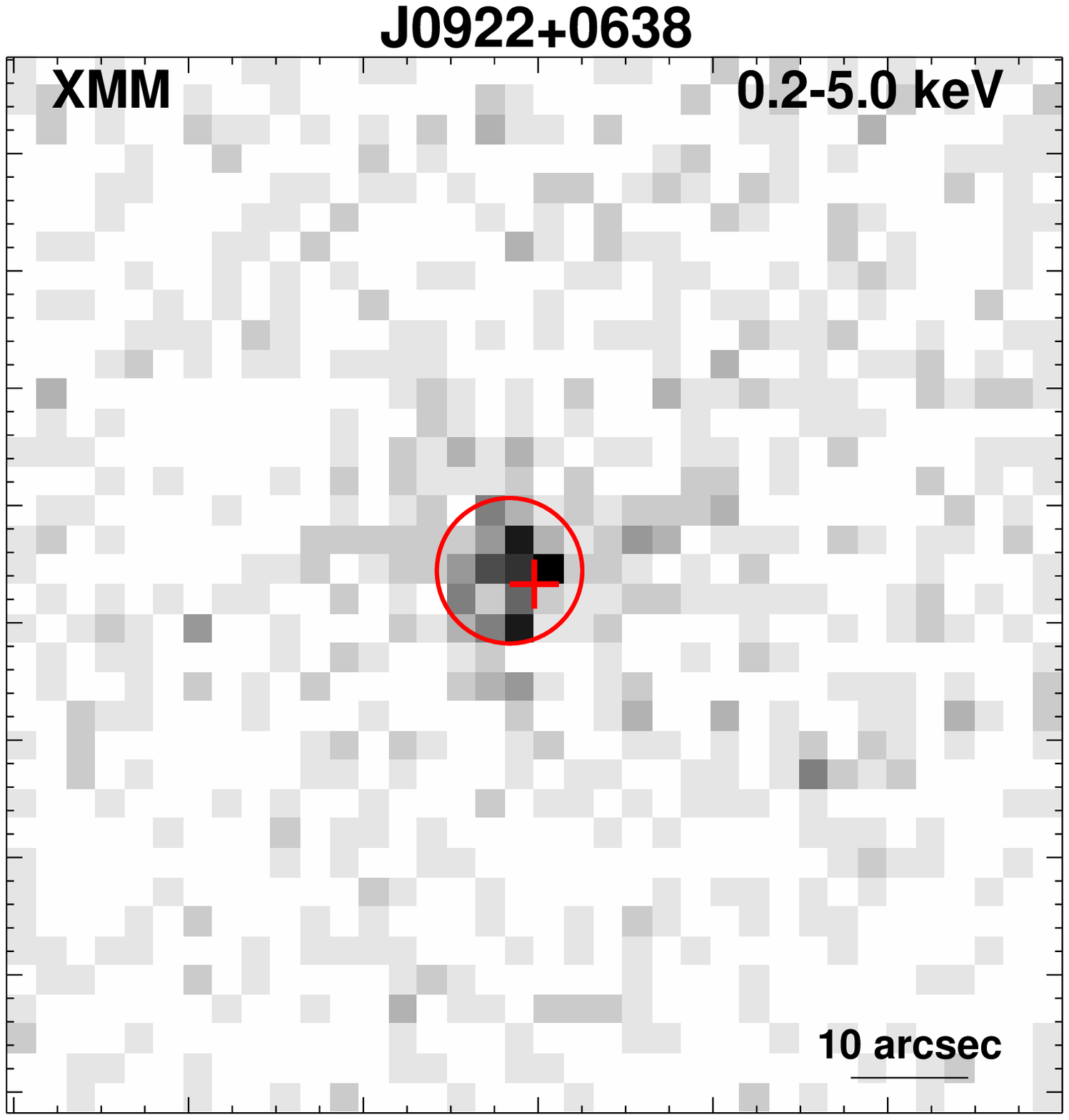}}
  \resizebox{0.215\hsize}{!}{\includegraphics[clip]{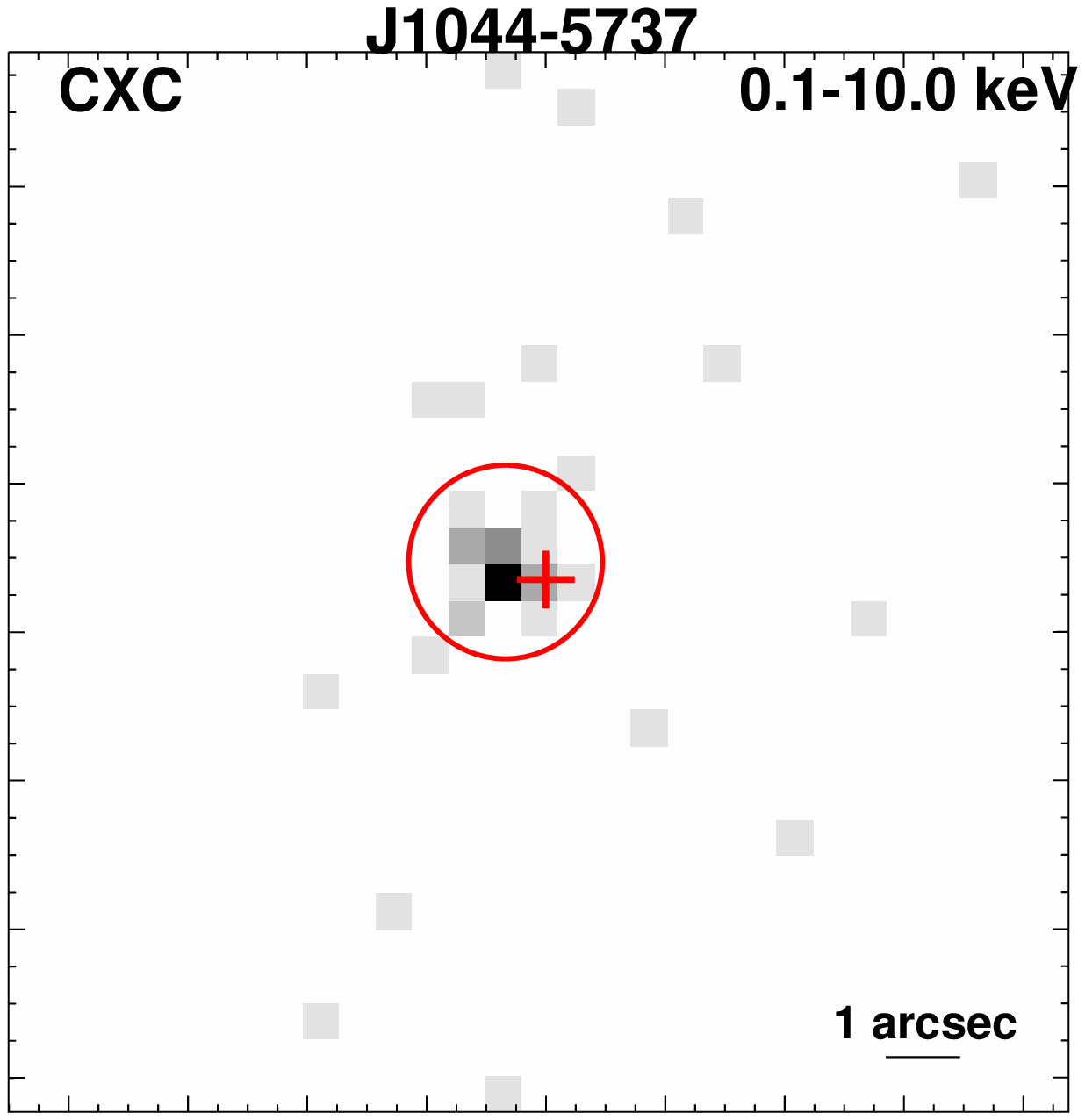}}
  \resizebox{0.215\hsize}{!}{\includegraphics[clip]{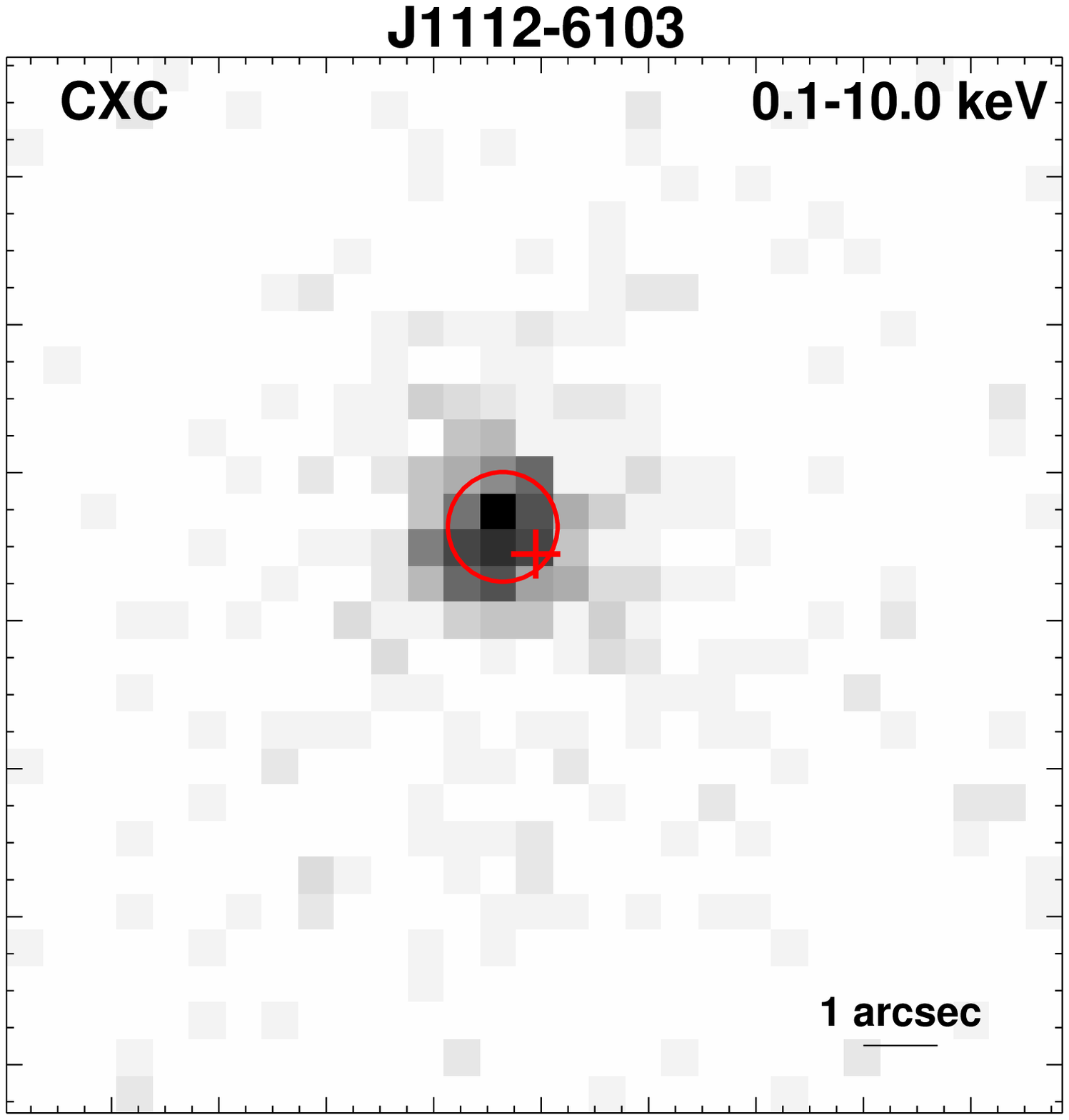}}
  \resizebox{0.215\hsize}{!}{\includegraphics[clip]{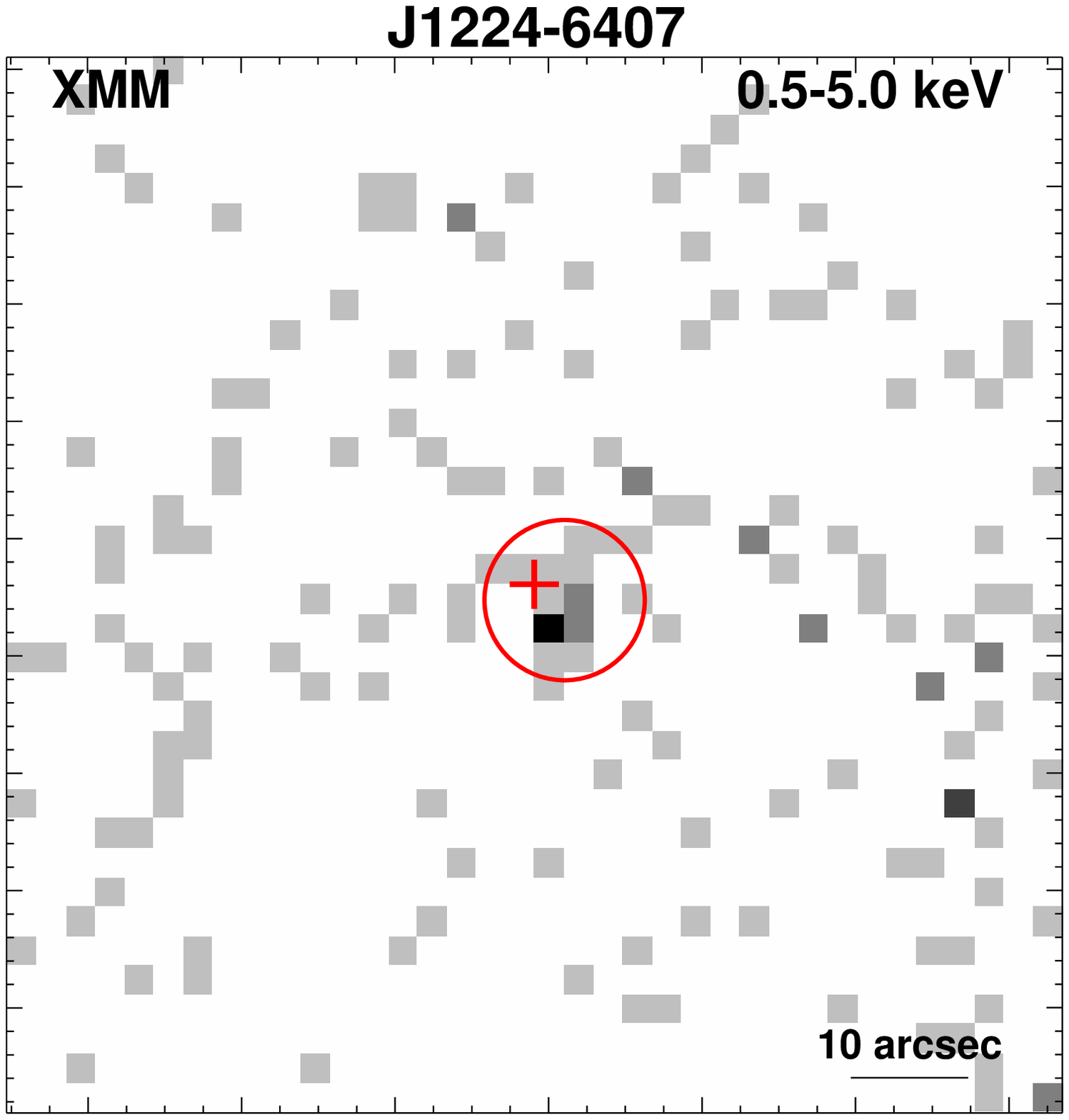}}
\\[0.5ex]
  \resizebox{0.215\hsize}{!}{\includegraphics[clip]{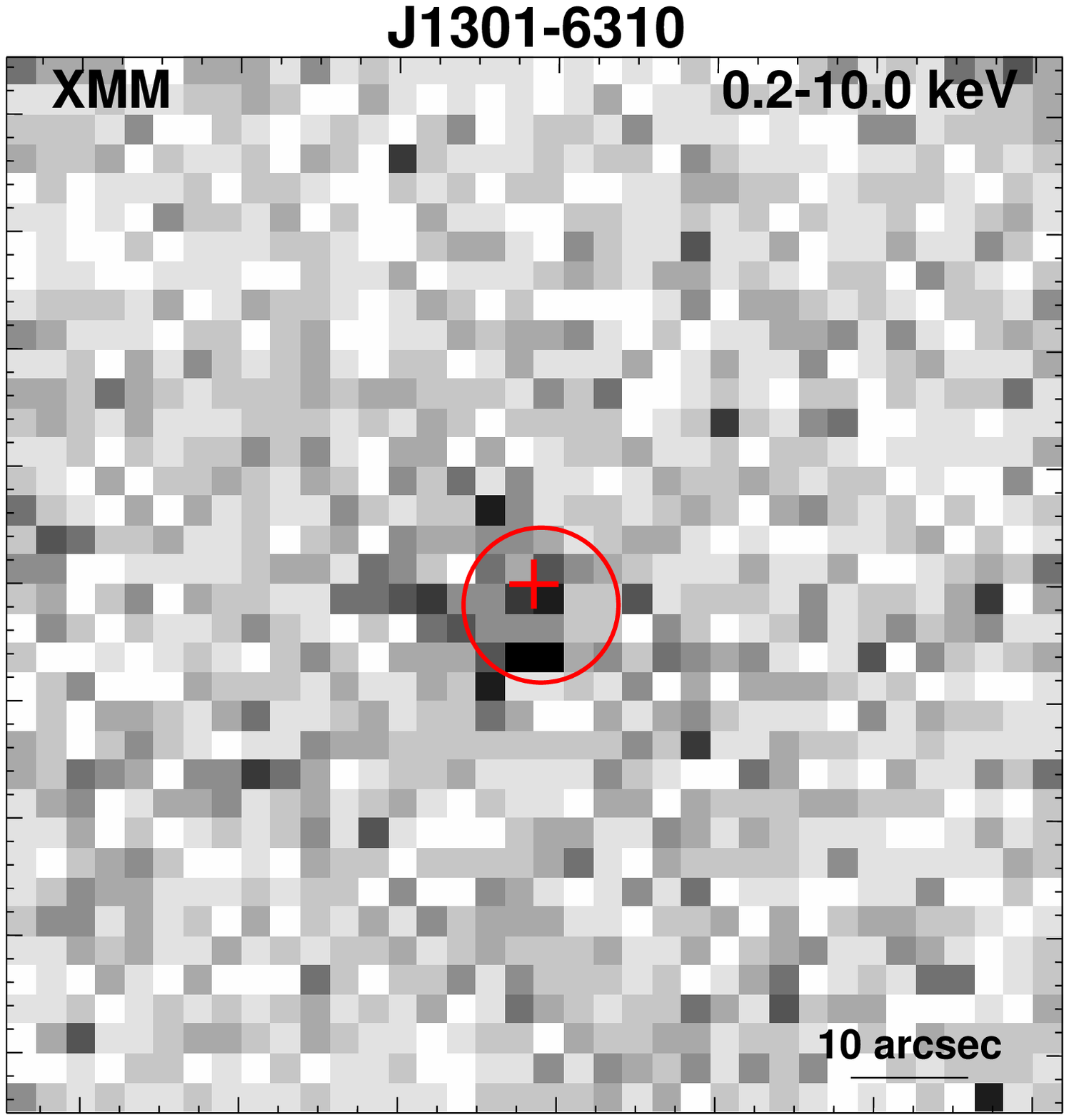}}
  \resizebox{0.215\hsize}{!}{\includegraphics[clip]{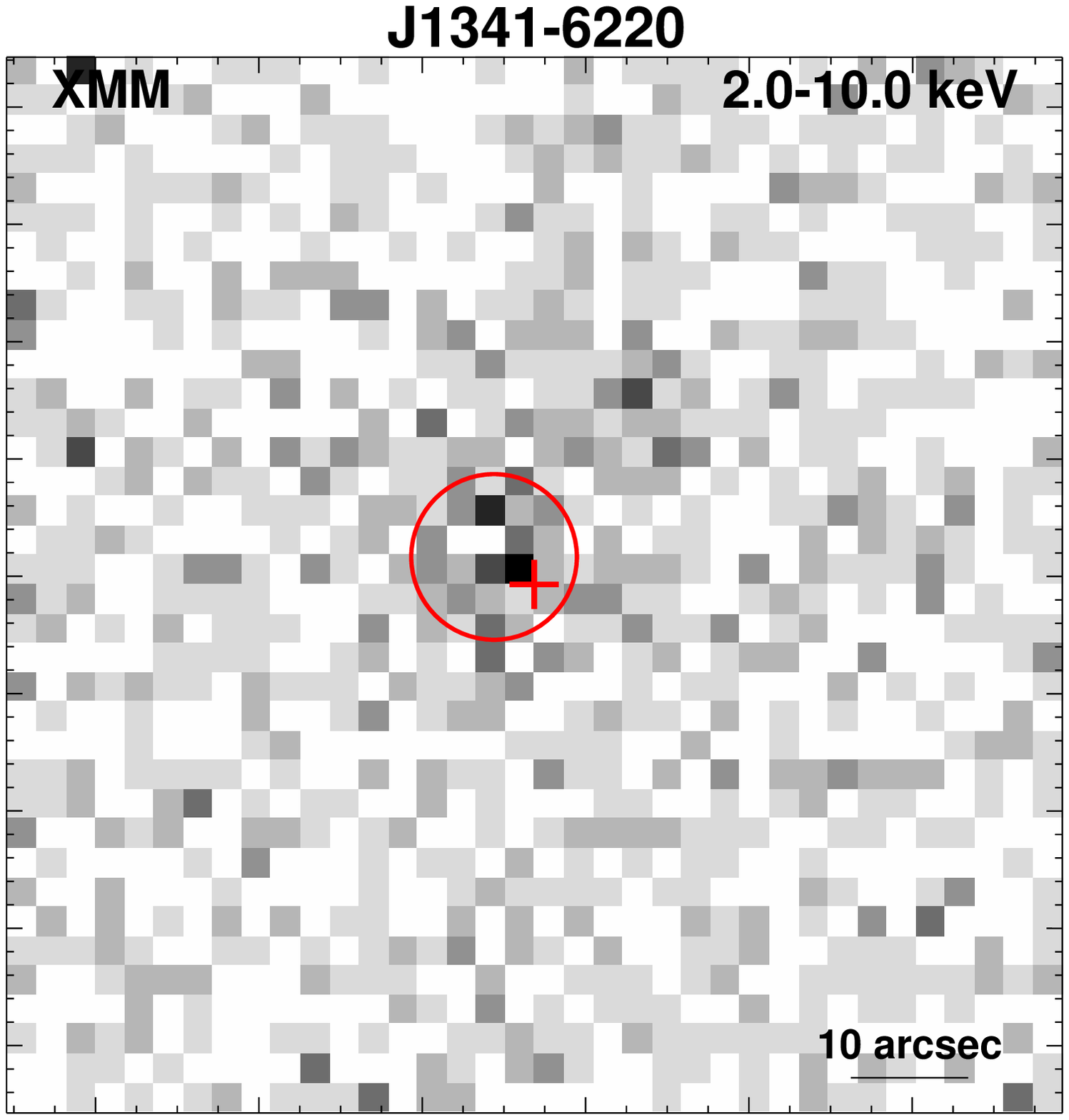}}
  \resizebox{0.215\hsize}{!}{\includegraphics[clip]{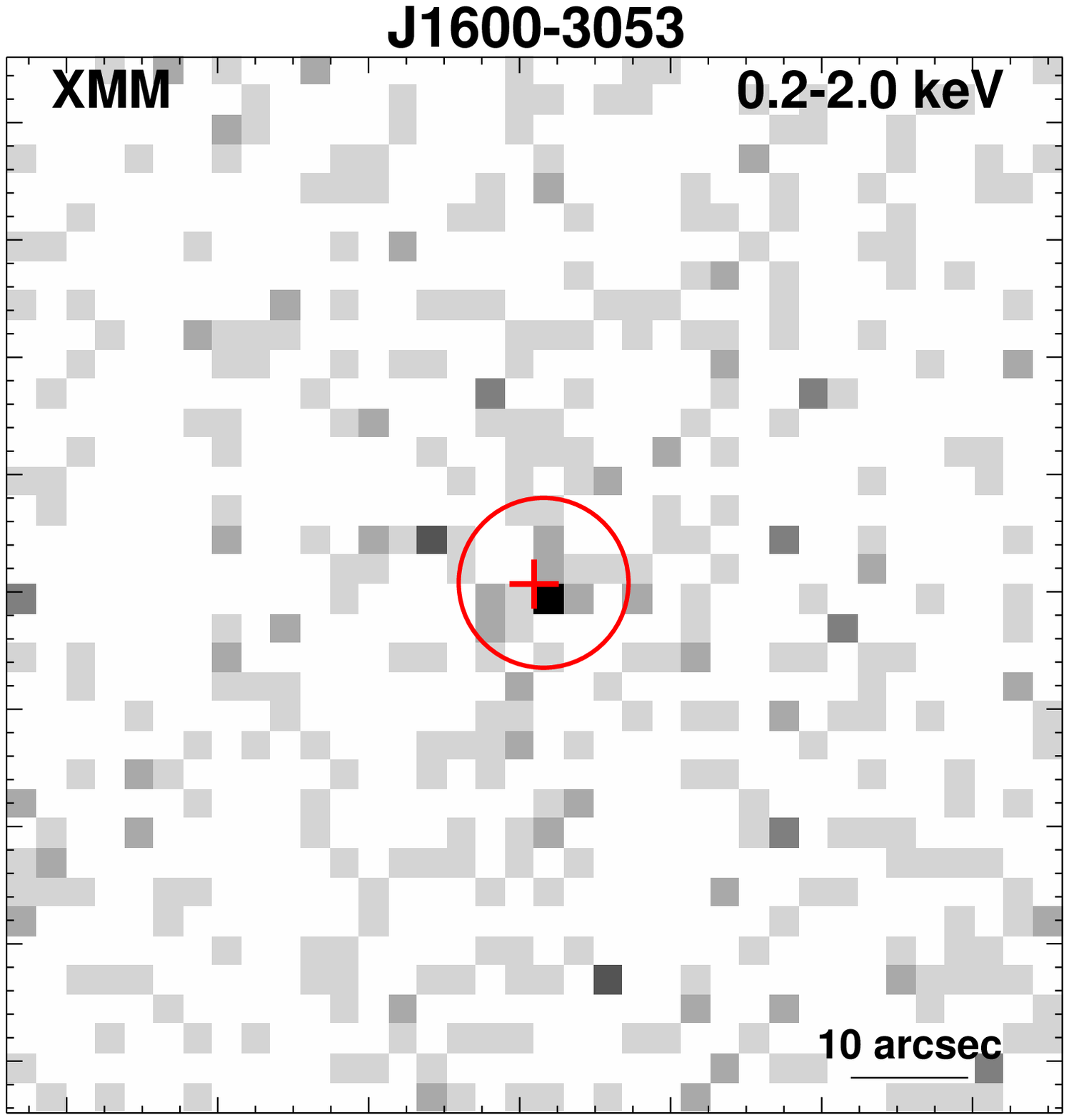}}
  \resizebox{0.215\hsize}{!}{\includegraphics[clip]{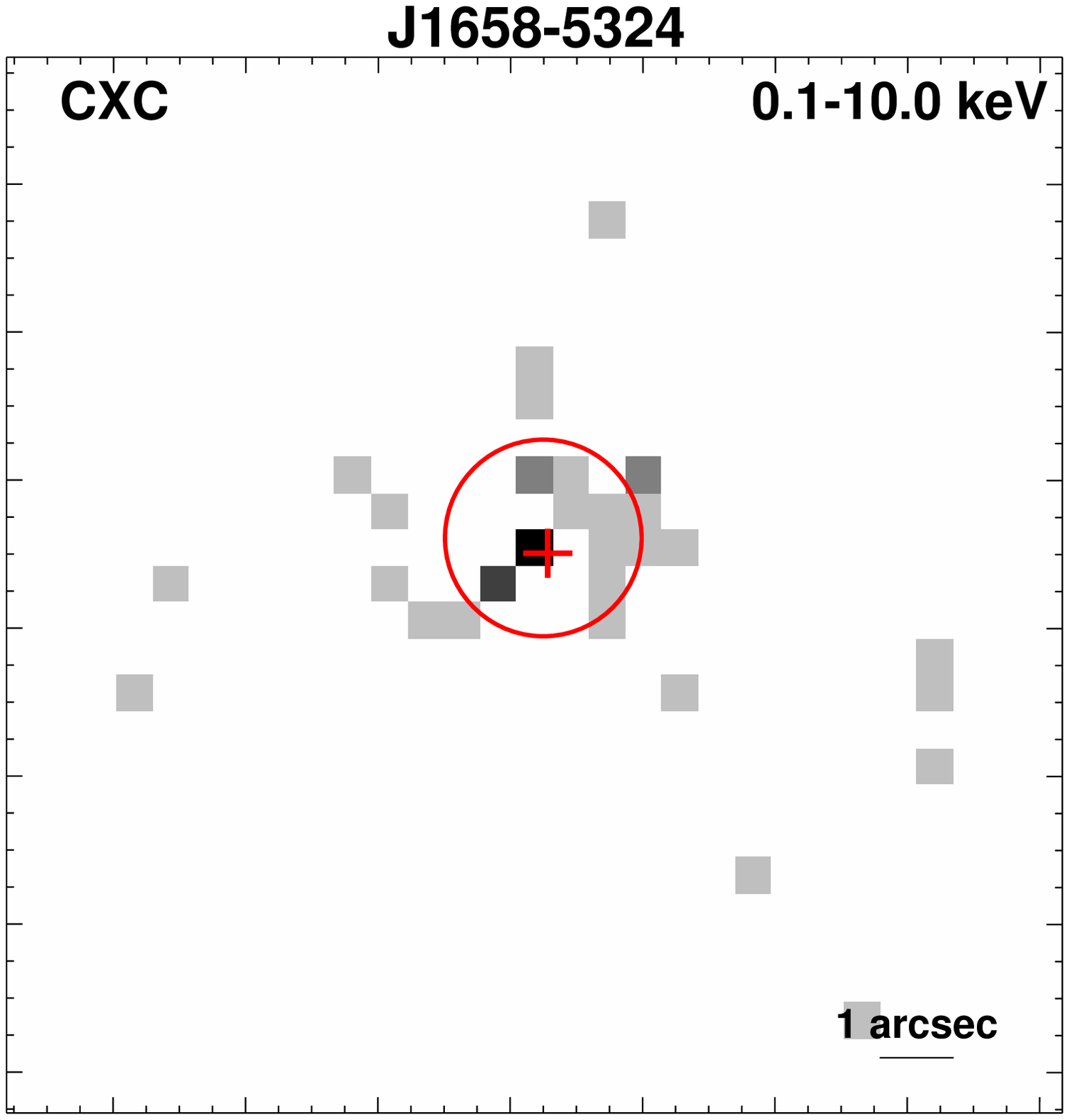}}
\\[0.5ex]
  \resizebox{0.215\hsize}{!}{\includegraphics[clip]{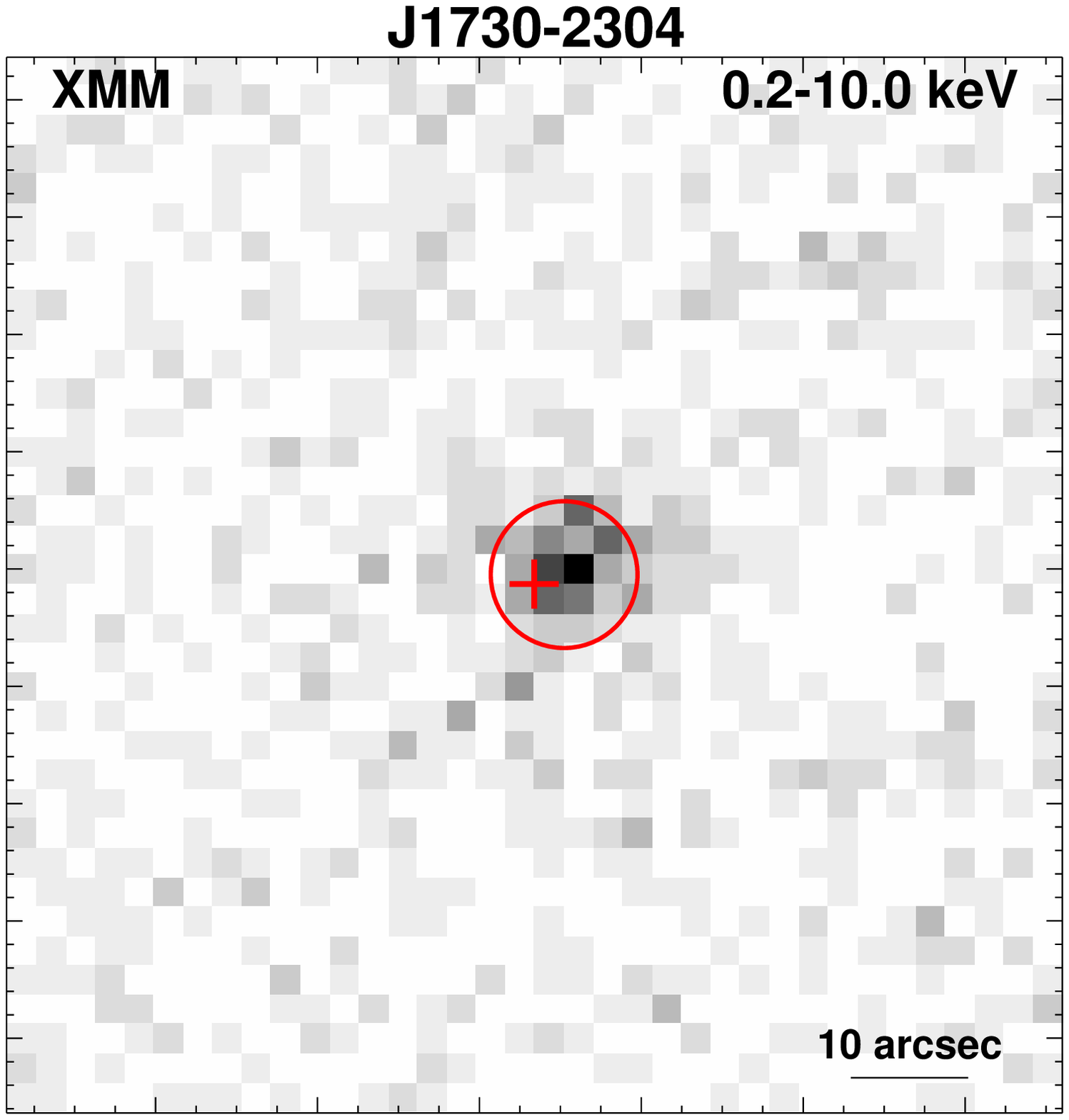}}
  \resizebox{0.215\hsize}{!}{\includegraphics[clip]{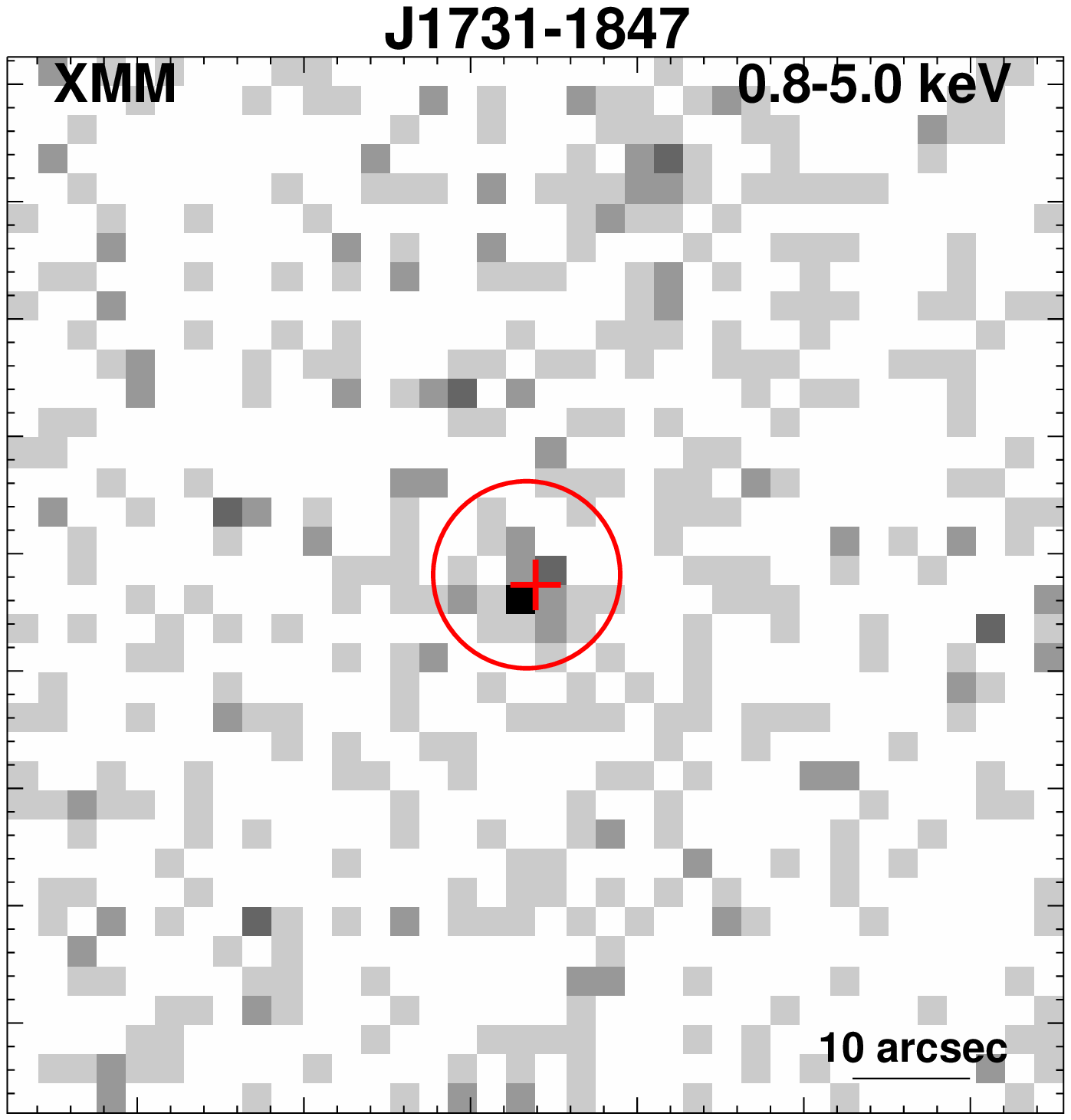}}
  \resizebox{0.215\hsize}{!}{\includegraphics[clip]{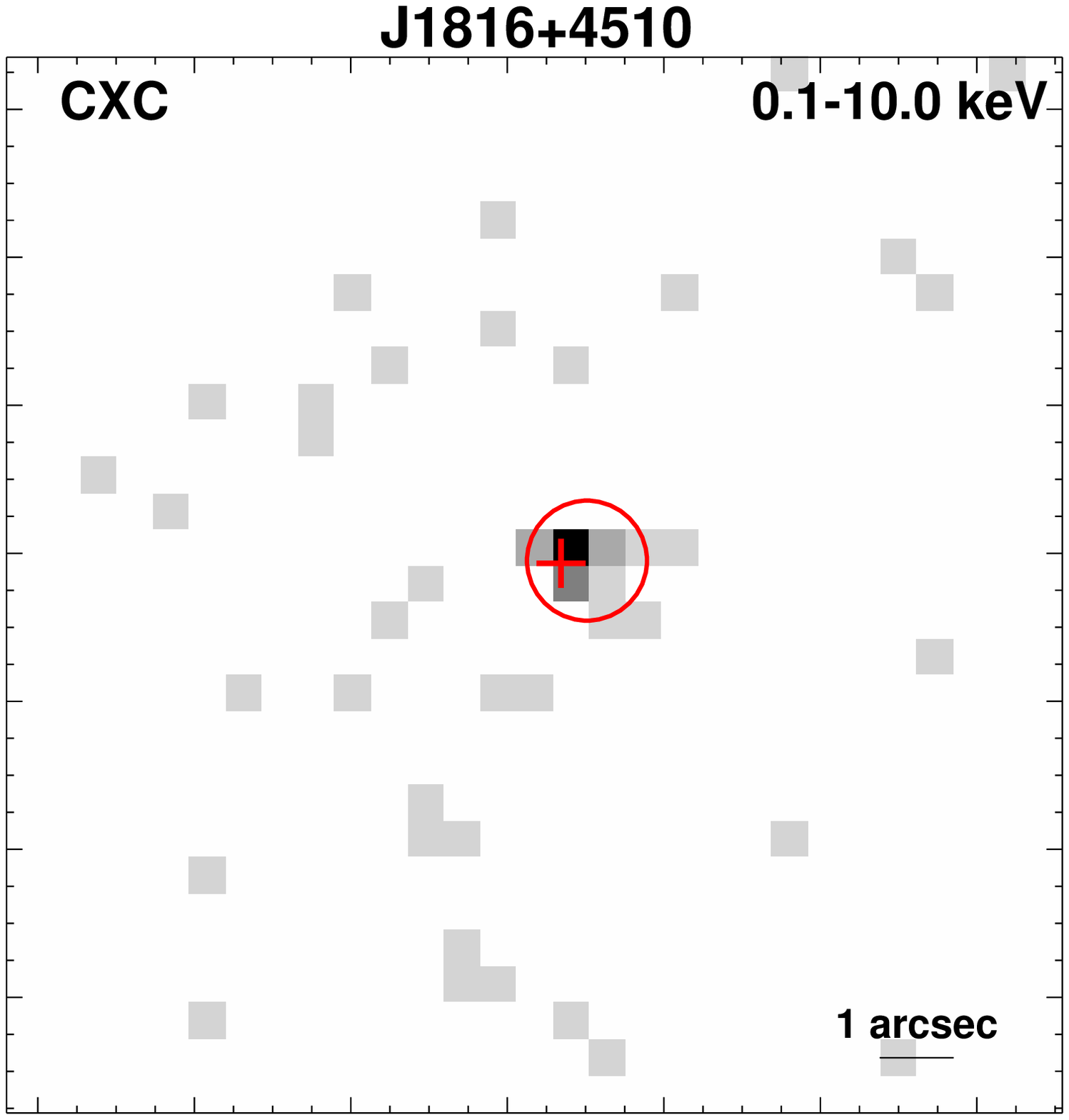}}
  \resizebox{0.215\hsize}{!}{\includegraphics[clip]{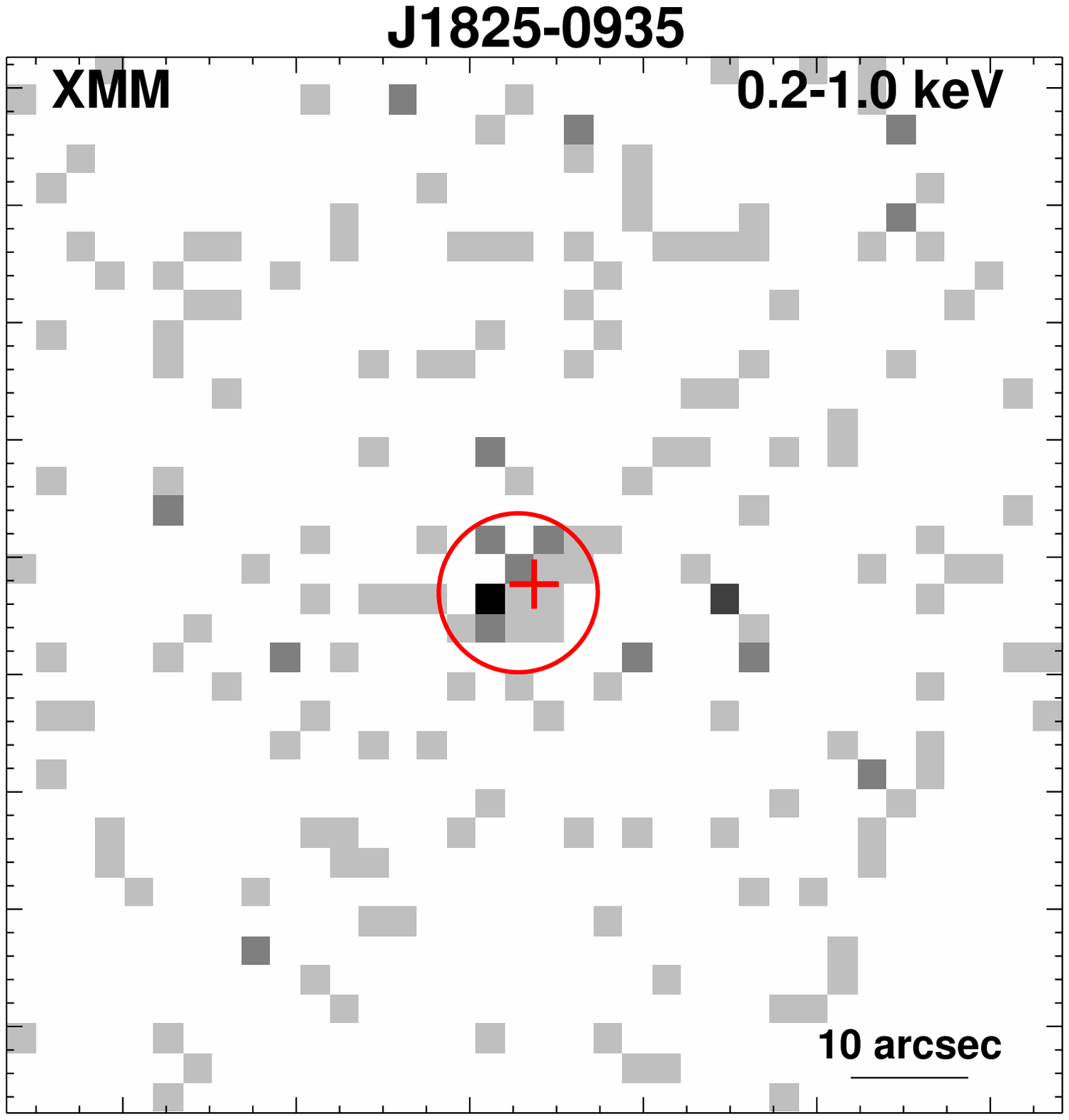}}
\\[0.5ex]
  \resizebox{0.215\hsize}{!}{\includegraphics[clip]{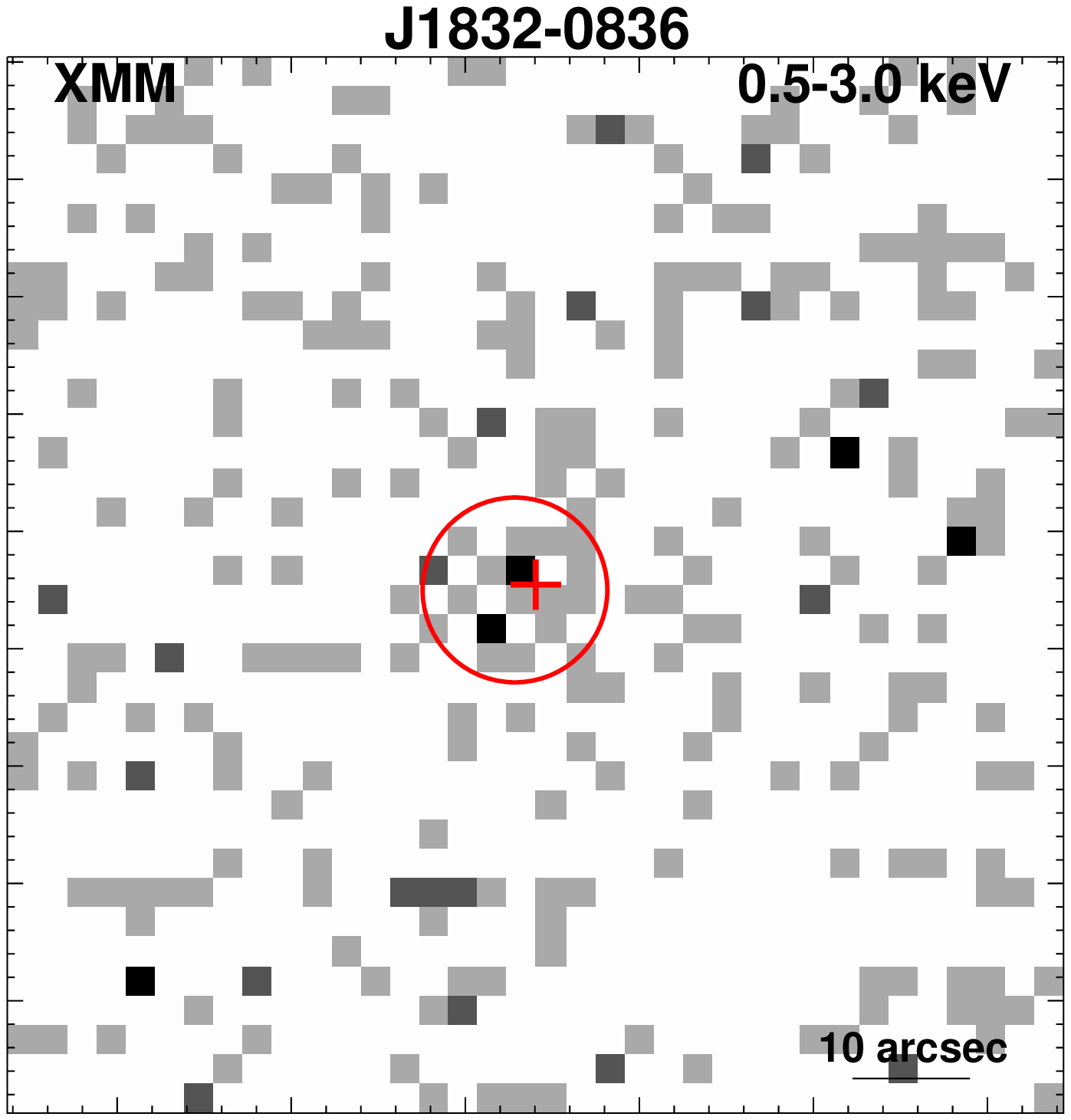}}
  \resizebox{0.215\hsize}{!}{\includegraphics[clip]{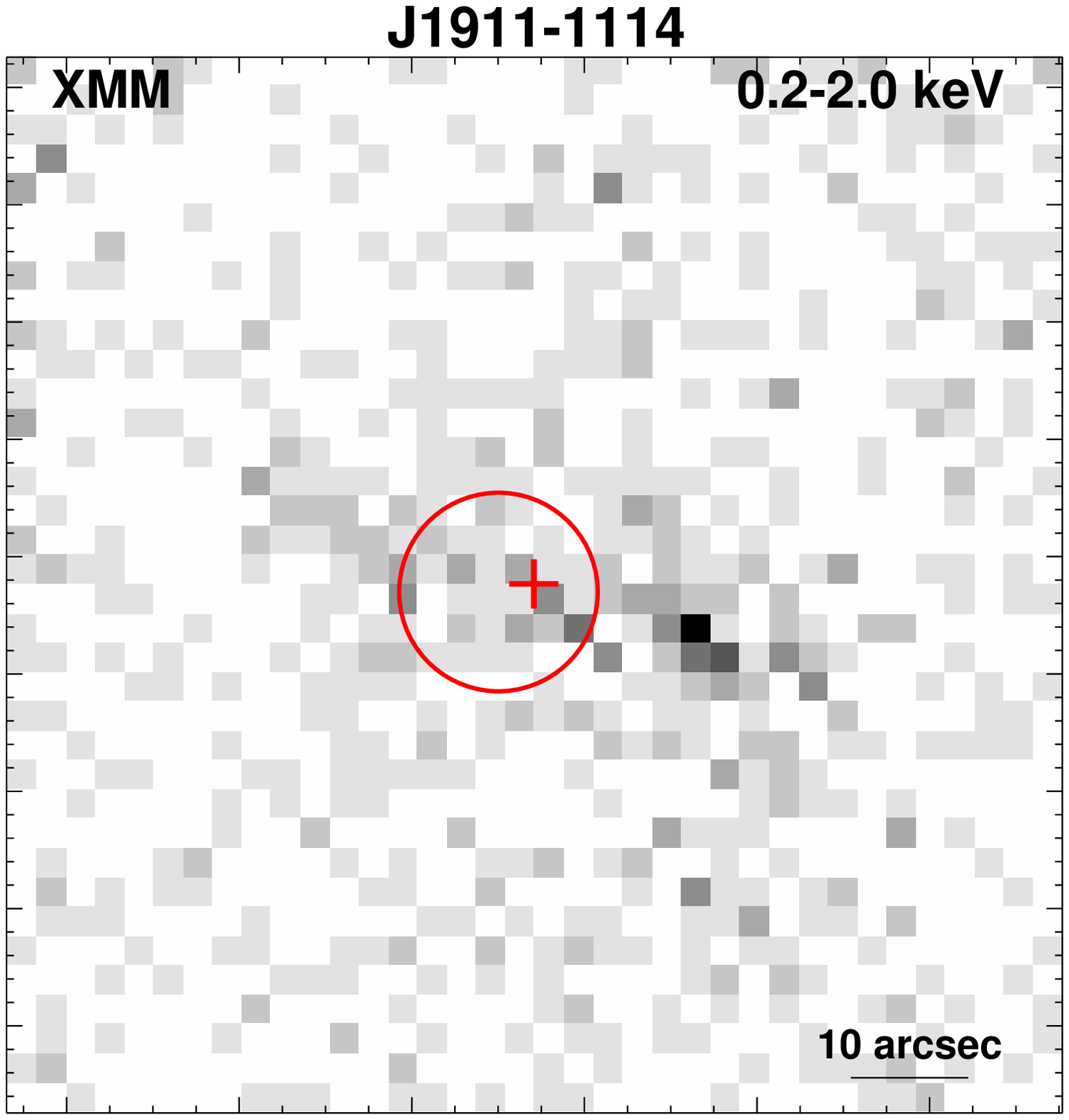}}
  \resizebox{0.215\hsize}{!}{\includegraphics[clip]{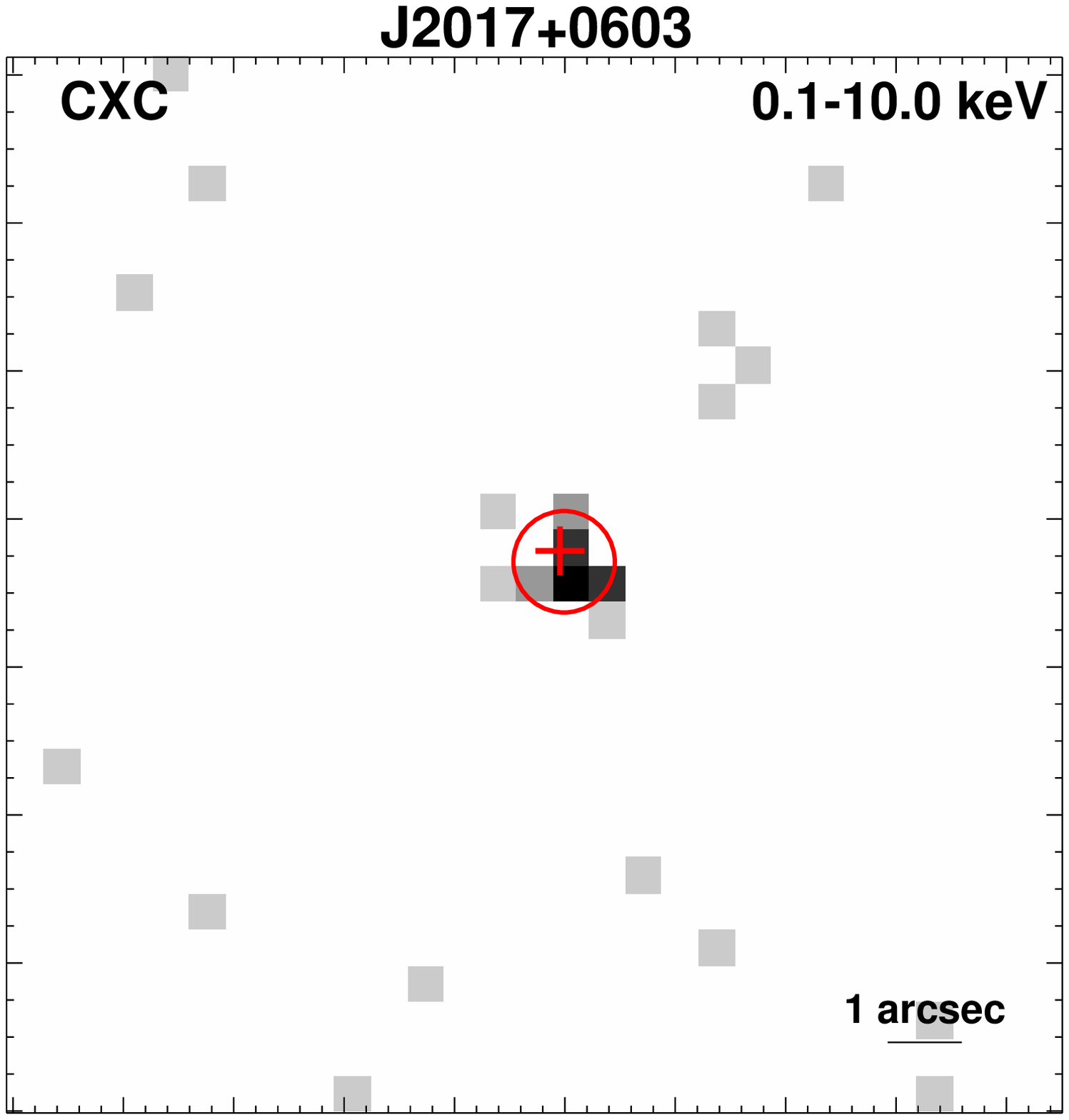}}
  \resizebox{0.215\hsize}{!}{\includegraphics[clip]{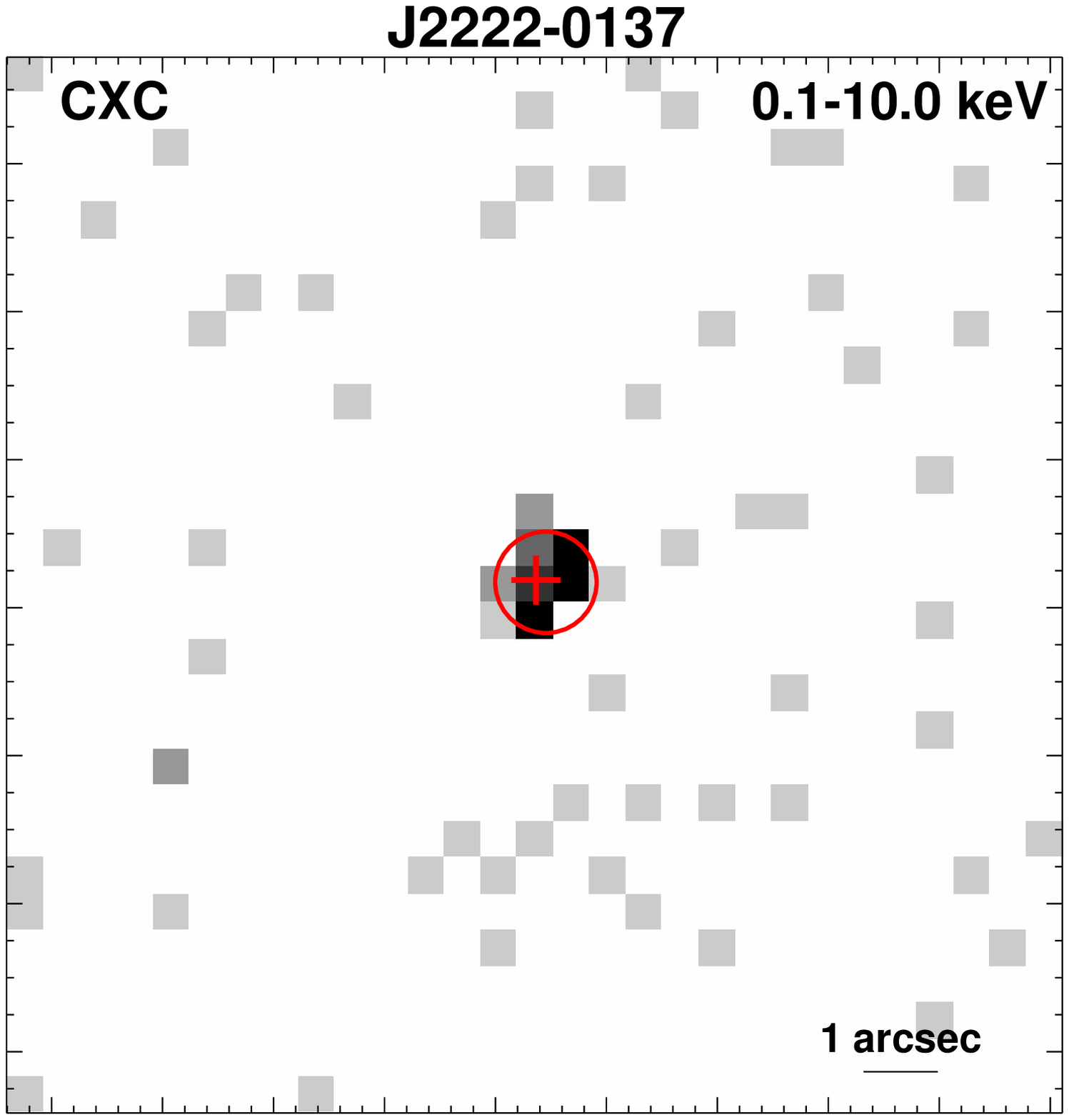}}
  \caption[X-ray images of all radio pulsars with newly detected X-ray
counterparts]{X-ray images of the region around all radio pulsars where an X-ray
counterpart was detected in our correlation. Each panel has a size of $1\farcm5 \times 1\farcm5$
for XMM-Newton observations and $15\arcsec \times 15\arcsec$ for Chandra
observations. North is up and east is to the left. The cross marks the radio
position of the pulsar and the red circle indicates the 3$\sigma$ uncertainty in
the measured X-ray position.}
  \label{fig:XraySources}
\end{figure*}

In our search for X-ray counterparts of radio pulsars we found \identified~X-ray 
point sources sufficiently close to the corresponding radio pulsar's timing 
positions to claim a match. The X-ray images of these sources are displayed in 
Figure \ref{fig:XraySources}. Their coordinates, the offset between their X-ray 
position and the radio pulsar's timing position, $\Delta$, as well as the chance 
probability $P_\text{coin}$ and the X-ray to visual flux ratio for each of this 
sources are displayed in Table \ref{tab2:detected_psrs_results}. 
Due to a limitation in the photon statistics a detailed spectral analysis was 
possible only for six of the \identified~counterparts. The results of this 
spectral analysis are summarized in Table \ref{tab3:spectral_psr_results}.  
In Table \ref{tab4:detected_psrs_list} the relevant parameters of the radio 
pulsars are listed along with the flux, the luminosity and the X-ray efficiency 
$\mu_\text{0.1--2 keV}$ of their potential counterparts.

The following subsections highlight the results for the 20 potential X-ray
counterparts and provide background information on the corresponding radio pulsars. 

\subsection{PSR J0101--6422}
The binary millisecond pulsar PSR J0101--6422 was discovered with the Parkes
radio telescope in an identification campaign for unidentified Fermi sources
\citep{2012ApJ...748L...2K}. It was in the focus of the Chandra telescope for 9.0 ksec. 
In this observation we detected a faint X-ray point source only $0\farcs2\pm0\farcs4$ 
off from the pulsar's radio position, making it a very likely X-ray
counter part of the pulsar. This point source has $34\pm6$
source counts.  Converting the photon flux to an energy flux yields
$F_\text{X}^\text{0.1--2 keV}=(3.0\pm 1.0) \times 10^{-14}~\text{erg s}^{-1}
~\text{cm}^{-2}$ in the 0.1 to 2 keV range.

\subsection{PSR B0114+58}
The pulsar PSR B0114+58 (PSR J0117+5914) was first detected in the Princeton-NRAO 
pulsar survey \citep{B0114Discovery}. In the XMM-Newton data archive we found a
9 ksec observation which had the pulsar in the focus of the telescope. Analyzing 
this data we detected a point source with $113\pm13$ source counts in the PN- and 
MOS1/2 cameras. It is located only $1\farcs6\pm2\farcs1$ off from the radio pulsar's 
timing position, making it a positional match with the pulsar. 
The detected photons allowed a very brief spectral analysis. To do so we extracted 
the source counts from a circle of radius $20\arcsec$ centered on the
counterpart. 
The background spectra were extracted from an annulus with inner and outer radius 
$45\arcsec$ and $75\arcsec$, respectively. The extracted counts were binned with at 
least 8 counts/bin in case of the PN data and 5 counts/bin for the MOS1/2 data sets. 
We first fitted the energy spectra with an absorbed power law and free varying 
$N_\text{H}$. This resulted in a fit with $\chi^2=12.1$ for 12 d.o.f. and 
$N_\text{H}=0.5_{-0.5}^{+0.7}\times 10^{22}~\text{cm}^{-2}$, in agreement with 
the $N_\text{H}=0.15\times 10^{22}$ cm$^{-2}$ as deduce from the radio dispersion 
measurement DM and the model for the galactic distribution of free electrons  by 
\citet{2002astro.ph..7156C}. To put better constraints on the remaining parameters 
we fixed $N_\text{H}$ to the $DM$-based value. This gave an acceptable fit 
($\chi^2=13.2$, 13 d.o.f.) with a photon index $\Gamma=3.3\pm 0.5$. Modeling the 
spectrum with a blackbody fitted the data as well and yielded a $\chi^2=12.2$ 
(13 d.o.f.). The resulting blackbody temperature is  $2.1_{-0.3}^{+0.4}\times 10^6$ K 
with a normalization factor $K=3.7_{-1.1}^{+1.3}\times 10^{-7}$. Using the 
normalization factor $K=L_{39}/D_{10}^2$ ($L_{39}=L/10^{39}$ erg/s and
$D_{10}=d/10$ kpc) we deduced the emitting blackbody radius $R_{BB}$ to be:
\begin{equation} R_{BB}=\sqrt{\frac{K \cdot D^2_{10} \times 10^{39}~\text{erg/s}}{4
\pi \sigma_B  T^4}} = (0.4\pm0.2)~\text{km.} \end{equation} Herein $\sigma_B$ is
the Stefan-Boltzmann constant.

\subsection{PSR J0337+1715}
J0337+1715 is a millisecond pulsar in a triple system \citep{2014Natur.505..520R}. 
In an $\sim 18$ ksec archival XMM-Newton observation which had the pulsar in the 
focus of the telescope we detected a central point source with $303\pm21$ source 
counts in the PN and MOS1/2 data. The position of this source matches the radio 
pulsar's timing position by only $1\farcs7\pm2\farcs1$ making it a positional 
counterpart of the pulsar. During the publication process of this paper a detailed analysis of the PSR J0337+1775 
was published on astro-ph by \citet{2016arXiv160200655S}. We therefore refer to this 
publication for further analysis details and source properties.

\subsection{PSR B0540+23}
PSR B0540+23 (PSR J0543+2329) was first detected by \citet{1972Natur.240..229D}.  
In archival XMM-Newton data of $\sim 20$ ksec observing time in which the pulsar 
was in the focus of the telescope we detected a faint point source $3\farcs4\pm4\farcs8$ 
near to the radio pulsar's timing position. The positional coincidence makes this X-ray 
source a very likely counterpart of the radio pulsar. It is detected with $17\pm4$ source 
counts in the PN and MOS1/2 data. Converting the photon flux to an energy 
flux yields $F_\text{X}^\text{0.1--2 keV}=(8\pm 4) \times 10^{-15}~\text{erg s}^{-1}~\text{cm}^{-2}$ 
in the 0.1 to 2 keV range.

\subsection{PSR B0919+06}
The pulsar PSR B0919+06 (PSR J0922+0638) was discovered in the second Molonglo 
pulsar survey \citep{B0919Discovery}. It has a measured parallax which
places it at a distance of $1.1\pm 0.2$ kpc \citep{2012ApJ...755...39V}.
In archival XMM-Newton data of $\sim 65$ ksec observing time, taken with 
the pulsar position in the focus of the telescope, we detected a point 
source $2\farcs4\pm2\farcs1$ near to the radio pulsar's timing position. 
Merging the PN and MOS1/2 data yielded $171\pm18$ source counts which 
supported a brief spectral analysis. 

The energy spectrum was extracted by selecting all events from within a circle of 
radius $15\arcsec$ centered on the potential pulsar counterpart. The background 
spectrum was extracted by selecting all events from an annulus with radii 
$5\arcsec$ and $75\arcsec$, respectively. The spectra were binned so as to 
have at least 20 counts/bin in the PN data and 10 counts/bin in the MOS1/2 data. 
Fitting a power law model to this spectral data gave a $\chi^2=4.0$ for 8 
d.o.f.~and an $N_\text{H}<1.2\times 10^{21} \text{cm}^{-2}$. The column 
absorption is in agreement with the value estimated from the radio pulsar's
dispersion measure, $N_\text{H}^{DM}= 8.42\times 10^{20}~\text{cm}^{-2}$, 
using the model of \citet{2002astro.ph..7156C}. The fitted photon index 
$\Gamma$ is $2.3_{-0.4}^{+0.8}$. We also tested a single blackbody model 
but did not find it to fit the data ($\chi^2=15.9$ for 8 d.o.f.).

\subsection{PSR J1044--5737}
PSR J1044--5737 is a young and energetic pulsar. It was discovered in a 
blind frequency search of Fermi LAT data \citep{2010ApJ...725..571S}.
No radio dispersion measure has been reported for this pulsar in the 
literature so far. Its distance is therefore still unconstrained.
The Chandra data archive has a 10 ksec deep observation with the pulsar
in the telescope focus. Analyzing this data we detected a point source matching the
pulsar's radio timing position by $0\farcs6\pm0\farcs9$. This source
is detected with $30\pm6$ source counts. Converting the photon flux 
to an energy flux yields $F_\text{X}^\text{0.1--2 keV}=2.4\pm 1.0 
\times 10^{-14}~\text{erg s}^{-1}~\text{cm}^{-2}$ in the 0.1 to 2 keV range.

\subsection{PSR J1112--6103}
PSR J1112--6103 was discovered in the Parkes multi-beam pulsar survey
\citep{J1112J1301_6305Discovery}. Two Chandra observations are covering the
radio position of the pulsar. In the first one (Obs.ID 6706) the pulsar position
 was in the focus of the telescope. The second observation (Obs.ID 8905) was
pointed on the HII region NGC 3576 in which the pulsar position was observed 
$2\farcs4$ off-axis. In the combined data we detected a point source 
$0\farcs6\pm0\farcs5$ near to the pulsar's timing position which makes it a
very likely counterpart of the pulsar. From both observations we have $331\pm19$ 
source counts which allowed a brief spatial and spectral analysis.

\begin{figure}
  \centering
  \resizebox{0.9\hsize}{!}{\includegraphics{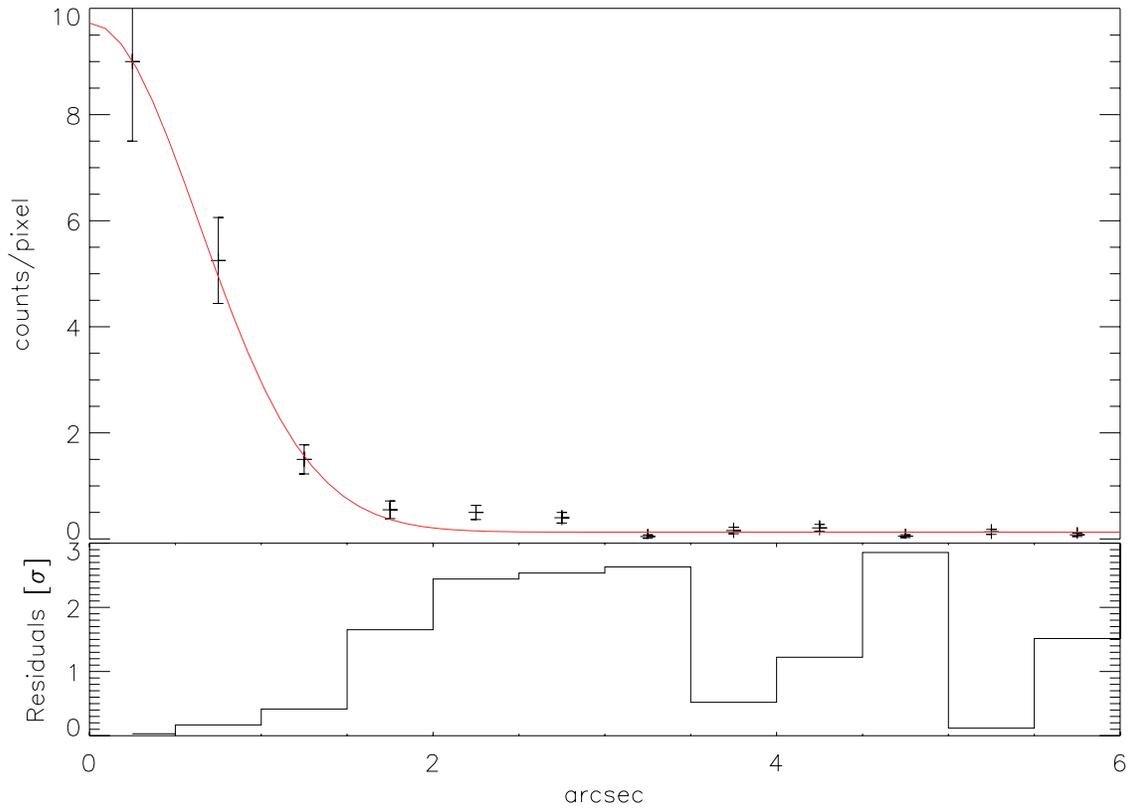}}
  \caption[Radial intensity distribution around PSR J1112--6103]{{\it Top:}
Radial intensity distribution around PSR J1112--6103 in obs.ID 6706. The red
line shows the simulated distribution for a point source convolved with a
Gaussian with a FWHM=$1\arcsec$. {\it Bottom:} Difference between this two
profiles, expressed in terms of the $1\sigma$ uncertainty in the derived counts
per annulus.}
  \label{fig:J1112.3}
\end{figure}

The CIAO-tool \sffamily srcextent\normalfont~indicated that the source is
extended at a 90\% confidential level. To explore the existence of a possible
PWN around PSR J1112--6103 we produced radial profiles centered on the deduced X-ray
position for both ACIS observations.
The profiles were obtained by integrating all counts in 12 concentric,
equally-spaced annuli with a minimum radius of $0\farcs5$ and maximum radius
$6\arcsec$ and dividing the selected counts by the respective ring area. This profiles were then
compared with the profile of a point source at the same off-axis angle,
simulated with MARX\footnote{See \url{http://space.mit.edu/cxc/marx/}} and
convolved with a Gaussian with a full width at half maximum (FWHM) of $1\arcsec$
as used by \citet{2003ApJ...585L..41R} to detect a faint PWN surrounding PSR
J0538+2817. Figure \ref{fig:J1112.3} shows the resulting diagram obtained for
obs.ID 6706. No extension of the X-ray source is visible with a significance of
$3\sigma$. Therefore, we see no evidence for a PWN surrounding J1112--6103. 

To extract the source counts for spectral analysis we selected all events within
an aperture of $3\arcsec$ centered on the counterpart. All counts were binned 
to have at least 20 cts/bin. 

\begin{figure}
  \centering
  \resizebox{0.9\hsize}{!}{\includegraphics[clip]{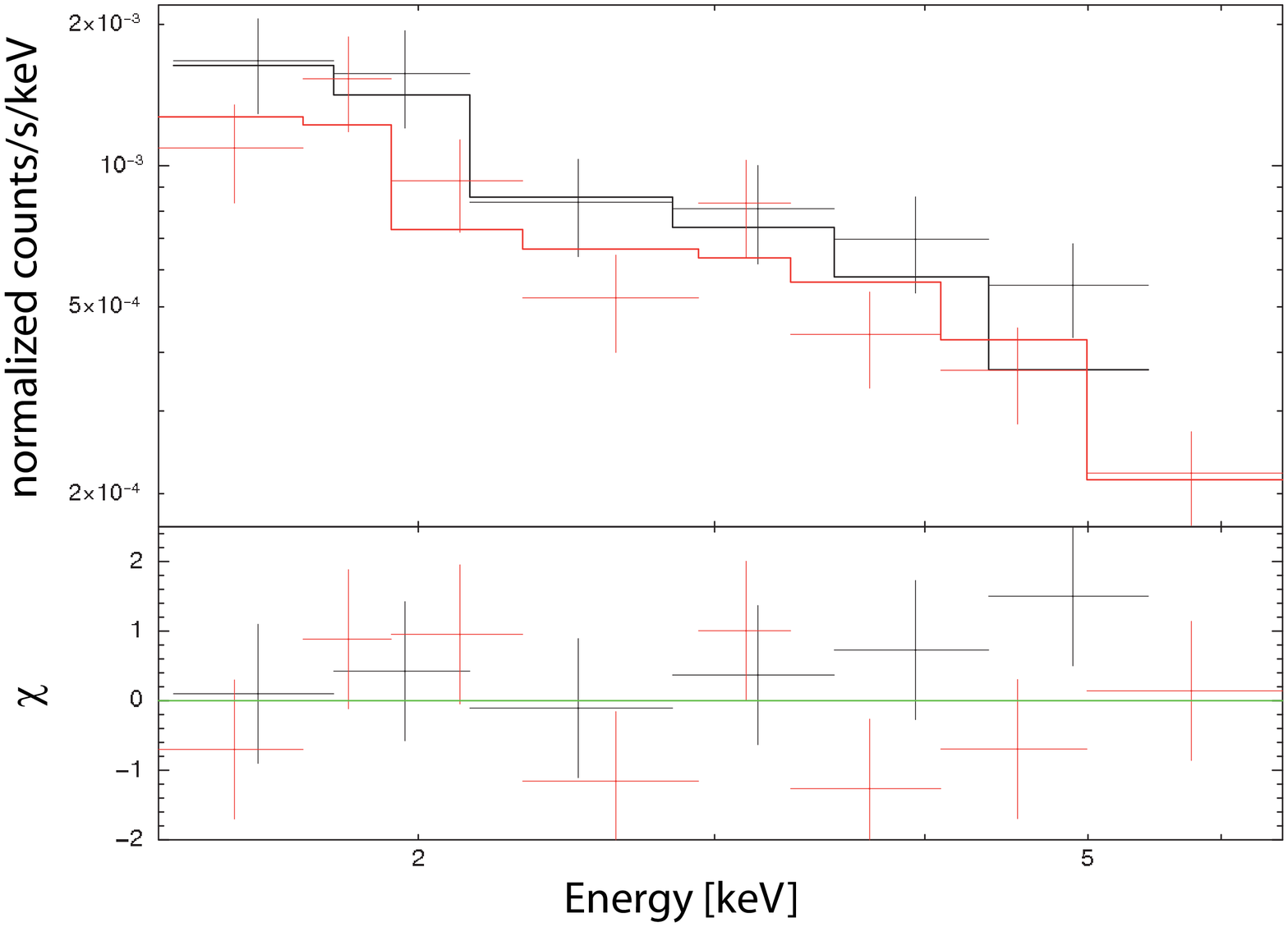}}
  \caption[X-ray spectrum of PSR J1112--6103]{X-ray spectrum of PSR J1112--6103
fitted with a power law model. The $\chi^2$ statistic is shown for comparison.}
  \label{fig:J1112.PL}
\end{figure}
 
Testing a power law model yields an excellent fit with $\chi^2=9.7$ for 11 d.o.f.
and a power law index of $1.5 \pm 0.7$ (see Figure \ref{fig:J1112.PL}). The fitted 
column density $N_\text{H}=1.2^{+1.1}_{-1.0}\times 10^{22}~\text{cm}^{-2}$ is in 
agreement with both, the results of the LAB Survey of Galactic HI 
\citep{2005A&A...440..775K}, where the column density in the direction
to PSR J1112--6103 is between $(1.0-1.3)\times 10^{22}~\text{cm}^{-2}$ and 
$1.8\times 10^{22}~\text{cm}^{-2}$ as estimated from the pulsar's radio
dispersion measure DM.

Although a blackbody model fit gives an acceptable description of the spectrum
as well ($\chi^2=12.7$ for 11 d.o.f.), the resulting blackbody temperature  
$T=(1.4\pm 0.2)\times 10^7$ K for an emitting radius $R_{\text{BB}} \sim 157\pm28$\,m 
is almost an order of magnitude higher than what is observed
in other X-ray detected rotation powered pulsars. The blackbody model is 
rejected for this reason.

\subsection{PSR B1221--63}
The radio pulsar PSR B1221--63 (PSR J1224--6407) was first detected by
\citet{B1221Discovery}. It was observed  on-axis by XMM-Newton with an
exposure time of $\sim 7$ ksec. Combining the MOS1/2 and PN data we 
detected a faint X-ray source which matches the pulsar position by 
$0\farcs6\pm0\farcs9$. In total we detected $(30\pm 8)$ source counts 
in the PN and MOS1/2 detectors. Using the net exposure time on this 
source of 3.9 and 6.4 ksec for the PN and the two MOS1/2 cameras, 
respectively, the photon flux converts to en energy flux of 
$F_\text{X}^\text{0.1--2 keV}=(1 \pm 0.6) \times 10^{-14}~\text{erg s}^{-1}~\text{cm}^{-2}$ in the 0.1 to 2 keV range.
 
\subsection{PSR J1301--6310}\label{J1301-6310}
The middle-aged pulsar PSR J1301--6310 was discovered in the Parkes multi-beam survey 
\citep{2003MNRAS.342.1299K}. XMM-Newton observed the pulsar position twice for a total
of $\sim 60$ ksec. In the combined data we detected a X-ray point source with $78\pm13$ 
source counts in the PN- and MOS1/2 cameras $1\farcs9\pm2\farcs3$ near to the radio pulsar's 
timing position. The recorded events allow for a very coarse spectral analysis for which we 
extracted the events from within a circle of $20\arcsec$ radius, centered on the position 
of the counterpart. Due to the small data statistics both a blackbody and power law spectrum 
fitted equally good. The fit of a power law results in $\chi^2=15.2$ for 15 d.o.f., an 
$N_\text{H}< 3.9 \times 10^{21}~\text{cm}^{-2}$ and a spectral index of $3.4^{+0.8}_{-0.6}$. 
Using a blackbody spectrum leads to $\chi^2=17.0$ for 15 d.o.f. and 
$N_\text{H} \leq 9\times 10^{20}~\text{cm}^{-2}$. After fixing $N_\text{H}$ to the $DM$ 
based value of $2.6\times 10^{21}~\text{cm}^{-2}$ both fits still appear to be reasonable 
(see Table \ref{tab3:spectral_psr_results}), albeit the blackbody radius computed from 
the blackbody normalization resulted in an unreasonable small value of $R_{BB}= 0.09 \pm 0.03$ km.

\subsection{PSR B1338--62}
PSR B1338--62 (PSR J1341-6220) is a young, Vela-like pulsar which was discovered in 
1985 at the Parkes radio telescope \citep{B1338Discovery}. The pulsar is associated
with the supernova remnant G308.8--0.1 \citep{1992Kaspi} and according to
\citet{2000Wang} among the most actively glitching pulsars known. The pulsar 
position was in the field of view of the XMM-Newton telescope for about 42 ksec
during two different epochs and with different off-axis angles. Merging the PN- 
and MOS1/2 data we detected two faint X-ray sources close to the radio pulsar's timing 
position: $S_1$ with the coordinates RA$=13^h 41^m 43.1^s  \pm 0.3^s$, 
DEC$=-62^\circ 20^m 18^s \pm 2^s$ and $S_2$ with RA$=13^h 41^m 43.3^s  \pm 0.3^s$, 
DEC$=-62^\circ 20^m 06^s \pm 2^s$. 

Source $S_1$ is close to the pulsar position, the difference is
$4\farcs2\pm 2\farcs4$. This source is only distinguishable from $S_2$ at
energies above $\sim 2$ keV (see Figure \ref{fig:XraySources}). Thus, it has a
hard spectrum. For $S_2$ we found a possible counterpart in the 2MASS All-Sky
Catalog of Point Sources, 2MASS J13414307-6220063 with a positional discrepancy
of only $1\farcs5 \pm 2\farcs2$.  

In order to fit the energy spectrum of the detected X-ray source $S_1$ we
extracted the counts from within a radius of $12\farcs5$, centered on the point 
source, and excluding events from the region around $S_2$. The background spectrum 
was extracted from a region close to $S_1$. Using a power law to fit the energy 
spectrum gave an acceptable fit ($\chi^2=3.1$ for 8 d.o.f.) with $N_H=4 \pm 3 
\times 10^{22}~\text{cm}^{-2}$. The column absorption is in agreement with the 
$DM$ based value of $2.2\times 10^{22}~\text{cm}^{-2}$ \citep{2002astro.ph..7156C}. 
Fixing $N_\text{H}$ to this value yields $\chi^2=3.2$ for 9 d.o.f. and a photon index 
$\Gamma=1.1 \pm 0.7$. A blackbody model fit gave $\chi^2=2.8$ for 7 d.o.f. with 
fixed $N_\text{H}$, but the resulting blackbody temperature $T=1.3_{-0.4}^{+1.8}\times 10^7$ K 
is too high to be consistent with neutron star cooling emission or coming from a hot
spot on the neutron star surface.

\subsection{PSR J1600--3053}
The high galactic latitude binary pulsar PSR J1600--3053 was detected by
\citet{2007ApJ...656..408J}. The XMM-Newton archive contains data from an
observation in which the pulsar was in the focus of the telescope for 
$\sim 25$ ksec. As in this observation the PN camera was operated in timing mode, 
not providing suitable spatial resolution for source detection, we analysed only the 
two MOS1/2 data sets and detected a faint X-ray source with $37 \pm 10$ 
source counts which matches the pulsar's radio timing position by $0\farcs8 \pm 2\farcs4$,
making it a very likely counterpart. Converting the photon flux to energy flux results in 
$F_\text{X}^\text{0.1--2 keV}=(6\pm 4) \times 10^{-15}~\text{erg s}^{-1}~\text{cm}^{-2}$.

\subsection{PSR J1658--5324}
This pulsar was detected in a survey of unidentified Fermi LAT sources with the
Parkes radio telescope \citep{2012ApJ...748L...2K}. It is an isolated millisecond 
pulsar. In the Chandra data archive we found an observation in which this pulsar 
was in the telescope focus for $\sim 10$ ksec. Analyzing this data we detected a faint
X-ray point source with $26\pm5$ counts which matches the pulsar position by 
$0\farcs22\pm0\farcs28$ what makes it a potential counterpart of the pulsar. Converting 
the photon flux of this counterpart to energy flux yields 
$F_\text{X}^\text{0.1--2 keV}=(1.6\pm 0.6) \times 10^{-14}~\text{erg s}^{-1}~\text{cm}^{-2}$. 

\subsection{PSR J1730--2304}
The solitary millisecond pulsar J1730--2304 was discovered by \citet{1995ApJ...439..933L}. 
The XMM-Newton data archive holds data from an observation in which the pulsar was in the
focus of the telescope for $\sim 23$ ksec.  Analyzing this data we detected an X-ray source 
with in total $248\pm21$ counts in the PN and MOS1/2 data. It matches the pulsar position 
by $2\farcs7\pm2\farcs1$ which makes it a potential X-ray counterpart of the pulsar.

The data statistics allows for a brief spectral analysis for which we extracted the source 
counts from a circle of radius $20\arcsec$, centered on the position of the point source.
For the spectral binning we used at least 20 cts/bin and 10 cts/bin for the PN and MOS1/2 
data, respectively. 

Fitting a single power law or blackbody model to the data showed systematic 
deviations  at $\approx 1.4$ keV and did not result in acceptable fits. Adding
a Gaussian emission line profile to the power law model improved the fit to
yield $\chi^2=17.0$ for 16 d.o.f., though the statistical justification for the 
need of such a spectral line component is barely significant due to the low photon
statistics in the spectrum. The line center was fitted to be $1.36_{-0.1}^{+0.08}$ keV. 
The blackbody model still showed systematic deviations even after adding a spectral 
emission line component to the model spectrum. Deeper Chandra observations which 
provide better data statistics are required to further explore the source spectrum 
with higher significance.

\subsection{PSR J1731--1847}
The binary pulsar PSR J1731--1847 was discovered in the High Time Resolution
Universe Pulsar survey \citep{2011MNRAS.416.2455B}. It was observed on axis,
both with XMM-Newton and Chandra for $\sim 40$ ksec and {$\sim 10$ ksec}, 
respectively. In the merged XMM-Newton MOS1/2 data a point source is clearly detected
$1\farcs2\pm2\farcs7$ arcsec from the radio position with in total $47\pm13$
counts. The photon flux corresponds to an energy flux of $F_\text{X}^\text{0.1--2 keV}=(6\pm
4) 
\times 10^{-15}~\text{erg s}^{-1}~\text{cm}^{-2}$. The XMM-Newton PN data were 
not usable for the analysis because of high background flares which strongly 
reduced the S/N ratio in that data. An optical source with 
visible magnitude $V=17.34^\text{mag}$ is within the $1\sigma$ error circle 
around the position obtained in the XMM-Newton data. However, using the Chandra 
data we can rule out the possibility that the optical source is the X-ray 
counterpart of the detected XMM source. The Chandra source position 
is more than 3\arcsec away from the optical source whereas the positional 
uncertainty is only $0\farcs5$.

\subsection{PSR J1816+4510}
This eclipsing millisecond pulsar was studied in detail in the optical, UV and $\gamma$-ray 
band by \citet{2012ApJ...753..174K}. The Chandra data archive holds an observation in which
the pulsar position was in the focus of the telescope for $\sim 34$ ksec. Using this 
observation we detected a faint potential X-ray counterpart $0\farcs4\pm0\farcs3$ 
near to the pulsars timing position. It has $16\pm4$ source counts. The photon flux
converts to an energy flux of $F_\text{X}^\text{0.1--2 keV}=(3.1\pm 1.6) \times 10^{-15}~\text{erg
s}^{-1}~\text{cm}^{-2}$.

\subsection{PSR B1822-09}
The middle-aged pulsar B1822-09 (PSR J1825--0935) was first mentioned in the
literature by \citet{1972Natur.240..229D}. In the XMM-Newton data archive we 
found a short $\sim 5$ ksec deep observation targeted on this pulsar. Using this 
data we detected an X-ray source $1\farcs5\pm2\farcs3$ near to the pulsar's timing 
position which makes it a potential counterpart of the pulsar. The X-ray source has
$46\pm11$ source counts in the merged PN and MOS1/2 data. The photon flux corresponds
to an energy flux of  $F_\text{X}^\text{0.1--2 keV}=(1.7\pm 0.8) \times
10^{-14}~\text{erg s}^{-1}~\text{cm}^{-2}$.

\subsection{PSR J1832--0836}
This isolated millisecond pulsar was discovered in the radio band during the
High Time Resolution Universe Pulsar survey \citep{2013MNRAS.433..259B}. In an
archival observation of AX J1832.3--0840, a low luminous X-ray pulsator
\citep{2010MNRAS.402.2388K}, the pulsar was observed $\sim 4\arcmin$ off-axis.
In the merged EPIC-PN and MOS1/2 data a faint X-ray source was detected
$1\farcs8\pm2\farcs6$ near to the radio timing position with $34\pm9$ counts,
making it a potential pulsar counterpart. Its photon flux converts to an energy
flux of $F_\text{X}^\text{0.1--2 keV}=(3.8\pm 2.0) \times 10^{-15}~\text{erg s}^{-1}~\text{cm}^{-2}$.

\subsection{PSR J1911--1114}
Radio emission from the binary millisecond pulsar PSR J1911--1114 was discovered
by \citet{1996MNRAS.283.1383L}. The XMM-Newton data archive holds a $\sim 50$ ksec deep 
observation of this pulsar in which the PN was operated in timing mode. Using the
EPIC MOS1/2 data we detected an X-ray source $3\farcs1\pm2\farcs8$ near to the pulsars
position. It as $54 \pm 13$ source counts in the merged MOS1 and MOS2 data. The photon
flux converts to an energy flux of $F_\text{X}^\text{0.1--2 keV}=3.7\pm 1.8 \times 10^{-15}~\text{erg
s}^{-1}~\text{cm}^{-2}$.

\subsection{PSR J2017+0603}
PSR J2017+0603 is a millisecond pulsar in a binary system. It was first detected
by its pulsed $\gamma$-ray emission with the Fermi satellite \citep{2011ApJ...732...47C}. 
In the discovery paper the authors derived an upper limit on the X-ray flux in the 
$0.5-8$ keV band of $F_\text{X}^\text{0.5--8 keV} <6\times10^{-14}~\text{erg s}^{-1}\text{cm} ^{-2}$. 
Using an archival Chandra ACIS-S observation aimed on this pulsar we detected an X-ray source
$0\farcs2\pm0\farcs2$ near to the radio pulsar's timing position, making it a potential X-ray
counterpart of the pulsar. It has $19\pm5$ source counts which we convert to an energy 
flux of $F_\text{X}^\text{0.1--2 keV}= (1.1\pm0.6)\times10^{-14}~\text{erg s}^{-1} \text{cm} ^{-2}$.

\subsection{PSR J2222--0137}
The binary millisecond pulsar PSR J2222--0137 \citep{2012arXiv1209.4293B} was observed on-axis 
for $\sim 30$ ksec with the Chandra ACIS-S camera. In this data we detected an X-ray source 
with $26\pm 5$ source counts $0.1\pm0.2$ arcsec near to the pulsar's radio position, making it
a potential pulsar counterpart. The photon flux corresponds to an energy flux of 
$F_\text{X}^\text{0.1--2 keV}= 4.5\pm 2.0\times10^{-15}~\text{erg s}^{-1}\text{cm} ^{-2}$.

\section{Upper limits on the X-ray counterparts of radio pulsars}

In Table \ref{tab5:upper_limits_results} the non-thermal flux $F_\text{0.1--2
keV}^\text{PL}$ and the upper limits on the blackbody temperature $T_\infty$, as
measured by a distant observer at infinity, are listed for all rotation-powered 
pulsars for which no X-ray counterpart was detected in the archival data. Furthermore, 
the non-thermal luminosity $L_\text{0.1--2 keV}^\text{PL}=4 \pi d^2_{U\!L} F_\text{0.1--2 keV}^\text{PL}$ 
and the X-ray efficiency $\mu_\text{0.1--2 keV}^\text{PL}= L_\text{0.1--2 keV}^\text{PL}/\dot
E$ are listed there as well. Therein $\dot E=-4 \pi^2 I \dot P P^{-3}$ is the spin-down
luminosity and $I= 10^{45}$ g cm$^2$ the canonical moment of inertia for a
neutron star with a mass of $1.4~M_\odot$ and a radius of 10 km. In Table 
\ref{tab6:psr_with_lowest_muX} the pulsars with the lowest X-ray efficiency
upper-limit are displayed separately. 

\begin{figure*}[!tbp]
  \resizebox{0.9\hsize}{!}{\includegraphics[clip]{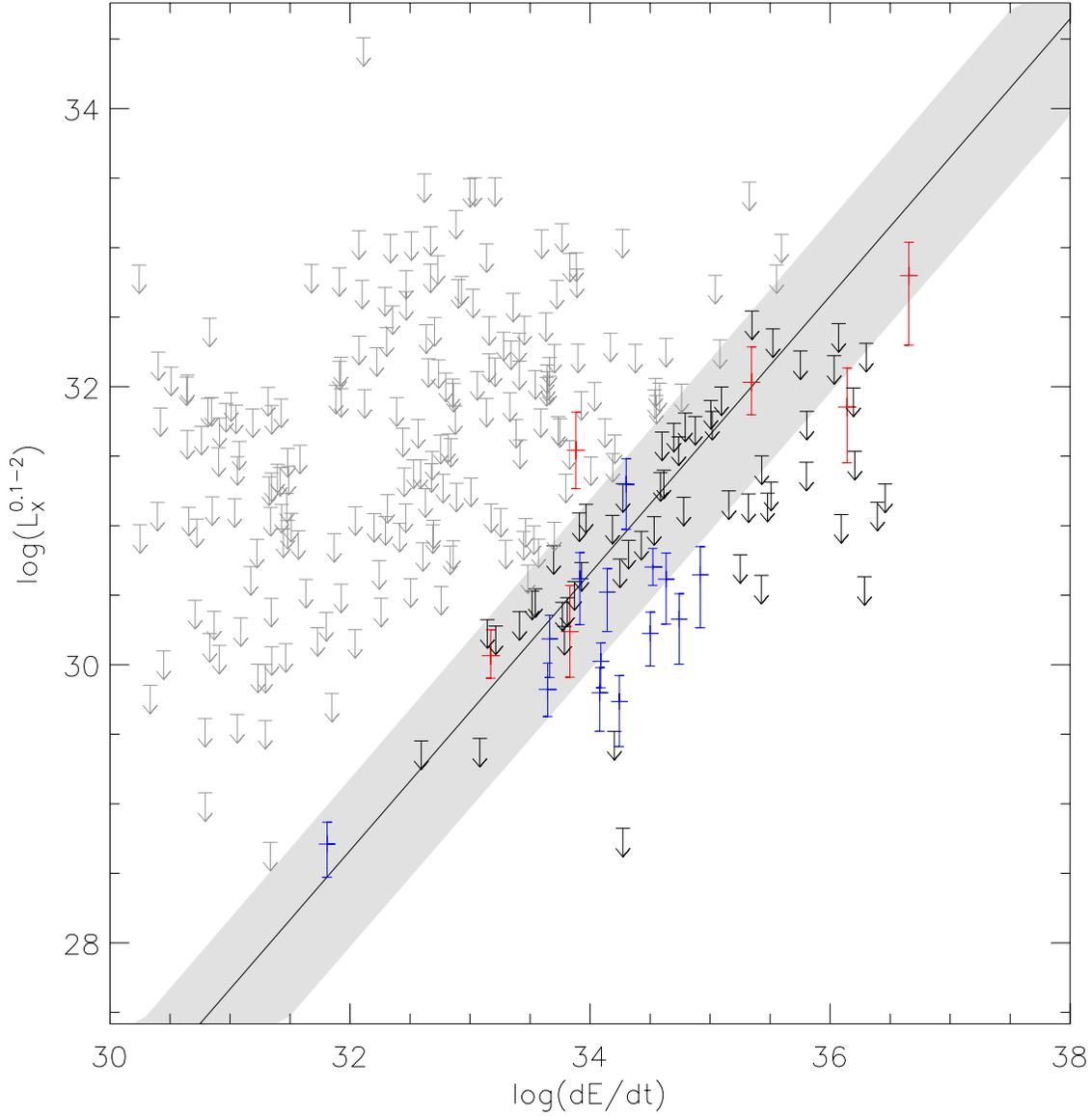}}
  \caption[Spin-down luminosity plotted against the isotropic X-ray
luminosity]{Spin-down luminosity $dE/dt$ plotted against the isotropic X-ray luminosity 
$L_X^{0.1-2}$ in the 0.1--2 keV band. The red error bars represent the $1\sigma$ confidence 
range and indicate those potential pulsar counterparts for which we could perform
a brief spectral analysis in order to obtain the energy flux and X-ray luminosity.
Values for the remaining potential counterparts are plotted in blue. 
In the case of the non-detected pulsars the $2\sigma$ upper limit is indicated by an arrow. 
The straight line represent the relation found by \citet{2009ASSL..357...91B} 
with an uncertainty range indicated by the gray shaded region.}
  \label{figure:Lx_Edot}
\end{figure*}

Figure \ref{figure:Lx_Edot} displays the correlation of the isotropic X-ray
luminosity $L_X^{0.1-2}$ in the 0.1--2 keV band with the pulsar's spin-down luminosity
$\dot E$. In red color we plotted the X-ray luminosity for all potential pulsar
counterparts for which we were able to perform a brief spectral analysis. 
Blue error bars indicate the X-ray luminosity of potential counterparts for which 
we had to assume the spectral shape to convert the photon flux to an energy flux
resp. X-ray luminosity. In addition we added the $2\sigma$ upper limits on the 
non-thermal X-ray luminosity for all pulsars, for which at least a dispersion 
measure based distance estimate is known. The gray shaded region in this graph 
shows the correlation between $L_X^{0.1-2}$ and $\dot E$ as found by \citet{2009ASSL..357...91B}:

\begin{equation} \label{eq:Lx_F}
L_X^{0.1-2}=10^{-3.24^{+0.26}_{-0.66}} \dot E^{-0.997^{+0.008}_{-0.001}}.
\end{equation}
For this fit the authors used $\approx 80$ rotation-powered pulsars for which
spectral information was obtained in XMM-Newton and Chandra observations, respectively. 

\begin{figure*}
  \centering
  \resizebox{0.75\hsize}{!}{\includegraphics[clip]{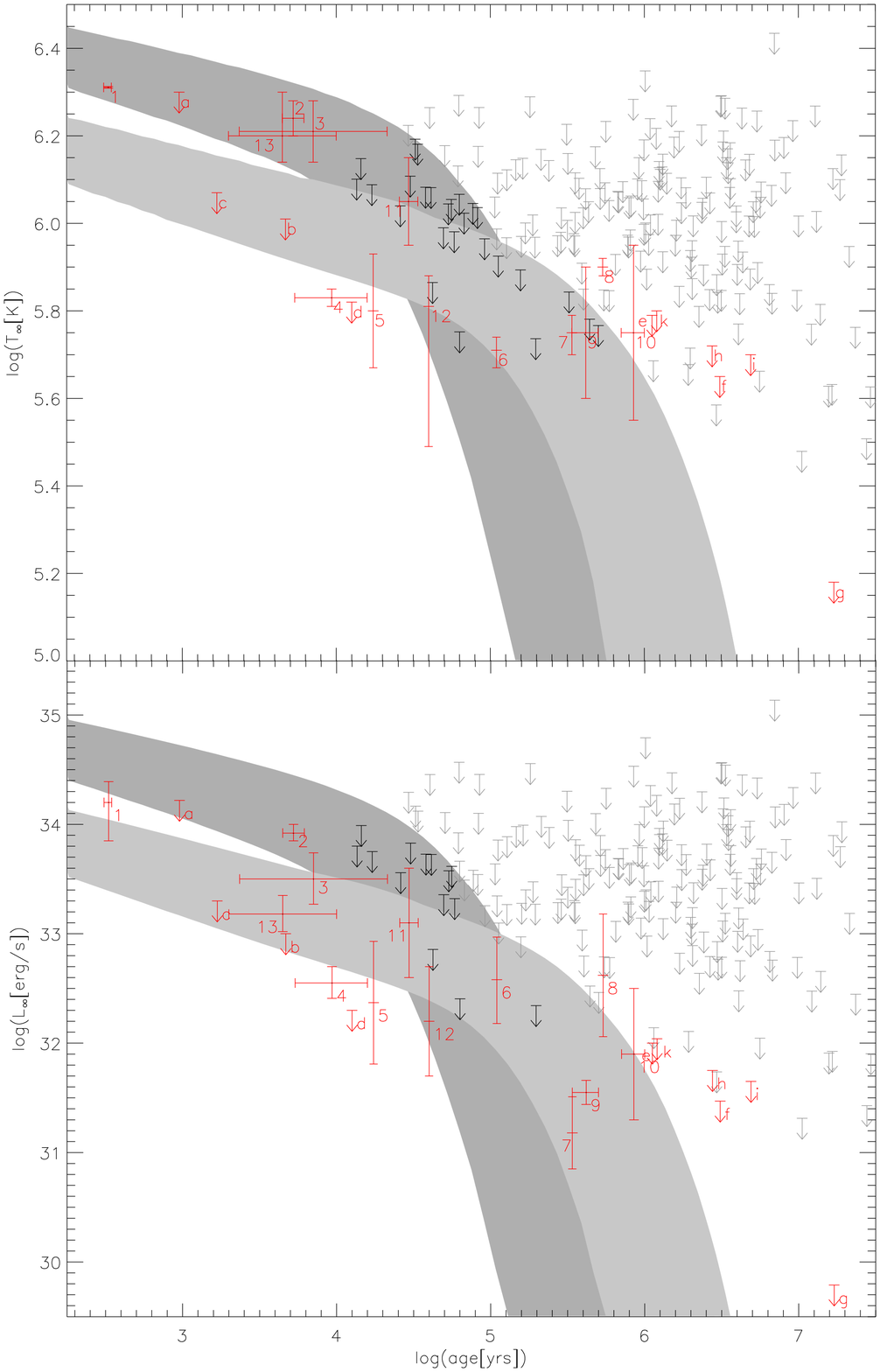}}
  \caption{Comparison of the derived upper limits on the temperature 
  (\textit{upper panel}) and bolometric luminosity (\textit{lower
  panel}) of the radio pulsars for which data where found in the XMM-Newton
  or Chandra archives. The plotted cooling curves were taken from \citet{2009ApJ...707.1131P} 
  (see text for more informations).
}
  \label{figure:Tau_Te}
\end{figure*}

If no association between a pulsar and e.g.~a hosting supernova remnant is
possible, the only way to estimate the age of a pulsar is by its spin-down 
age $\tau=2P/\dot P$. The Crab pulsar is the only pulsar for which the true
historic age is known \citep{2003LNP...598....7G}.  In this case the
difference between the spin-down age $\tau$ and the true historic age is of the
order of $\approx 20\%$. In Figure \ref{figure:Tau_Te} we plotted the 
derived $3\sigma$ temperature upper limits $T_\infty$ against the spin-down 
age of the pulsars. In this figure we show an error bar for the pulsar age 
only if kinematic informations are available, i.e., the age is known with 
accurately determined errors. 

To compare the data with actual cooling models we used the cooling curves of
\citet{2009ApJ...707.1131P}. For the computation of the curves \citet{2009ApJ...707.1131P} 
used a $1.4~M_\odot$ neutron star with the equation of state taken from 
\citet{1998PhRvC..58.1804A}, different neutron $^1$S$_0$, proton $^1$S$_0$, and
neutron $^3$P$_2$ gap models were taken from \citet{2004ApJS..155..623P}. The authors 
showed that the cooling of a neutron star in the minimal cooling paradigm is only
slightly dependent on the neutron star mass and the used neutron and proton
$^1$S$_0$ gap model. The main uncertainties in the minimal cooling paradigm are
a possible evolution of the chemical composition of the envelope of the
neutron star and hence its opacity, as well as the number of different gap
models that are allowed by current measurements. In Figure \ref{figure:Tau_Te}
we therefore compare our data with the allowed temperature range of these
models.

The lower panel in Figure \ref{figure:Tau_Te} shows the bolometric luminosity plotted 
against the pulsars' spindown age. We added all pulsars from the literature for which 
thermal emission from their surface has been detected or upper limits on the temperature 
have been published. In this plot the upper limits derived in this paper are indicated by 
gray arrows. 
The cooling models shown are computed for a $1.4~M_\odot$ star with the equation of state 
adopted from \citet{1998PhRvC..58.1804A}, different neutron $^1$S$_0$, proton $^1$S$_0$, and
neutron $^3$P$_2$ gap models as introduced by \citet{2004ApJS..155..623P}. The
area in dark and light gray contain cooling models computed with heavy- and light-element 
dominated envelopes, respectively. The error bars indicate the error range in temperature 
or bolometric luminosity of all pulsars with detected thermal emission, e.g. (1) Cas A: \citet{2011MNRAS.412L.108S,2006ApJ...645..283F};
(2) J0822--4300: \citet{1999ApJ...525..959Z,2012ApJ...755..141B};
(3) J1210--5226: \citet{1998A&A...331..821Z,1988ApJ...332..940R};
(4) B0833--45: \citet{2001ApJ...552L.129P,1995Natur.373..587A};
(5) B1706--44: \citet{2004ApJ...600..343M,1995ApJ...441..756K};
(6) B0656+14: \citet{2002ApJ...574..377M,2003ApJ...593L..89B};
(7) J0633+1748: \citet{1997ApJ...477..905H,2001ApJ...561..930C};
(8) B1055--52: \citet{2005ApJ...623.1051D,2003MNRAS.342.1299K};
(9) J1856--3754: \citet{2003A&A...399.1109B,2013MNRAS.429.3517M};
(10) J0720--3125: \citet{2003ApJ...590.1008K,2011MNRAS.417..617T};
(11) J0538+2817: \citet{2004MmSAI..75..458Z,2003ApJ...593L..31K};
(12) B2334+61: \citet{2006ApJ...639..377M,2004ApJ...616..247Y};
(13) J1119--6127: \citet{2008ApJ...684..532S,2012ApJ...754...96K}.

All pulsars with derived temperature upper limit are indicated by a red arrow, e.g.
(a) Crab: \citet{2011ApJ...743..139W};
(b) J0205+6449: \citet{2004ApJ...616..403S,2008ApJS..174..379F};
(c) J1124--5916: \citet{2003ApJ...591L.139H,2002ApJ...567L..71C};
(d) RX J0007.0+7302: \citet{2004ApJ...612..398H,2004ApJ...601.1045S};
(e) B2224+65: \citet{2004ApJ...610L..37G};
(f) B1929+10: \citet{2007A&A...467.1209H};
(g) B0950+08: \citet{2006ApJ...645.1421B,2004ApJ...604..339C};
(h) B0628--28: \citet{2004ApJ...616..452Z,2002ApJ...571..906B};
(i) B0823+26: \citet{2005ApJ...633..367B};
(k) J2043+2740: \citet{2004ApJ...615..908B,2002ApJ...571..906B}.
We added the error in the age for pulsars with known kinematic age.
These data are taken from \citet{2004ApJS..155..623P,2009ApJ...707.1131P},
\citet{2013RPPh...76a6901O}, and \citet{2011ApJ...734...44Z}. Additionally, for
the central compact object in Puppis-A we adopted the new age estimate of
$(5.2\pm1.0)\times 10^3$ years \citep{2012ApJ...755..141B}. 
In Table \ref{tab7:psr_with_lowest_temp} we summarize the neutron stars which
have the lowest $3\sigma$ upper limit on the surface temperature in our sample.
They are indicated by a black arrow in the upper panel of Figure \ref{figure:Tau_Te}.

The ATNF Pulsar Catalog (Catalogue version number = 1.54) which we used to searched 
for potential X-ray counterparts in the XMM-Newton or Chandra data archives has 
currently about 2500 rotation-powered pulsars in its database \citep{2005AJ....129.1993M}.
However, only about 2000 of them have published period derivatives and spindown energies. 
For some others no dispersion measure has been published so far, so that it is not possible 
to get an estimate on their distance and hence no estimate on their X-ray luminosity or
luminosity upper limit and/or X-ray efficiency. 

\section{Conclusion}\label{sec:discussion_psrSearch}

We have systematically searched the archives of the X-ray observatories Chandra 
and XMM-Newton for serendipitous datasets which include a radio pulsar in the 
field of view. All \identified~detected X-ray counterparts of radio
pulsars fulfill the following criteria (see also Section \ref{sec:obs}): The
discrepancy between the radio and X-ray source position is less than $3\sigma$,
the probability that an X-ray source is by-chance at the pulsar's position is
below $3\sigma$ and no optical source having a X-ray-to-visual flux ratio
typical for stars is found within the $3\sigma$ error box around the pulsar's 
counterpart.

The non-thermal X-ray efficiency of all potential counterparts and or their 
upper limits are in agreement with the predictions made by \citet{2009ASSL..357...91B}. 
Only a small scatter is noticeable which is in agreement with statistical fluctuations 
\citep[e.g.][figure 18]{2010arXiv1006.0335B} and the uncertainties introduced
by e.g.~assuming the spectral parameters for the pulsars with less than 70 detected
counts. The uncertainties that can cause the scatter in equation \ref{eq:Lx_F} is
addressed separately in the next section. 

Four of the \identified~potential counterpart which we detected in our correlation 
were mentioned in the literature during the analysis and publication process of this paper. 
J1341--6220 is listed as 
radio pulsar with X-ray counterpart in \citet{2004ApJS..153..269K} and \citet{2008Smith}, 
but no detailed analysis of the X-ray properties of PSR J1341--6220 was presented by 
the authors. For the radio pulsars PSR J1112--6103 and PSR J2222--0137 a detection 
is claimed by \citet{2011ApJS..194...16T} and \citet{2011AIPC.1357...32B}, respectively,
but these authors did not investigate the X-ray data in detail too. In any case, we consider
this as a confirmation of our findings which strengthens the detections of the other potential
counterparts too.

\subsection{Non-thermal X-ray emission}

In a first step, the consistency of the measured upper limits on the non-thermal X-ray
emission was verified. This was done by comparing the values derived in Section
\ref{sec:upper_limits} with upper limits found in the literature. The limits on
$\mu_\text{X}$ of \citet{2012ApJS..201...37K} were confirmed, even though a
different energy band and a larger confidential level of 95.4~\% was used in
this work. Only in the case of PSR J1837--0604 our derived upper limit on the
flux is three times higher. However, this can be explained by the fact that in
\citet{2012ApJS..201...37K} the counts were not corrected for the encircled
energy fraction.

Thereafter, the non-thermal X-ray luminosity for the \identified~X-ray
counterparts and the upper limits derived for the \psrLxUL~ remaining pulsars 
were used to further study the scatter in the relation of the non-thermal X-ray 
luminosity to the spin-down power. Eleven sources were found with upper limits 
on $\mu_\text{0.1--2 keV}$ below $10^{-4}$.
This confirms the trend for finding smaller X-ray efficiencies with increasing 
sample size and higher sensitivity of X-ray observatories. The X-ray efficiency 
of a rotation-powered pulsar can be as high as $1.9\times 10^{-3}$, e.g., for the 
Crab pulsar \citep{2009ASSL..357...91B}, but also be below $2.2\times 10^{-6}$ as 
for PSR J0940--5428. 

It should be noted that young pulsars with $\dot E > \dot E_c \approx 4\times
10^{36}$ erg/s could manifest a distinct pulsar wind nebula
\citep[PWN,][]{2004Gotthelf}. If such a PWN exists and/or thermal emission from
the pulsar's surface is hot enough to contribute to the X-ray emission, the
non-thermal X-ray efficiency of several pulsars could be even lower than the
determined upper limit.

Theory predicts a distinct relation between the X-ray efficiency and the
spin-down power \citep[e.g.][]{2004ApJ...601.1038W}. The detected large scatter contradicts such a tight relation
and favors the interpretation of \citet{2009ASSL..357...91B} that the
efficiency of $\mu_\text{X} \sim 10^{-3}$ represents more the maximum
efficiency rather than a fixed correlation. It might be seen for pulsars
which are seen with an optimal viewing geometry. Because the orientation 
of the pulsar's magnetic axes with respect to the observer's line of
sight might not have been optimal for observation and no beaming correction
can be applied to the observed luminosity, the X-ray efficiency of pulsars
observed with an unfavorable viewing geometry appears to be smaller than 
that for pulsars observed with an optimal viewing geometry. A reason for 
the large dispersion in $\mu_\text{X}$ then would be mostly caused by the
viewing geometry and the inability to apply a beaming correction 
\citep[][and references therein]{2009ASSL..357...91B}.  

\citet{2010AIPC.1248...25K} reviewed all known PWNe and investigated
their X-ray efficiency. They found a scatter in the measured luminosities
too, despite of the fact that PWN emission is thought to be unbeamed 
\citep{2006ARA&A..44...17G}. Therefore, the beaming angle might be only 
one of several parameters which might cause the observed scatter in $\mu_\text{X}$ 
for pulsars. 

Certainly, erroneous distances contribute to the uncertainty in the derived 
X-ray luminosities too. Although the errors in the distance measurement were 
included in our $\mu_\text{X}$ calculations, most distances used in this work
are based on the radio dispersion measure $DM$ and large fluctuations in
it are not included in the model calculations. This could easily introduce a 
scatter of roughly one order of magnitude in the derived X-ray luminosity for 
some sources.

Finally, another reason for the scatter in $\mu_\text{X}$ could be a wrongly determined 
spin-down luminosity. In the last years several intermittent pulsars have been 
discovered \citep[e.g.,][]{2006Sci...312..549K,2012ApJ...746...63C,2012ApJ...758..141L},
which at first seemed to be ordinary radio pulsars, but later have been seen
to switch their radio emission off quasi-periodically, then showing different spin-down 
rates at time scales from months to years. In the so called on-state the spin-down 
rate increases by a factor of 2 to 3 compared to the off-state. Because the 
spin-down luminosity is proportional to the spin-down rate and for some radio
pulsars only a few observations exist, it can not be excluded that some sources have 
indeed a lower $\dot E$ than calculated from their period and period
derivatives. 

\subsection{Thermal X-ray emission}

As for the non-thermal X-ray emission, the consistency of the measured upper
limits were tested. Temperature upper limits for eight pulsars can be found in
the literature \citep{2004ApJ...610L..37G, 2009MNRAS.400.1445K,
2013ApJ...764....1O, 2013ApJ...764..180K}, including the coolest pulsar in the
sample presented in this work, PSR J2144--3933. For six pulsars similar values
for the temperature upper limit were found, with a minor deviation of $\pm
10~\%$. This small discrepancy may be due to the fact that
\citet{2013ApJ...764....1O} did not include the distance error in their
calculations. Additionally, it is difficult to compare the results in detail
because all aforementioned authors used slightly different methods to calculate
the upper limits and no detailed information is given in
\citet{2013ApJ...764....1O} about the effective exposure time and radius of
the annulus they used to compute the upper limits for PSRs J1814--1744 
and J1847--0130. However, the method used in this work makes use of all 
archival observations of a pulsar and, thus, in general derives deeper upper 
limits. In the case of PSR B1845--19 \citep{2013ApJ...764....1O} and 
PSR J0157+6212 \citep{2004ApJ...610L..37G} the temperature upper limits we 
deduced in this work are $32\%$ and $36\%$ lower than what is reported in
the literature. For PSR B1845--19 this is because two archival XMM-Newton
observations were used rather than one as in \citet{2013ApJ...764....1O}. In 
the case of PSR J0157+6212 the difference appears partly due to a different 
$N_\text{H}$ value, though a large, and yet unexplained, difference remains 
between the measured upper limit in our work and that by \citet{2004ApJ...610L..37G}. 

In total, $3\sigma$ upper limits on the blackbody temperature for \psrTempUL~pulsars 
were derived, ranging from $2.2\times 10^5$ K for PSR J2144--3933 to 
$2.7\times 10^6$ K for PSR J1822--1617. All measured temperature upper limits are in agreement
with current cooling models, which assume only nuclear matter in the core of a
neutron star and Cooper-pair breaking and formation as the main source of
neutrino emission. These models include the uncertainties regarding the allowed
amounts of light elements in the envelope of the star and the different possible
neutron $^3$P$_2$ gap models. A comparison of the upper limits with other cooling 
curves \citep[e.g.][]{1998PhR...292....1T, 2004ARA&A..42..169Y, 2009ASSL..357..289T} 
shows that they are in agreement with models that assume standard cooling with 
Cooper-pair breaking and formation too.

\section{Future prospects}\label{sec:summary_psrSearch}

In Table \ref{tab6:psr_with_lowest_muX} and \ref{tab7:psr_with_lowest_temp} we
list those pulsars which appear to have a rather low X-ray efficiency or $3\sigma$ 
temperature upper limit. Further observations of these sources will allow
to explore them in more detail which might result in even tighter
constraints on the cooling models or the theories that try 
to explain the pulsar's non-thermal X-ray efficiency. Detecting, for example, 
thermal X-ray emission from a neutron star with a surface temperature lower 
than what is predicted by the minimal cooling paradigm would require to include 
enhanced cooling mechanisms in those models to explain that observation.

In fall 2017 the eROSITA Observatory (extended ROentgen Survey with an 
Imaging Telescope Array) is scheduled for launch on the Russian Spectrum X-Gamma
Satellite. Three month after the start it will begin to survey the X-ray sky 
with an XMM-Newton like sensitivity \citep{2012arXiv1209.3114M}. This data will enable us to study 
all $\approx 2500$ known pulsars in the X-ray band with three times better 
spatial resolution and $\approx 20$ times higher sensitivity than  
possible in the ROSAT All-Sky Survey. The eROSITA survey will
provide unique X-ray data which enables us to investigate a larger and even 
more unbiased pulsar sample than possible with the data available in the 
XMM-Newton and Chandra data archives.

\acknowledgments
\noindent{\bf Acknowledgments}\linebreak
We acknowledge the use of the XMM-Newton and Chandra data archive. T.P.
acknowledges support from and participation in the International Max Planck
Research School on Astrophysics at the Ludwig Maximilian University of Munich,
Germany, IMPRS. 

\bibliographystyle{apj} 
\bibliography{literatur} 

\begin{thebibliography}{}
\expandafter\ifx\csname natexlab\endcsname\relax\def\natexlab#1{#1}\fi

\bibitem[{{Akmal} {et~al.}(1998){Akmal}, {Pandharipande}, \&
  {Ravenhall}}]{1998PhRvC..58.1804A}
{Akmal}, A., {Pandharipande}, V.~R., \& {Ravenhall}, D.~G. 1998, \prc, 58, 1804

\bibitem[{{Arzoumanian} {et~al.}(2011){Arzoumanian}, {Gotthelf}, {Ransom},
  {Safi-Harb}, {Kothes}, \& {Landecker}}]{2011ApJ...739...39A}
{Arzoumanian}, Z., {Gotthelf}, E.~V., {Ransom}, S.~M., {et~al.} 2011, \apj,
  739, 39

\bibitem[{{Aschenbach} {et~al.}(1995){Aschenbach}, {Egger}, \&
  {Tr{\"u}mper}}]{1995Natur.373..587A}
{Aschenbach}, B., {Egger}, R., \& {Tr{\"u}mper}, J. 1995, \nat, 373, 587

\bibitem[{{Bates} {et~al.}(2011){Bates}, {Bailes}, {Bhat}, {Burgay},
  {Burke-Spolaor}, {D'Amico}, {Jameson}, {Johnston}, {Keith}, {Kramer},
  {Levin}, {Lyne}, {Milia}, {Possenti}, {Stappers}, \& {van
  Straten}}]{2011MNRAS.416.2455B}
{Bates}, S.~D., {Bailes}, M., {Bhat}, N.~D.~R., {et~al.} 2011, \mnras, 416,
  2455

\bibitem[{{Becker}(1995)}]{1995PhDTBecker}
{Becker}, W. 1995, PhD thesis, LMU, Germany

\bibitem[{{Becker}(2009)}]{2009ASSL..357...91B}
{Becker}, W. 2009, in Astrophysics and Space Science Library, Vol. 357,
  Astrophysics and Space Science Library, ed. W.~{Becker}, 91

\bibitem[{{Becker} {et~al.}(2010){Becker}, {Huang}, \&
  {Prinz}}]{2010arXiv1006.0335B}
{Becker}, W., {Huang}, H.~H., \& {Prinz}, T. 2010, ArXiv e-prints,
  arXiv:1006.0335

\bibitem[{{Becker} {et~al.}(2005){Becker}, {Jessner}, {Kramer}, {Testa}, \&
  {Howaldt}}]{2005ApJ...633..367B}
{Becker}, W., {Jessner}, A., {Kramer}, M., {Testa}, V., \& {Howaldt}, C. 2005,
  \apj, 633, 367

\bibitem[{{Becker} {et~al.}(2012){Becker}, {Prinz}, {Winkler}, \&
  {Petre}}]{2012ApJ...755..141B}
{Becker}, W., {Prinz}, T., {Winkler}, P.~F., \& {Petre}, R. 2012, \apj, 755,
  141

\bibitem[{{Becker} \& {Tr{\"u}mper}(1997)}]{1997A&A...326..682B}
{Becker}, W., \& {Tr{\"u}mper}, J. 1997, \aap, 326, 682

\bibitem[{{Becker} \& {Tr{\"u}mper}(1999)}]{1999A&A...341..803B}
---. 1999, \aap, 341, 803

\bibitem[{{Becker} {et~al.}(1993){Becker}, {Tr{\"u}mper}, \&
  {{\"O}gelman}}]{1993ispu.conf..104B}
{Becker}, W., {Tr{\"u}mper}, J., \& {{\"O}gelman}, H. 1993, in Isolated
  Pulsars, ed. K.~A. {van Riper}, R.~I. {Epstein}, \& C.~{Ho}, 104

\bibitem[{{Becker} {et~al.}(2004){Becker}, {Weisskopf}, {Tennant}, {Jessner},
  {Dyks}, {Harding}, \& {Zhang}}]{2004ApJ...615..908B}
{Becker}, W., {Weisskopf}, M.~C., {Tennant}, A.~F., {et~al.} 2004, \apj, 615,
  908

\bibitem[{{Becker} {et~al.}(2006){Becker}, {Kramer}, {Jessner}, {Taam}, {Jia},
  {Cheng}, {Mignani}, {Pellizzoni}, {de Luca}, {S{\l}owikowska}, \&
  {Caraveo}}]{2006ApJ...645.1421B}
{Becker}, W., {Kramer}, M., {Jessner}, A., {et~al.} 2006, \apj, 645, 1421

\bibitem[{{Boyles} {et~al.}(2011){Boyles}, {Lorimer}, {McLaughlin}, {Ransom},
  {Lynch}, {Kaspi}, {Archibald}, {Stairs}, {McPhee}, {Roberts}, {Kondratiev},
  {Hessels}, {van Leeuwen}, {Champion}, {Deller}, \&
  {Dunlap}}]{2011AIPC.1357...32B}
{Boyles}, J., {Lorimer}, D.~R., {McLaughlin}, M.~A., {et~al.} 2011, in American
  Institute of Physics Conference Series, Vol. 1357, American Institute of
  Physics Conference Series, ed. M.~{Burgay}, N.~{D'Amico}, P.~{Esposito},
  A.~{Pellizzoni}, \& A.~{Possenti}, 32--35

\bibitem[{{Boyles} {et~al.}(2012){Boyles}, {Lynch}, {Ransom}, {Stairs},
  {Lorimer}, {McLaughlin}, {Hessels}, {Kaspi}, {Kondratiev}, {Archibald},
  {Berndsen}, {Cardoso}, {Cherry}, {Epstein}, {Karako-Argaman}, {McPhee},
  {Pennucci}, {Roberts}, {Stovall}, \& {van Leeuwen}}]{2012arXiv1209.4293B}
{Boyles}, J., {Lynch}, R.~S., {Ransom}, S.~M., {et~al.} 2012, ArXiv e-prints,
  arXiv:1209.4293

\bibitem[{{Brisken} {et~al.}(2002){Brisken}, {Benson}, {Goss}, \&
  {Thorsett}}]{2002ApJ...571..906B}
{Brisken}, W.~F., {Benson}, J.~M., {Goss}, W.~M., \& {Thorsett}, S.~E. 2002,
  \apj, 571, 906

\bibitem[{{Brisken} {et~al.}(2003){Brisken}, {Thorsett}, {Golden}, \&
  {Goss}}]{2003ApJ...593L..89B}
{Brisken}, W.~F., {Thorsett}, S.~E., {Golden}, A., \& {Goss}, W.~M. 2003,
  \apjl, 593, L89

\bibitem[{{Burgay} {et~al.}(2013){Burgay}, {Bailes}, {Bates}, {Bhat},
  {Burke-Spolaor}, {Champion}, {Coster}, {D'Amico}, {Johnston}, {Keith},
  {Kramer}, {Levin}, {Lyne}, {Milia}, {Ng}, {Possenti}, {Stappers}, {Thornton},
  {Tiburzi}, {van Straten}, \& {Bassa}}]{2013MNRAS.433..259B}
{Burgay}, M., {Bailes}, M., {Bates}, S.~D., {et~al.} 2013, \mnras, 433, 259

\bibitem[{{Burwitz} {et~al.}(2003){Burwitz}, {Haberl}, {Neuh{\"a}user},
  {Predehl}, {Tr{\"u}mper}, \& {Zavlin}}]{2003A&A...399.1109B}
{Burwitz}, V., {Haberl}, F., {Neuh{\"a}user}, R., {et~al.} 2003, \aap, 399,
  1109

\bibitem[{{Camilo} {et~al.}(2002){Camilo}, {Manchester}, {Gaensler}, {Lorimer},
  \& {Sarkissian}}]{2002ApJ...567L..71C}
{Camilo}, F., {Manchester}, R.~N., {Gaensler}, B.~M., {Lorimer}, D.~R., \&
  {Sarkissian}, J. 2002, \apjl, 567, L71

\bibitem[{{Camilo} {et~al.}(2012){Camilo}, {Ransom}, {Chatterjee}, {Johnston},
  \& {Demorest}}]{2012ApJ...746...63C}
{Camilo}, F., {Ransom}, S.~M., {Chatterjee}, S., {Johnston}, S., \& {Demorest},
  P. 2012, \apj, 746, 63

\bibitem[{{Caraveo} {et~al.}(2001){Caraveo}, {De Luca}, {Mignani}, \&
  {Bignami}}]{2001ApJ...561..930C}
{Caraveo}, P.~A., {De Luca}, A., {Mignani}, R.~P., \& {Bignami}, G.~F. 2001,
  \apj, 561, 930

\bibitem[{{Chatterjee} {et~al.}(2004){Chatterjee}, {Cordes}, {Vlemmings},
  {Arzoumanian}, {Goss}, \& {Lazio}}]{2004ApJ...604..339C}
{Chatterjee}, S., {Cordes}, J.~M., {Vlemmings}, W.~H.~T., {et~al.} 2004, \apj,
  604, 339

\bibitem[{{Cognard} {et~al.}(2011){Cognard}, {Guillemot}, {Johnson}, {Smith},
  {Venter}, {Harding}, {Wolff}, {Cheung}, {Donato}, {Abdo}, {Ballet}, {Camilo},
  {Desvignes}, {Dumora}, {Ferrara}, {Freire}, {Grove}, {Johnston}, {Keith},
  {Kramer}, {Lyne}, {Michelson}, {Parent}, {Ransom}, {Ray}, {Romani}, {Saz
  Parkinson}, {Stappers}, {Theureau}, {Thompson}, {Weltevrede}, \&
  {Wood}}]{2011ApJ...732...47C}
{Cognard}, I., {Guillemot}, L., {Johnson}, T.~J., {et~al.} 2011, \apj, 732, 47

\bibitem[{{Cordes} \& {Lazio}(2002)}]{2002astro.ph..7156C}
{Cordes}, J.~M., \& {Lazio}, T.~J.~W. 2002, ArXiv Astrophysics e-prints,
  arXiv:astro-ph/0207156

\bibitem[{{Davies} {et~al.}(1972){Davies}, {Lyne}, \&
  {Seiradakis}}]{1972Natur.240..229D}
{Davies}, J.~G., {Lyne}, A.~G., \& {Seiradakis}, J.~H. 1972, \nat, 240, 229

\bibitem[{{de Jager} {et~al.}(1989){de Jager}, {Raubenheimer}, \&
  {Swanepoel}}]{1989A&A...221..180D}
{de Jager}, O.~C., {Raubenheimer}, B.~C., \& {Swanepoel}, J.~W.~H. 1989, \aap,
  221, 180

\bibitem[{{De Luca} {et~al.}(2005){De Luca}, {Caraveo}, {Mereghetti},
  {Negroni}, \& {Bignami}}]{2005ApJ...623.1051D}
{De Luca}, A., {Caraveo}, P.~A., {Mereghetti}, S., {Negroni}, M., \& {Bignami},
  G.~F. 2005, \apj, 623, 1051

\bibitem[{{Fesen} {et~al.}(2008){Fesen}, {Rudie}, {Hurford}, \&
  {Soto}}]{2008ApJS..174..379F}
{Fesen}, R., {Rudie}, G., {Hurford}, A., \& {Soto}, A. 2008, \apjs, 174, 379

\bibitem[{{Fesen} {et~al.}(2006){Fesen}, {Hammell}, {Morse}, {Chevalier},
  {Borkowski}, {Dopita}, {Gerardy}, {Lawrence}, {Raymond}, \& {van den
  Bergh}}]{2006ApJ...645..283F}
{Fesen}, R.~A., {Hammell}, M.~C., {Morse}, J., {et~al.} 2006, \apj, 645, 283

\bibitem[{{Gaensler} \& {Slane}(2006)}]{2006ARA&A..44...17G}
{Gaensler}, B.~M., \& {Slane}, P.~O. 2006, \araa, 44, 17

\bibitem[{{Gonzalez} {et~al.}(2004){Gonzalez}, {Kaspi}, {Lyne}, \&
  {Pivovaroff}}]{2004ApJ...610L..37G}
{Gonzalez}, M.~E., {Kaspi}, V.~M., {Lyne}, A.~G., \& {Pivovaroff}, M.~J. 2004,
  \apjl, 610, L37

\bibitem[{{Gotthelf}(2004)}]{2004Gotthelf}
{Gotthelf}, E.~V. 2004, in IAU Symposium, Vol. 218, Young Neutron Stars and
  Their Environments, ed. {F.~Camilo \& B.~M.~Gaensler}, 225

\bibitem[{{Green} \& {Stephenson}(2003)}]{2003LNP...598....7G}
{Green}, D.~A., \& {Stephenson}, F.~R. 2003, in Lecture Notes in Physics,
  Berlin Springer Verlag, Vol. 598, Supernovae and Gamma-Ray Bursters, ed.
  K.~{Weiler}, 7--19

\bibitem[{{Halpern} {et~al.}(2004){Halpern}, {Gotthelf}, {Camilo}, {Helfand},
  \& {Ransom}}]{2004ApJ...612..398H}
{Halpern}, J.~P., {Gotthelf}, E.~V., {Camilo}, F., {Helfand}, D.~J., \&
  {Ransom}, S.~M. 2004, \apj, 612, 398

\bibitem[{{Halpern} \& {Wang}(1997)}]{1997ApJ...477..905H}
{Halpern}, J.~P., \& {Wang}, F.~Y.-H. 1997, \apj, 477, 905

\bibitem[{{Hughes} {et~al.}(2003){Hughes}, {Slane}, {Park}, {Roming}, \&
  {Burrows}}]{2003ApJ...591L.139H}
{Hughes}, J.~P., {Slane}, P.~O., {Park}, S., {Roming}, P.~W.~A., \& {Burrows},
  D.~N. 2003, \apjl, 591, L139

\bibitem[{{Hui} \& {Becker}(2007)}]{2007A&A...467.1209H}
{Hui}, C.~Y., \& {Becker}, W. 2007, \aap, 467, 1209

\bibitem[{{Jacoby} {et~al.}(2007){Jacoby}, {Bailes}, {Ord}, {Knight}, \&
  {Hotan}}]{2007ApJ...656..408J}
{Jacoby}, B.~A., {Bailes}, M., {Ord}, S.~M., {Knight}, H.~S., \& {Hotan}, A.~W.
  2007, \apj, 656, 408

\bibitem[{{Kalberla} {et~al.}(2005){Kalberla}, {Burton}, {Hartmann}, {Arnal},
  {Bajaja}, {Morras}, \& {P{\"o}ppel}}]{2005A&A...440..775K}
{Kalberla}, P.~M.~W., {Burton}, W.~B., {Hartmann}, D., {et~al.} 2005, \aap,
  440, 775

\bibitem[{{Kaplan} {et~al.}(2009){Kaplan}, {Esposito}, {Chatterjee},
  {Possenti}, {McLaughlin}, {Camilo}, {Chakrabarty}, \&
  {Slane}}]{2009MNRAS.400.1445K}
{Kaplan}, D.~L., {Esposito}, P., {Chatterjee}, S., {et~al.} 2009, \mnras, 400,
  1445

\bibitem[{{Kaplan} {et~al.}(2004){Kaplan}, {Frail}, {Gaensler}, {Gotthelf},
  {Kulkarni}, {Slane}, \& {Nechita}}]{2004ApJS..153..269K}
{Kaplan}, D.~L., {Frail}, D.~A., {Gaensler}, B.~M., {et~al.} 2004, \apjs, 153,
  269

\bibitem[{{Kaplan} {et~al.}(2003){Kaplan}, {van Kerkwijk}, {Marshall},
  {Jacoby}, {Kulkarni}, \& {Frail}}]{2003ApJ...590.1008K}
{Kaplan}, D.~L., {van Kerkwijk}, M.~H., {Marshall}, H.~L., {et~al.} 2003, \apj,
  590, 1008

\bibitem[{{Kaplan} {et~al.}(2012){Kaplan}, {Stovall}, {Ransom}, {Roberts},
  {Kotulla}, {Archibald}, {Biwer}, {Boyles}, {Dartez}, {Day}, {Ford}, {Garcia},
  {Hessels}, {Jenet}, {Karako}, {Kaspi}, {Kondratiev}, {Lorimer}, {Lynch},
  {McLaughlin}, {Rohr}, {Siemens}, {Stairs}, \& {van
  Leeuwen}}]{2012ApJ...753..174K}
{Kaplan}, D.~L., {Stovall}, K., {Ransom}, S.~M., {et~al.} 2012, \apj, 753, 174

\bibitem[{{Kargaltsev} {et~al.}(2012){Kargaltsev}, {Durant}, {Pavlov}, \&
  {Garmire}}]{2012ApJS..201...37K}
{Kargaltsev}, O., {Durant}, M., {Pavlov}, G.~G., \& {Garmire}, G. 2012, \apjs,
  201, 37

\bibitem[{{Kargaltsev} \& {Pavlov}(2010)}]{2010AIPC.1248...25K}
{Kargaltsev}, O., \& {Pavlov}, G.~G. 2010, X-ray Astronomy 2009; Present
  Status, Multi-Wavelength Approach and Future Perspectives, 1248, 25

\bibitem[{{Kaspi} {et~al.}(1992){Kaspi}, {Manchester}, {Johnston}, {Lyne}, \&
  {D'Amico}}]{1992Kaspi}
{Kaspi}, V.~M., {Manchester}, R.~N., {Johnston}, S., {Lyne}, A.~G., \&
  {D'Amico}, N. 1992, \apjl, 399, L155

\bibitem[{{Kaur} {et~al.}(2010){Kaur}, {Wijnands}, {Paul}, {Patruno}, \&
  {Degenaar}}]{2010MNRAS.402.2388K}
{Kaur}, R., {Wijnands}, R., {Paul}, B., {Patruno}, A., \& {Degenaar}, N. 2010,
  \mnras, 402, 2388

\bibitem[{{Keane} {et~al.}(2013){Keane}, {McLaughlin}, {Kramer}, {Stappers},
  {Bassa}, {Purver}, \& {Weltevrede}}]{2013ApJ...764..180K}
{Keane}, E.~F., {McLaughlin}, M.~A., {Kramer}, M., {et~al.} 2013, \apj, 764,
  180

\bibitem[{{Kerr} {et~al.}(2012){Kerr}, {Camilo}, {Johnson}, {Ferrara},
  {Guillemot}, {Harding}, {Hessels}, {Johnston}, {Keith}, \& {et
  al.}}]{2012ApJ...748L...2K}
{Kerr}, M., {Camilo}, F., {Johnson}, T.~J., {et~al.} 2012, \apjl, 748, L2

\bibitem[{{Komesaroff} {et~al.}(1973){Komesaroff}, {Ables}, {Cooke},
  {Hamilton}, \& {McCulloch}}]{B1221Discovery}
{Komesaroff}, M.~M., {Ables}, J.~G., {Cooke}, D.~J., {Hamilton}, P.~A., \&
  {McCulloch}, P.~M. 1973, \aplett, 15, 169

\bibitem[{{Koribalski} {et~al.}(1995){Koribalski}, {Johnston}, {Weisberg}, \&
  {Wilson}}]{1995ApJ...441..756K}
{Koribalski}, B., {Johnston}, S., {Weisberg}, J.~M., \& {Wilson}, W. 1995,
  \apj, 441, 756

\bibitem[{{Kraft} {et~al.}(1991){Kraft}, {Burrows}, \&
  {Nousek}}]{1991ApJ...374..344K}
{Kraft}, R.~P., {Burrows}, D.~N., \& {Nousek}, J.~A. 1991, \apj, 374, 344

\bibitem[{{Kramer} {et~al.}(2003{\natexlab{a}}){Kramer}, {Lyne}, {Hobbs},
  {L{\"o}hmer}, {Carr}, {Jordan}, \& {Wolszczan}}]{2003ApJ...593L..31K}
{Kramer}, M., {Lyne}, A.~G., {Hobbs}, G., {et~al.} 2003{\natexlab{a}}, \apjl,
  593, L31

\bibitem[{{Kramer} {et~al.}(2006){Kramer}, {Lyne}, {O'Brien}, {Jordan}, \&
  {Lorimer}}]{2006Sci...312..549K}
{Kramer}, M., {Lyne}, A.~G., {O'Brien}, J.~T., {Jordan}, C.~A., \& {Lorimer},
  D.~R. 2006, Science, 312, 549

\bibitem[{{Kramer} {et~al.}(2003{\natexlab{b}}){Kramer}, {Bell}, {Manchester},
  {Lyne}, {Camilo}, {Stairs}, {D'Amico}, {Kaspi}, {Hobbs}, {Morris},
  {Crawford}, {Possenti}, {Joshi}, {McLaughlin}, {Lorimer}, \&
  {Faulkner}}]{2003MNRAS.342.1299K}
{Kramer}, M., {Bell}, J.~F., {Manchester}, R.~N., {et~al.} 2003{\natexlab{b}},
  \mnras, 342, 1299

\bibitem[{{Kuiper} \& {Hermsen}(2015)}]{2015MNRAS.449.3827K}
{Kuiper}, L., \& {Hermsen}, W. 2015, \mnras, 449, 3827

\bibitem[{{Kumar} {et~al.}(2012){Kumar}, {Safi-Harb}, \&
  {Gonzalez}}]{2012ApJ...754...96K}
{Kumar}, H.~S., {Safi-Harb}, S., \& {Gonzalez}, M.~E. 2012, \apj, 754, 96

\bibitem[{{Li} {et~al.}(2008){Li}, {Lu}, \& {Li}}]{2008ApJ...682.1166L}
{Li}, X.-H., {Lu}, F.-J., \& {Li}, Z. 2008, \apj, 682, 1166

\bibitem[{{Lorimer} {et~al.}(1996){Lorimer}, {Lyne}, {Bailes}, {Manchester},
  {D'Amico}, {Stappers}, {Johnston}, \& {Camilo}}]{1996MNRAS.283.1383L}
{Lorimer}, D.~R., {Lyne}, A.~G., {Bailes}, M., {et~al.} 1996, \mnras, 283, 1383

\bibitem[{{Lorimer} {et~al.}(2012){Lorimer}, {Lyne}, {McLaughlin}, {Kramer},
  {Pavlov}, \& {Chang}}]{2012ApJ...758..141L}
{Lorimer}, D.~R., {Lyne}, A.~G., {McLaughlin}, M.~A., {et~al.} 2012, \apj, 758,
  141

\bibitem[{{Lorimer} {et~al.}(1995){Lorimer}, {Nicastro}, {Lyne}, {Bailes},
  {Manchester}, {Johnston}, {Bell}, {D'Amico}, \&
  {Harrison}}]{1995ApJ...439..933L}
{Lorimer}, D.~R., {Nicastro}, L., {Lyne}, A.~G., {et~al.} 1995, \apj, 439, 933

\bibitem[{{Maccacaro} {et~al.}(1988){Maccacaro}, {Gioia}, {Wolter}, {Zamorani},
  \& {Stocke}}]{1988Maccacaro}
{Maccacaro}, T., {Gioia}, I.~M., {Wolter}, A., {Zamorani}, G., \& {Stocke},
  J.~T. 1988, \apj, 326, 680

\bibitem[{{Manchester} {et~al.}(1985){Manchester}, {Damico}, \&
  {Tuohy}}]{B1338Discovery}
{Manchester}, R.~N., {Damico}, N., \& {Tuohy}, I.~R. 1985, \mnras, 212, 975

\bibitem[{{Manchester} {et~al.}(2005){Manchester}, {Hobbs}, {Teoh}, \&
  {Hobbs}}]{2005AJ....129.1993M}
{Manchester}, R.~N., {Hobbs}, G.~B., {Teoh}, A., \& {Hobbs}, M. 2005, \aj, 129,
  1993

\bibitem[{{Manchester} {et~al.}(1978){Manchester}, {Lyne}, {Taylor}, {Durdin},
  {Large}, \& {Little}}]{B0919Discovery}
{Manchester}, R.~N., {Lyne}, A.~G., {Taylor}, J.~H., {et~al.} 1978, \mnras,
  185, 409

\bibitem[{{Manchester} {et~al.}(2001){Manchester}, {Lyne}, {Camilo}, {Bell},
  {Kaspi}, {D'Amico}, {McKay}, {Crawford}, {Stairs}, {Possenti}, {Kramer}, \&
  {Sheppard}}]{J1112J1301_6305Discovery}
{Manchester}, R.~N., {Lyne}, A.~G., {Camilo}, F., {et~al.} 2001, \mnras, 328,
  17

\bibitem[{{Marshall} \& {Schulz}(2002)}]{2002ApJ...574..377M}
{Marshall}, H.~L., \& {Schulz}, N.~S. 2002, \apj, 574, 377

\bibitem[{{McGowan} {et~al.}(2004){McGowan}, {Zane}, {Cropper}, {Kennea},
  {C{\'o}rdova}, {Ho}, {Sasseen}, \& {Vestrand}}]{2004ApJ...600..343M}
{McGowan}, K.~E., {Zane}, S., {Cropper}, M., {et~al.} 2004, \apj, 600, 343

\bibitem[{{McGowan} {et~al.}(2006){McGowan}, {Zane}, {Cropper}, {Vestrand}, \&
  {Ho}}]{2006ApJ...639..377M}
{McGowan}, K.~E., {Zane}, S., {Cropper}, M., {Vestrand}, W.~T., \& {Ho}, C.
  2006, \apj, 639, 377

\bibitem[{{Merloni} {et~al.}(2012){Merloni}, {Predehl}, {Becker},
  {B{\"o}hringer}, {Boller}, {Brunner}, {Brusa}, {Dennerl}, {Freyberg},
  {Friedrich}, {Georgakakis}, {Haberl}, {Hasinger}, {Meidinger}, {Mohr},
  {Nandra}, {Rau}, {Reiprich}, {Robrade}, {Salvato}, {Santangelo}, {Sasaki},
  {Schwope}, {Wilms}, \& t.~{German eROSITA Consortium}}]{2012arXiv1209.3114M}
{Merloni}, A., {Predehl}, P., {Becker}, W., {et~al.} 2012, ArXiv e-prints,
  arXiv:1209.3114

\bibitem[{{Mignani} {et~al.}(2013){Mignani}, {Vande Putte}, {Cropper},
  {Turolla}, {Zane}, {Pellizza}, {Bignone}, {Sartore}, \&
  {Treves}}]{2013MNRAS.429.3517M}
{Mignani}, R.~P., {Vande Putte}, D., {Cropper}, M., {et~al.} 2013, \mnras, 429,
  3517

\bibitem[{{Monet} {et~al.}(2003){Monet}, {Levine}, {Canzian}, {Ables}, {Bird},
  {Dahn}, {Guetter}, {Harris}, {Henden}, {Leggett}, {Levison}, {Luginbuhl},
  {Martini}, {Monet}, {Munn}, {Pier}, {Rhodes}, {Riepe}, {Sell}, {Stone},
  {Vrba}, {Walker}, {Westerhout}, {Brucato}, {Reid}, {Schoening}, {Hartley},
  {Read}, \& {Tritton}}]{2003AJ....125..984M}
{Monet}, D.~G., {Levine}, S.~E., {Canzian}, B., {et~al.} 2003, \aj, 125, 984

\bibitem[{{Morrison} \& {McCammon}(1983)}]{1983ApJ...270..119M}
{Morrison}, R., \& {McCammon}, D. 1983, \apj, 270, 119

\bibitem[{{Olausen} {et~al.}(2013){Olausen}, {Zhu}, {Vogel}, {Kaspi}, {Lyne},
  {Espinoza}, {Stappers}, {Manchester}, \& {McLaughlin}}]{2013ApJ...764....1O}
{Olausen}, S.~A., {Zhu}, W.~W., {Vogel}, J.~K., {et~al.} 2013, \apj, 764, 1

\bibitem[{{{\"O}zel}(2013)}]{2013RPPh...76a6901O}
{{\"O}zel}, F. 2013, Reports on Progress in Physics, 76, 016901

\bibitem[{{Page} {et~al.}(2004){Page}, {Lattimer}, {Prakash}, \&
  {Steiner}}]{2004ApJS..155..623P}
{Page}, D., {Lattimer}, J.~M., {Prakash}, M., \& {Steiner}, A.~W. 2004, \apjs,
  155, 623

\bibitem[{{Page} {et~al.}(2009){Page}, {Lattimer}, {Prakash}, \&
  {Steiner}}]{2009ApJ...707.1131P}
---. 2009, \apj, 707, 1131

\bibitem[{{Pavlov} {et~al.}(2001){Pavlov}, {Zavlin}, {Sanwal}, {Burwitz}, \&
  {Garmire}}]{2001ApJ...552L.129P}
{Pavlov}, G.~G., {Zavlin}, V.~E., {Sanwal}, D., {Burwitz}, V., \& {Garmire},
  G.~P. 2001, \apjl, 552, L129

\bibitem[{{Possenti} {et~al.}(2002){Possenti}, {Cerutti}, {Colpi}, \&
  {Mereghetti}}]{2002A&A...387..993P}
{Possenti}, A., {Cerutti}, R., {Colpi}, M., \& {Mereghetti}, S. 2002, \aap,
  387, 993

\bibitem[{{Ransom} {et~al.}(2014){Ransom}, {Stairs}, {Archibald}, {Hessels},
  {Kaplan}, {van Kerkwijk}, {Boyles}, {Deller}, {Chatterjee},
  {Schechtman-Rook}, {Berndsen}, {Lynch}, {Lorimer}, {Karako-Argaman}, {Kaspi},
  {Kondratiev}, {McLaughlin}, {van Leeuwen}, {Rosen}, {Roberts}, \&
  {Stovall}}]{2014Natur.505..520R}
{Ransom}, S.~M., {Stairs}, I.~H., {Archibald}, A.~M., {et~al.} 2014, \nat, 505,
  520

\bibitem[{{Roger} {et~al.}(1988){Roger}, {Milne}, {Kesteven}, {Wellington}, \&
  {Haynes}}]{1988ApJ...332..940R}
{Roger}, R.~S., {Milne}, D.~K., {Kesteven}, M.~J., {Wellington}, K.~J., \&
  {Haynes}, R.~F. 1988, \apj, 332, 940

\bibitem[{{Romani} \& {Ng}(2003)}]{2003ApJ...585L..41R}
{Romani}, R.~W., \& {Ng}, C.-Y. 2003, \apjl, 585, L41

\bibitem[{{Safi-Harb} \& {Kumar}(2008)}]{2008ApJ...684..532S}
{Safi-Harb}, S., \& {Kumar}, H.~S. 2008, \apj, 684, 532

\bibitem[{{Saz Parkinson} {et~al.}(2010){Saz Parkinson}, {Dormody}, {Ziegler},
  {Ray}, {Abdo}, {Ballet}, {Baring}, {Belfiore}, {Burnett}, {Caliandro},
  {Camilo}, {Caraveo}, {de Luca}, {Ferrara}, {Freire}, {Grove}, {Gwon},
  {Harding}, {Johnson}, {Johnson}, {Johnston}, {Keith}, {Kerr},
  {Kn{\"o}dlseder}, {Makeev}, {Marelli}, {Michelson}, {Parent}, {Ransom},
  {Reimer}, {Romani}, {Smith}, {Thompson}, {Watters}, {Weltevrede}, {Wolff}, \&
  {Wood}}]{2010ApJ...725..571S}
{Saz Parkinson}, P.~M., {Dormody}, M., {Ziegler}, M., {et~al.} 2010, \apj, 725,
  571

\bibitem[{{Schartel}(2012)}]{2012MmSAI..83...97S}
{Schartel}, N. 2012, \memsai, 83, 97

\bibitem[{{Seward} \& {Wang}(1988)}]{1988ApJ...332..199S}
{Seward}, F.~D., \& {Wang}, Z.-R. 1988, \apj, 332, 199

\bibitem[{{Shternin} {et~al.}(2011){Shternin}, {Yakovlev}, {Heinke}, {Ho}, \&
  {Patnaude}}]{2011MNRAS.412L.108S}
{Shternin}, P.~S., {Yakovlev}, D.~G., {Heinke}, C.~O., {Ho}, W.~C.~G., \&
  {Patnaude}, D.~J. 2011, \mnras, 412, L108

\bibitem[{{Slane} {et~al.}(2004{\natexlab{a}}){Slane}, {Helfand}, {van der
  Swaluw}, \& {Murray}}]{2004ApJ...616..403S}
{Slane}, P., {Helfand}, D.~J., {van der Swaluw}, E., \& {Murray}, S.~S.
  2004{\natexlab{a}}, \apj, 616, 403

\bibitem[{{Slane} {et~al.}(2004{\natexlab{b}}){Slane}, {Zimmerman}, {Hughes},
  {Seward}, {Gaensler}, \& {Clarke}}]{2004ApJ...601.1045S}
{Slane}, P., {Zimmerman}, E.~R., {Hughes}, J.~P., {et~al.} 2004{\natexlab{b}},
  \apj, 601, 1045

\bibitem[{{Smith} {et~al.}(2008){Smith}, {Guillemot}, {Camilo}, {Cognard},
  {Dumora}, {Espinoza}, {Freire}, {Gotthelf}, {Harding}, {Hobbs}, {Johnston},
  {Kaspi}, {Kramer}, {Livingstone}, {Lyne}, {Manchester}, {Marshall},
  {McLaughlin}, {Noutsos}, {Ransom}, {Roberts}, {Romani}, {Stappers},
  {Theureau}, {Thompson}, {Thorsett}, {Wang}, \& {Weltevrede}}]{2008Smith}
{Smith}, D.~A., {Guillemot}, L., {Camilo}, F., {et~al.} 2008, \aap, 492, 923

\bibitem[{{Spiewak} {et~al.}(2016){Spiewak}, {Kaplan}, {Archibald}, {Gentile},
  {Hessels}, {Lorimer}, {Lynch}, {McLaughlin}, {Ransom}, {Stairs}, \&
  {Stovall}}]{2016arXiv160200655S}
{Spiewak}, R., {Kaplan}, D.~L., {Archibald}, A., {et~al.} 2016, ArXiv e-prints,
  arXiv:1602.00655

\bibitem[{{Stokes} {et~al.}(1985){Stokes}, {Taylor}, {Welsberg}, \&
  {Dewey}}]{B0114Discovery}
{Stokes}, G.~H., {Taylor}, J.~H., {Welsberg}, J.~M., \& {Dewey}, R.~J. 1985,
  \nat, 317, 787

\bibitem[{{Tananbaum} {et~al.}(2014){Tananbaum}, {Weisskopf}, {Tucker},
  {Wilkes}, \& {Edmonds}}]{2014RPPh...77f6902T}
{Tananbaum}, H., {Weisskopf}, M.~C., {Tucker}, W., {Wilkes}, B., \& {Edmonds},
  P. 2014, Reports on Progress in Physics, 77, 066902

\bibitem[{{Taylor} \& {Cordes}(1993)}]{1993ApJ...411..674T}
{Taylor}, J.~H., \& {Cordes}, J.~M. 1993, \apj, 411, 674

\bibitem[{{Tetzlaff} {et~al.}(2011){Tetzlaff}, {Eisenbeiss}, {Neuh{\"a}user},
  \& {Hohle}}]{2011MNRAS.417..617T}
{Tetzlaff}, N., {Eisenbeiss}, T., {Neuh{\"a}user}, R., \& {Hohle}, M.~M. 2011,
  \mnras, 417, 617

\bibitem[{{Townsley} {et~al.}(2011){Townsley}, {Broos}, {Chu}, {Gruendl},
  {Oey}, \& {Pittard}}]{2011ApJS..194...16T}
{Townsley}, L.~K., {Broos}, P.~S., {Chu}, Y.-H., {et~al.} 2011, \apjs, 194, 16

\bibitem[{{Tsuruta}(1998)}]{1998PhR...292....1T}
{Tsuruta}, S. 1998, \physrep, 292, 1

\bibitem[{{Tsuruta}(2009)}]{2009ASSL..357..289T}
{Tsuruta}, S. 2009, in Astrophysics and Space Science Library, Vol. 357,
  Astrophysics and Space Science Library, ed. W.~{Becker}, 289

\bibitem[{{Verbiest} {et~al.}(2012){Verbiest}, {Weisberg}, {Chael}, {Lee}, \&
  {Lorimer}}]{2012ApJ...755...39V}
{Verbiest}, J.~P.~W., {Weisberg}, J.~M., {Chael}, A.~A., {Lee}, K.~J., \&
  {Lorimer}, D.~R. 2012, \apj, 755, 39

\bibitem[{{Wang} {et~al.}(2000){Wang}, {Manchester}, {Pace}, {Bailes}, {Kaspi},
  {Stappers}, \& {Lyne}}]{2000Wang}
{Wang}, N., {Manchester}, R.~N., {Pace}, R.~T., {et~al.} 2000, \mnras, 317, 843

\bibitem[{{Wang} \& {Zhao}(2004)}]{2004ApJ...601.1038W}
{Wang}, W., \& {Zhao}, Y. 2004, \apj, 601, 1038

\bibitem[{{Weisskopf} {et~al.}(2011){Weisskopf}, {Tennant}, {Yakovlev},
  {Harding}, {Zavlin}, {O'Dell}, {Elsner}, \& {Becker}}]{2011ApJ...743..139W}
{Weisskopf}, M.~C., {Tennant}, A.~F., {Yakovlev}, D.~G., {et~al.} 2011, \apj,
  743, 139

\bibitem[{{Xmm-Newton Survey Science Centre}(2013)}]{2013yCat.9044....0X}
{Xmm-Newton Survey Science Centre}, C. 2013, VizieR Online Data Catalog, 9044,
  0

\bibitem[{{Yakovlev} \& {Pethick}(2004)}]{2004ARA&A..42..169Y}
{Yakovlev}, D.~G., \& {Pethick}, C.~J. 2004, \araa, 42, 169

\bibitem[{{Yar-Uyaniker} {et~al.}(2004){Yar-Uyaniker}, {Uyaniker}, \&
  {Kothes}}]{2004ApJ...616..247Y}
{Yar-Uyaniker}, A., {Uyaniker}, B., \& {Kothes}, R. 2004, \apj, 616, 247

\bibitem[{{Zavlin} \& {Pavlov}(2004{\natexlab{a}})}]{2004ApJ...616..452Z}
{Zavlin}, V.~E., \& {Pavlov}, G.~G. 2004{\natexlab{a}}, \apj, 616, 452

\bibitem[{{Zavlin} \& {Pavlov}(2004{\natexlab{b}})}]{2004MmSAI..75..458Z}
---. 2004{\natexlab{b}}, \memsai, 75, 458

\bibitem[{{Zavlin} {et~al.}(1998){Zavlin}, {Pavlov}, \&
  {Trumper}}]{1998A&A...331..821Z}
{Zavlin}, V.~E., {Pavlov}, G.~G., \& {Trumper}, J. 1998, \aap, 331, 821

\bibitem[{{Zavlin} {et~al.}(1999){Zavlin}, {Tr{\"u}mper}, \&
  {Pavlov}}]{1999ApJ...525..959Z}
{Zavlin}, V.~E., {Tr{\"u}mper}, J., \& {Pavlov}, G.~G. 1999, \apj, 525, 959

\bibitem[{{Zhu} {et~al.}(2011){Zhu}, {Kaspi}, {McLaughlin}, {Pavlov}, {Ng},
  {Manchester}, {Gaensler}, \& {Woods}}]{2011ApJ...734...44Z}
{Zhu}, W.~W., {Kaspi}, V.~M., {McLaughlin}, M.~A., {et~al.} 2011, \apj, 734, 44

\end{thebibliography}

\onecolumn

\setlength{\smallskipamount}{0.8mm}
\scriptsize
\begin{longtable}{l c c c c c c c c c l}
\noalign{\medskip}
\caption{Used observations.}\label{tab1:used_observations} \\
\hline\hline \noalign{\smallskip} \multicolumn{1}{c}{PSR} & \multicolumn{1}{c}{satellite$^{(\alpha)}$} & \multicolumn{1}{c}{obsID} & \multicolumn{1}{c}{date} & \multicolumn{1}{c}{instrument$^{(\beta)}$} & \multicolumn{1}{c}{filter} & \multicolumn{1}{c}{$t_\text{on-axis}$} & \multicolumn{1}{c}{$t(\theta)$} & \multicolumn{1}{c}{$\theta^{(\gamma)}$} & \multicolumn{1}{c}{count rate$^{(\delta)}$} & \multicolumn{1}{c}{flag$^{(\delta)}$} \\ \noalign{\smallskip}
\multicolumn{1}{c}{ } & \multicolumn{1}{c}{ } & \multicolumn{1}{c}{ } & \multicolumn{1}{c}{} & \multicolumn{1}{c}{} & \multicolumn{1}{c}{} & \multicolumn{1}{c}{ksec} & \multicolumn{1}{c}{ksec} & \multicolumn{1}{c}{arcmin} & \multicolumn{1}{c}{cts/ksec} & \multicolumn{1}{c}{} \\ \noalign{\smallskip} \hline \noalign{\smallskip}
\endfirsthead
\multicolumn{11}{c}%
{{\bfseries \tablename\ \thetable{} -- continued from previous page}} \\
\hline\hline \noalign{\smallskip} \multicolumn{1}{c}{PSR} & \multicolumn{1}{c}{satellite$^{(\alpha)}$} & \multicolumn{1}{c}{obsID} & \multicolumn{1}{c}{date} & \multicolumn{1}{c}{instrument$^{(\beta)}$} & \multicolumn{1}{c}{filter} & \multicolumn{1}{c}{$t_\text{on-axis}$} & \multicolumn{1}{c}{$t(\theta)$} & \multicolumn{1}{c}{$\theta^{(\gamma)}$} & \multicolumn{1}{c}{count rate$^{(\delta)}$} & \multicolumn{1}{c}{flag$^{(\delta)}$} \\ \noalign{\smallskip} \hline \noalign{\smallskip} 
\endhead
\noalign{\smallskip} \hline \multicolumn{11}{|r|}{{Continued on next page}} \\ \hline
\endfoot
\hline
\noalign{\smallskip}
\noalign{\smallskip}
\multicolumn{11}{p{0.8\textwidth}}{{\bf Notes.}\hspace{3mm}$^{(\alpha)}$ C:Chandra X:XMM-Newton;
$^{(\beta)}$ in the case of XMM-Newton EPIC-PN observations the     used observation mode is added (EFF: Extended Full Frame; FF: Full Frame; LW: Large Window; SW: Small Window)
$^{(\gamma)}$ off-axis angle;
$^{(\delta)}$ If flagged with {\it ul} than 2$\sigma$ upper limit on count rate, otherwise see text.}
\endlastfoot
J0051+0423 & C & \dataset[ADS/Sa.CXO#obs/15968]{15968} & 2013-12-22 & ACIS-S & - & 3.6 & 3.5 & 0.2 & 0.83 & ul \\
J0101--6422 & C & \dataset[ADS/Sa.CXO#obs/15363]{15363} & 2013-02-17 & ACIS-I & - & 9.8 & 9.0 & 0.7 & - & d \\
J0117+5914 & X & 0112200201 & 2002-07-09 & PN/FF & Medium & 5.8 & 5.1 & 1.7 & - & d \\
 & X & 0112200201 & 2002-07-09 & MOS1 & Medium & 7.9 & 7.7 & 1.7 & - & d \\
 & X & 0112200201 & 2002-07-09 & MOS2 & Medium & 8.1 & 7.9 & 1.7 & - & d \\
J0134--2937 & X & 0654800501 & 2010-12-30 & PN/FF & Medium & 30.4 & 11.8 & 11.4 & 3.44 & ul \\
 & X & 0654800501 & 2010-12-30 & MOS1 & Medium & 39.0 & 2.1 & 11.4 & 1.44 & ul \\
 & X & 0654800501 & 2010-12-30 & MOS2 & Medium & 39.5 & 19.1 & 11.4 & 1.55 & ul \\
\end{longtable}
\normalsize

\setlength{\smallskipamount}{0.8mm}
\begin{deluxetable}{l l l c c c c}
\tabletypesize{\scriptsize}
\tablecolumns{7}
\tablewidth{0pc}
\tablecaption{Pulsars and coordinates of their potential X-ray counterparts along with their
 positional discrepancy, chance probability for a false detection, signal-to-noise and X-ray to visual flux
 ratio.\label{tab2:detected_psrs_results}}
\tablehead{
\colhead{PSR} & \colhead{RA} & \colhead{DEC} & \colhead{$\Delta$} & \colhead{S/N} & \colhead{$P_\text{coin}$} & $\log(f_X/f_V)$ \\
\colhead{} & \colhead{hh:mm:ss} & \colhead{dd:mm:ss} & \colhead{arcsec} & \colhead{$\sigma_\text{G}^{(\alpha)}$} & \colhead{} &  }
\startdata
J0101--6422 & $01:01:11.11\pm 0.06$ & $-64:22:30.4\pm 0.4$ & $0.2\pm 0.4$ & 15.9 & $5.5\times 10^{-6}$ & $>0.2$ \\
J0117+5914 & $01:17:38.5\pm 0.3$ & $+59:14:38\pm 2$ & $1.6\pm 2.1$ & 16.9 & $1.3\times 10^{-4}$ & $>0.2$ \\
J0337+1715 & $03:37:43.7\pm 0.1$ & $+17:15:13\pm 2$ & $1.7\pm 2.1$ & 33.0 & $3.0\times 10^{-4}$ & $>-1.1$ \\
J0543+2329 & $05:43:09.5\pm 0.2$ & $+23:29:07\pm 2$ & $3.4\pm 4.8$ & 5.9 & $2.8\times 10^{-4}$ & $>-0.4$ \\
J0922+0638 & $09:22:14.16\pm 0.14$ & $+06:38:24\pm 2$ & $2.4\pm 2.1$ & 17.8 & $6.3\times 10^{-4}$ & $>-0.7$ \\
J1044--5737 & $10:44:32.87\pm 0.05$ & $-57:37:19.1\pm 0.4$ & $0.6\pm 0.9$ & 12.6 & $5.9\times 10^{-5}$ & $>0.1$ \\
J1112--6103 & $11:12:14.87\pm 0.03$ & $-61:03:30.7\pm 0.2$ & $0.6\pm 0.5$ & 43.0 & $4.1\times 10^{-4}$ & $>0.0$ \\
J1224--6407 & $12:24:21.8\pm 0.3$ & $-64:07:55\pm 2$ & $2.9\pm 2.3$ & 6.1 & $2.3\times 10^{-4}$ & $>-0.3$ \\
\enddata
\begin{scriptsize}
\tablenotetext{\alpha}{$\sigma_G=1+\sqrt{c_\text{bg}+0.75}$, where $c_\text{bg}$ are the background counts.}
\end{scriptsize}
\end{deluxetable}

\setlength{\smallskipamount}{0.8mm}
\begin{deluxetable}{l c c c c c c}
\tabletypesize{\scriptsize}
\tablecolumns{7}
\tablewidth{0pc}
\tablecaption{Spectral parameters of model spectra fitted to potential pulsar counterparts
which were detected with sufficient photon statistics to allow for a brief
spectral analysis.\label{tab3:spectral_psr_results}}
\tablehead{
\colhead{PSR} & \colhead{model} & \colhead{$\chi^2$/d.o.f.} & \colhead{$N_\text{H}$} & \colhead{$\Gamma / T_\text{emit}$} & \colhead{Norm\tablenotemark{\beta}} & $R_\text{emit}$ \\
\colhead{} & \colhead{} & \colhead{} & \colhead{$10^{22}$ cm$^{-2}$} & \colhead{-/keV} & \colhead{} & km }
\startdata
J0117+5914 & BB\tablenotemark{\alpha} & 12.2/13 & $<0.48$ & $0.18_{-0.03}^{+0.04}$ & $3.7^{+1.3}_{-1.1} \times 10^{-7}$ & $0.4\pm0.2$ \\
\noalign{\smallskip}
  {}       & PL\tablenotemark{\alpha} & 13.2/13 & $0.5_{-0.5}^{+0.7}$ & $3.3\pm0.5$ & $(8\pm2) \times 10^{-6}$ & - \\
\noalign{\smallskip}
J0922+0638 & PL & 4.0/8 & $<0.12$ & $2.3_{-0.4}^{+0.8}$ & $1.9^{+0.9}_{-0.4} \times 10^{-6}$ & - \\
\noalign{\smallskip}
J1301-6310 & BB\tablenotemark{\alpha} & 23.8/16 & $<0.09$ & $0.25_{-0.07}^{+0.08}$ & $9^{+4}_{-2} \times 10^{-8}$ & $0.08_{-0.03}^{+0.04}$ \\
\noalign{\smallskip}
 {}        & PL\tablenotemark{\alpha} & 16.4/16 & $<0.39$ & $3.4_{-0.6}^{+0.8}$ & $(3.2\pm0.7) \times 10^{-6}$ & - \\
\noalign{\smallskip}
J1341-6220 & BB\tablenotemark{\alpha} & 12.8/23 &  $<2.2$  & $1.2_{-0.4}^{+1.6}$  & $3^{+4}_{-2} \times 10^{-7}$ & $0.04_{-0.02}^{+0.06}$ \\
 {}        & PL\tablenotemark{\alpha} & 13.6/23 & $4\pm3$ & $1.1\pm0.7$ & $1.5^{+2.1}_{-1.0} \times 10^{-6}$ & - \\
\noalign{\smallskip}
J1730-2304 & BB\tablenotemark{\alpha} & 12.2/18 & $<0.53$ & $0.26_{-0.21}^{+0.06}$ & $1.5^{+0.4}_{-0.7} \times 10^{-7}$ & $0.026_{-0.023}^{+0.009}$ \\
\enddata
\begin{scriptsize}
\tablenotetext{\alpha}{Spectral parameters were obtained with $N_\text{H}$ fixed to DM based value and we added the result of $N_\text{H}$ before fixing.}
\tablenotetext{\beta}{PL: photons keV$^{-1}$ cm$^{-2}$ s$^{-1}$ at 1 keV; BB: $L[10^{39} \text{erg/s}]/(D[10 \text{kpc}])^2$, where $L$ is the source luminosity and $D$ the distance to the source.}
\end{scriptsize}
\end{deluxetable}

6.43873E-07,-1.10052e-07,1.29864e-07

\setlength{\smallskipamount}{0.8mm}
\begin{deluxetable}{l c c c c c c c c c c}
\tabletypesize{\scriptsize}
\tablecolumns{11}
\tablewidth{0pc}
\tablecaption{Essential radio pulsar parameters along with the flux, luminosity and X-ray efficiency of the potential 
X-ray counterparts for thermal (BB) and/or non-thermal (PL) emission.\label{tab4:detected_psrs_list}}
\tablehead{
\colhead{PSR} & \colhead{DM} & \colhead{P} & \colhead{$\log(\text{Age})$} & \colhead{$\log(\dot E)$} & \colhead{$n_\text{H}$} & \colhead{d} & \colhead{model} & \colhead{$f_\text{0.1--2 keV}$} & \colhead{$L_\text{0.1--2 keV}$} & $\mu_\text{0.1--2 keV}$ \\
\colhead{} & \colhead{pc cm$^{-3}$} & \colhead{ms} & \colhead{yrs} & \colhead{erg s$^{-1}$} & \colhead{$10^{21}$ cm$^{-2}$} & \colhead{kpc} & \colhead{} & \colhead{erg s$^{-1}$ cm$^{-2}$} & \colhead{erg s$^{-1}$} &  }
\startdata
J0117+5914 & 49.423 & 101 & 5.4 & 35.3 & 1.5 & 2.2 & BB & $(3.1\pm0.6) \times 10^{-14}$ & $(1.8\pm0.4) \times 10^{31}$ & $(8\pm2) \times 10^{-5}$ \\
\noalign{\smallskip}
 &  &  &  &  &  &  & PL & $1.8^{+1.4}_{-0.8} \times 10^{-13}$ & $1.1^{+0.9}_{-0.4} \times 10^{32}$ & $(5\pm2) \times 10^{-4}$ \\
\noalign{\smallskip}
J0337+1715 & - & 2 & 9.4 & 34.5 & 0.7 & 1.4 & PL\tablenotemark{\alpha} & $(2.17\pm0.15) \times 10^{-14}$ & $5.1^{+0.3}_{-0.4} \times 10^{30}$ & $(1.49\pm0.10) \times 10^{-4}$ \\
\noalign{\smallskip}
J0543+2329 & 77.7115 & 245 & 5.4 & 34.6 & 2.4 & 2.1 & PL\tablenotemark{\alpha} & $(8\pm2) \times 10^{-15}$ & $(4.1\pm1.1) \times 10^{30}$ & $(10\pm3) \times 10^{-5}$ \\
\noalign{\smallskip}
J0922+0638 & 27.271 & 430 & 5.7 & 33.8 & 0.8 & 1.1 & PL & $1.2^{+1.4}_{-0.6} \times 10^{-14}$ & $1.7^{+2.0}_{-0.9} \times 10^{30}$ & $(2.5\pm1.4) \times 10^{-4}$ \\
\noalign{\smallskip}
J1224-6407 & 97.47 & 216 & 5.8 & 34.3 & 3.0 & 4.0 & PL\tablenotemark{\alpha} & $(10\pm3) \times 10^{-15}$ & $(1.9\pm0.5) \times 10^{31}$ & $(10\pm3) \times 10^{-4}$ \\
\noalign{\smallskip}
J1301-6310 & 86.1 & 663 & 5.3 & 33.9 & 2.7 & 1.9 & BB & $(7.4\pm1.4) \times 10^{-15}$ & $(3.0\pm0.6) \times 10^{30}$ & $(4.0\pm0.8) \times 10^{-4}$ \\
\noalign{\smallskip}
 &  &  &  &  &  &  & PL & $9^{+7}_{-4} \times 10^{-14}$ & $3^{+3}_{-2} \times 10^{31}$ & $(5\pm2) \times 10^{-3}$ \\
\enddata
\begin{scriptsize}
\tablenotetext{\alpha}{Flux derived using PIMMS with a power law model and a photon index of 1.7.}
\end{scriptsize}
\end{deluxetable}

\setlength{\smallskipamount}{0.8mm}
\scriptsize
\begin{longtable}{l c c c c c c c c c c}
\noalign{\medskip}
\caption{Pulsar parameters, flux upper limits and X-ray efficiency upper limits 
for pulsars for which no X-ray counterpart was detected in the XMM-Newton and Chandra 
archival data by June 1$^{st}$ 2016.\label{tab5:upper_limits_results}} \\
\hline\hline \noalign{\smallskip} \multicolumn{1}{c}{PSR} & \multicolumn{1}{c}{DM} & \multicolumn{1}{c}{P} & \multicolumn{1}{c}{$\log(\text{Age})$} & \multicolumn{1}{c}{$\log(\dot E)$} & \multicolumn{1}{c}{$N_\text{H}/N_\text{H}^\text{LAB}$} & \multicolumn{1}{c}{d} & \multicolumn{1}{c}{$f_\text{0.1--2 keV}^\text{PL}$} & \multicolumn{1}{c}{$\log(L_\text{0.1--2 keV}^\text{PL})$} & \multicolumn{1}{c}{$\mu_\text{0.1--2 keV}^\text{PL}$} & \multicolumn{1}{c}{$\log(T_\infty)$} \\ \noalign{\smallskip}
\multicolumn{1}{c}{} & \multicolumn{1}{c}{pc cm$^{-3}$} & \multicolumn{1}{c}{ms} & \multicolumn{1}{c}{yrs} & \multicolumn{1}{c}{erg s$^{-1}$} & \multicolumn{1}{c}{$10^{21}$ cm$^{-2}$} & \multicolumn{1}{c}{kpc} & \multicolumn{1}{c}{erg s$^{-1}$ cm$^{-2}$} & \multicolumn{1}{c}{erg s$^{-1}$} & \multicolumn{1}{c}{} & \multicolumn{1}{c}{K} \\ \noalign{\smallskip} \hline \noalign{\smallskip}
\endfirsthead
\multicolumn{11}{c}%
{{\bfseries \tablename\ \thetable{} -- continued from previous page}} \\
\hline\hline \noalign{\smallskip} \multicolumn{1}{c}{PSR} & \multicolumn{1}{c}{DM} & \multicolumn{1}{c}{P} & \multicolumn{1}{c}{$\log(\text{Age})$} & \multicolumn{1}{c}{$\log(\dot E)$} & \multicolumn{1}{c}{$N_\text{H}/N_\text{H}^\text{LAB}$} & \multicolumn{1}{c}{d} & \multicolumn{1}{c}{$f_\text{0.1--2 keV}^\text{PL}$} & \multicolumn{1}{c}{$\log(L_\text{0.1--2 keV}^\text{PL})$} & \multicolumn{1}{c}{$\mu_\text{0.1--2 keV}^\text{PL}$} & \multicolumn{1}{c}{$\log(T_\infty)$} \\ \noalign{\smallskip} \hline \noalign{\smallskip} 
\endhead
\noalign{\smallskip} \hline \multicolumn{11}{|r|}{{Continued on next page}} \\ \hline
\endfoot
\hline
\noalign{\smallskip}
\noalign{\smallskip}
\multicolumn{11}{p{0.8\textwidth}}{{\bf References.}\hspace{3mm}[1] \citet{2012ApJ...755...39V}; [2] \citet{2002astro.ph..7156C}; [3] \citet{1993ApJ...411..674T}.}
\endlastfoot
J0051+0423 & 13.9 & 354 & 8.9 & 30.8 & 0.4/7.5 & 0.6 [2] & $4.6\times 10^{-15}$ & 29.6 & $<6.6\times 10^{-2}$ & 5.61 \\
J0134--2937 & 21.806 & 136 & 7.4 & 33.1 & 0.7/5.0 & 0.6 [2] & $4.0\times 10^{-15}$ & 29.5 & $<2.5\times 10^{-4}$ & 5.51 \\
J0137+1654 & 26.6 & 414 & 8.7 & 30.8 & 0.8/0.5 & 1.2 [2] & $4.8\times 10^{-15}$ & 30.2 & $<2.5\times 10^{-1}$ & 5.70 \\
J0139+5814 & 73.779 & 272 & 5.6 & 34.3 & 2.3/0.5 & 2.6 [1] & $7.8\times 10^{-15}$ & 30.9 & $<3.8\times 10^{-4}$ & 5.85 \\
J0157+6212 & 30.21 & 2351 & 5.3 & 32.8 & 0.9/8.2 & 1.7 [2] & $5.3\times 10^{-15}$ & 30.6 & $<6.4\times 10^{-3}$ & 5.74 \\
J0248+6021 & 370 & 217 & 4.8 & 35.3 & 11.4/7.5 & 23.2 [3] & $2.0\times 10^{-14}$ & 33.5 & $<1.4\times 10^{-2}$ & 6.29 \\
J0335+4555 & 47.153 & 269 & 8.8 & 31.2 & 1.5/4.2 & 1.6 [2] & $6.2\times 10^{-15}$ & 30.7 & $<3.4\times 10^{-1}$ & 5.85 \\
J0520--2553 & 33.77 & 241 & 8.1 & 31.9 & 1.0/4.4 & 1.7 [2] & $5.3\times 10^{-15}$ & 30.6 & $<4.5\times 10^{-2}$ & 5.81 \\
\end{longtable}
\normalsize

\setlength{\smallskipamount}{0.8mm}
\begin{deluxetable}{l c c c c c c}
\tabletypesize{\scriptsize}
\tablecolumns{7}
\tablewidth{0pc}
\tablecaption{The pulsars with an X-ray efficiency upper limit below $5\times 10^{-5}$.\label{tab6:psr_with_lowest_muX}}
\tablehead{
\colhead{PSR} & \colhead{$\frac{\dot E}{4\pi d^2}$} & \colhead{$\log(\dot E)$} & \colhead{$\mu^\text{PL}_\text{0.1--2 keV}$} & \colhead{$d$} & \colhead{$\log(\tau)$} & $B_\text{surf}$ \\
\colhead{} & \colhead{erg s$^{-1}$ cm$^{-2}$} & \colhead{erg s$^{-1}$} & \colhead{} & \colhead{kpc} & \colhead{yrs} & G }
\startdata
J0940--5428 & $1.9\times 10^{-09}$ & 36.3 & $2.2\times 10^{-6}$ & 3.0 & 4.6 & $1.7\times 10^{+12}$ \\
\noalign{\smallskip}
J1300+1240 & $4.4\times 10^{-10}$ & 34.3 & $3.5\times 10^{-6}$ & 0.6 & 8.9 & $8.5\times 10^{+08}$ \\
\noalign{\smallskip}
J1105--6107 & $8.3\times 10^{-10}$ & 36.4 & $6.0\times 10^{-6}$ & 5.0 & 4.8 & $1.0\times 10^{+12}$ \\
\noalign{\smallskip}
J1913+1011 & $1.1\times 10^{-09}$ & 36.5 & $6.9\times 10^{-6}$ & 4.8 & 5.2 & $3.5\times 10^{+11}$ \\
\noalign{\smallskip}
J1730--3350 & $8.2\times 10^{-10}$ & 36.1 & $9.7\times 10^{-6}$ & 3.5 & 4.4 & $3.5\times 10^{+12}$ \\
\noalign{\smallskip}
J1906+0746 & $7.7\times 10^{-11}$ & 35.4 & $1.6\times 10^{-5}$ & 5.4 & 5.1 & $1.7\times 10^{+12}$ \\
\noalign{\smallskip}
J2129--5721 & $8.3\times 10^{-10}$ & 34.2 & $2.1\times 10^{-5}$ & 0.4 & 9.5 & $2.8\times 10^{+08}$ \\
\noalign{\smallskip}
J1928+1746 & $3.9\times 10^{-10}$ & 36.2 & $2.1\times 10^{-5}$ & 5.8 & 4.9 & $9.6\times 10^{+11}$ \\
\noalign{\smallskip}
J1835--1106 & $1.9\times 10^{-10}$ & 35.3 & $3.5\times 10^{-5}$ & 2.8 & 5.1 & $1.9\times 10^{+12}$ \\
\noalign{\smallskip}
J1702--4310 & $2.0\times 10^{-10}$ & 35.8 & $4.5\times 10^{-5}$ & 5.2 & 4.2 & $7.4\times 10^{+12}$ \\
\noalign{\smallskip}
\enddata
\end{deluxetable}

\setlength{\smallskipamount}{0.8mm}
\begin{deluxetable}{l c c}
\tabletypesize{\scriptsize}
\tablecolumns{7}
\tablewidth{0pc}
\tablecaption{The pulsars with the lowest $3\sigma$ temperature upper limit (indicated 
by a black arrow in the upper panel of Figure \ref{figure:Tau_Te}). {\it Top:} Young 
neutron stars with $\tau \le 5\times 10^4$ yrs. {\it Bottom:} Neutron stars with $\tau > 5\times 10^4$
yrs.\label{tab7:psr_with_lowest_temp}}
\tablehead{
\colhead{PSR} & \colhead{$\log(\text{age})$} & \colhead{$T_\infty$} \\
\colhead{} & \colhead{yrs} & \colhead{$10^5$ K} }
\startdata
J0940--5428 & 4.63 & 7.3 \\
J1524--5706 & 4.70 & 9.8 \\
J1730--3350 & 4.42 & 11.0 \\
J1637--4642 & 4.62 & 12.1 \\
J1907+0918 & 4.58 & 12.1 \\
J1702--4310 & 4.23 & 12.3 \\
J1413--6141 & 4.13 & 12.6 \\
J1841--0524 & 4.48 & 12.8 \\
J1726--3530 & 4.16 & 14.1 \\
J1837--0604 & 4.53 & 15.2 \\
J1646--4346 & 4.51 & 15.6 \\
\hline \noalign{\smallskip}
J0157+6212 & 5.29 & 5.5 \\
J1105--6107 & 4.80 & 5.7 \\
J1457--5902 & 5.70 & 5.8 \\
J1846--0257 & 5.65 & 6.0 \\
J0742--2822 & 5.20 & 7.8 \\
J1906+0746 & 5.05 & 8.4 \\
J1913+0446 & 4.96 & 9.2 \\
J1801--2304 & 4.77 & 9.6 \\
J1850--0026 & 4.83 & 10.6 \\
J1928+1746 & 4.92 & 10.9 \\
J1055--6028 & 4.73 & 11.1 \\
J1828--1101 & 4.89 & 11.1 \\
J1841--0345 & 4.75 & 11.3 \\
J1806--2125 & 4.80 & 11.7 \\
\enddata
\end{deluxetable}

\end{document}